\documentclass[structabstract]{aa}  
\usepackage{graphicx}
\usepackage{subcaption}
\usepackage{color}
\usepackage{url}
\usepackage{amssymb}
\usepackage{amsmath}
\usepackage{txfonts}
\usepackage{multirow}
\usepackage{lscape}
\usepackage[flushleft]{threeparttable}
\usepackage{float}
\usepackage{natbib}
\usepackage{stmaryrd}
\usepackage{supertabular}
\usepackage{rotating}
\usepackage{longtable}
\usepackage{chngcntr}
\usepackage{changes}
\usepackage{pdflscape}
\usepackage{booktabs}
\usepackage{dcolumn}
\usepackage{gensymb}
\usepackage{bm}

\bibpunct{(}{)}{;}{a}{}{,} 

\makeatletter
\def\hlinewd#1{%
\noalign{\ifnum0=`}\fi\hrule \@height #1 %
\futurelet\reserved@a\@xhline}
\makeatother

\newfont{\gwpfont}{cmssq8 scaled 1000}

\begin{document}
   \title{From the core to the outskirts: structure analysis of three massive galaxy clusters
   \thanks{Based on observations from the Very Large Telescope at Paranal, Chile}
   \thanks{The full Table \ref{table:app} is only available in electronic form at the CDS via anonymous ftp to {\ttfamily cdsarc.u-strasbg.fr (130.79.128.5)} or via http://cdsweb.u-strasbg.fr/cgi-bin/qcat?J/A+A/}
   }

   \author{
          G. Fo\"ex\inst{1}
          \and
          G. Chon\inst{1}
          \and
          H. B\"ohringer\inst{1}
          }
   \institute{
          Max Planck Institute for Extraterrestrial Physics, Giessenbachstrasse, 85748 Garching, Germany
             }

   \date{Received ; accepted }

  \abstract  
 {} 
  {The hierarchical model of structure formation is a key prediction of the $\Lambda$ cold dark matter model, which can be tested by studying the large-scale environment and the substructure content of massive galaxy clusters. We present here a detailed analysis of the clusters RXCJ0225.9-4154, RXCJ0528.9-3927, and RXCJ2308.3-0211, as part of a sample of massive X-ray luminous clusters located at intermediate redshifts.}
  {We used a multiwavelength analysis, combining WFI photometric observations, VIMOS spectroscopy, and the X-ray surface brightness maps. We investigated the optical morphology of the clusters, we looked for significant counterparts in the residual X-ray emission, and we ran several statistical tests to assess their dynamical state. We correlated the results to define various substructure features, to study their properties, and to quantify their influence on simple dynamical mass estimators.}
  {RXCJ0225.9-4154 has a bi-modal core, and two massive galaxy groups are located in its immediate surroundings; they are aligned in an elongated structure that is also detected in X-rays at the $1\sigma$ level. RXCJ0528.9-3927 is located in a poor environment; an X-ray centroid shift and the presence of two central BCGs provide mild evidence for a recent and active dynamical history. RXCJ2308.3-0211 has complex central dynamics, and it is found at the core of a superstes-cluster.}
  {The complexity of the cluster's central dynamics reflects the richness of its large-scale environment: RXCJ0225 and RXCJ2308 present a mass fraction in substructures larger than the typical $5\%-15\%$, whereas the isolated cluster RXCJ0528 does not have any major substructures within its virial radius. The largest substructures are found in the cluster outskirts. The optical morphology of the clusters correlates with the orientation of their BCG, and with the position of the main axes of accretion.}


   \maketitle

%

\section{Introduction}

In the framework of the $\Lambda$ cold dark matter ($\Lambda$CDM) paradigm, galaxy clusters form hierarchically through accretion and mergers of smaller scale structures (e.g. \citealt{colberg99,moore99,evrard02,springel05,springel06}). The physical properties of substructures have been investigated with semi-analytical models (e.g. \citealt{taylorJE04,vandenbosch05,giocoli08,jiang14}) and numerical simulations, leading to several predictions that can be tested observationally. For instance, the tidal stripping of subhaloes is more effective in the high-density regions, thus the most massive substructures are expected to be preferentially found in the outskirts of a cluster (e.g. \citealt{ghigna00,delucia04,diemand04,nagai05}). Since more massive objects form later, hence giving less time for the destruction of subhaloes via tidal forces, they should contain a larger mass fraction within substructures, with typical values of $5\%-15\%$ (e.g. \citealt{delucia04,gao04,giocoli10,gao12,contini12}). The mass growth of cluster-scale haloes is driven at the $\sim60\%$ level by mergers, with a fraction of $\sim20\%$ due to major mergers of mass ratio $\le1/3$ (e.g. \citealt{genel10}). 

From the observational point of view, most sample studies have focused on estimating the fraction of clusters with significant substructures, either via X-ray (e.g. \citealt{mohr95,schuecker01,jeltema05,santos08,boehringer10,chon12b,mann12}), optical (e.g. \citealt{plionis02,flin06,ramella07,einasto12,foex13,wen13}), lensing (e.g. \citealt{dahle02,smith05,martinet16}), or dynamical analyses (e.g. \citealt{girardi97,solanes99,oegerle01,aguerri10,einasto12}). The results point towards a large fraction $30\%-70\%$ of clusters with a substantial substructure level, hence not having reached yet a relaxed state. Fewer works have been dedicated to studying the statistical properties of substructures, giving nonetheless important results about their galaxy content (e.g. \citealt{biviano02}), their radial and mass distribution (e.g. \citealt{grillo15,balestra16,caminha16,mohammed16,sebasta16}), their total mass fraction with respect to their cluster host (e.g. \citealt{guennou14,jee14}), or the growth of clusters as a function of merger mass ratios (e.g. \citealt{lemze13}). Alternatively, detailed studies targeting single objects have investigated the mass assembly history of clusters via mergers, the physics taking place during these events, their impact on the overall dynamics of the clusters, and possible correlations with the clusters' large-scale environment (e.g. \citealt{markevitch04,girardi06,owers09,girardi10,barrena11,maurogordato11,owers11,ziparo12,girardi15}). Extreme systems are of particular interest, since they can be used to challenge the $\Lambda$CDM predictions regarding the amount of substructure (e.g. \citealt{jee14,jauzac16,schwinn16}).

Clusters are now a standard tool for testing cosmological models (e.g. \citealt{allen11,bohringer14}). The main ingredient is the clusters mass, which can only be determined individually on small samples. Therefore, to obtain cosmological constraints from large cluster samples, one needs to rely on mass-observable scaling relations. Considerable effort has been made in recent years to calibrate these relations with various techniques (e.g. \citealt{giodini13} for a review). An important aspect that must be accounted for is the dynamical state of the clusters, in particular for scaling laws involving X-ray measurements. A direct consequence of a non-relaxed state is a possible mass bias for estimates relying on the hypothesis of hydrostatic equilibrium. Moreover, it has been shown that regular and substructured clusters, classified by morphological criteria, are characterised by significantly different scaling relations (e.g. \citealt{chon12b}).

In view of testing the $\Lambda$CDM predictions on substructures properties, and of calibrating scaling relations, we present here a combined photometric, X-ray, and dynamical study of RXCJ0225.9-4154 (z=0.2189), RXCJ0528.9-3927 (z=0.2837), and RXCJ2308.3-0211 (z=0.2968) as part of a larger sample of distant X-ray luminous galaxy clusters. Our main goals are to draw an unbiased picture of each cluster, to characterise their substructure content, and to investigate how the latter can affect dynamical mass estimators. 

The paper is organised as follows. In Section 2, we introduce the cluster sample, recall briefly the main results obtained in previous works, and describe the data sets used for this study. The methodology employed for the photometric, dynamical, and X-ray analyses are presented in Sects. 3, 4, and 5, respectively. We draw a global picture of each cluster in Section 6, before concluding in Section 7. All our results are scaled to a flat $\Lambda$CDM cosmology with $\Omega_{m}=0.3$, $\Omega_{\Lambda}=0.7$, and a Hubble constant $H_{0}=70\,\mathrm{km\,s^{-1}\,Mpc^{-1}}$.

\section{Data: description and reduction}

\subsection{Sample description}

Starting from the ROSAT-ESO Flux Limited X-ray survey (REFLEX, \citealt{bohringer01,bohringer04}), 13 distant X-ray galaxy clusters with luminosities $L_X^{\mathrm{bol}}=0.5-4\times10^{45}\mathrm{erg\,s^{-1}}$ were selected to form a statistically complete sample (DXL; see e.g. \citealt{zhang04b} for more details). The DXL sample contains the clusters that are the most X-ray luminous in the redshift interval $z=0.26-0.31$. Its volume completeness can be estimated with the well-known selection function of the REFLEX survey \citep{bohringer04}. The sample covers a mass range $M_{500}=0.48-1.1\times10^{15}\mathrm{M_{\odot}}$, \citep{zhang05}. In addition to the X-ray observations, wide-field photometric and spectroscopic follow-ups (see below) were conducted to allow for a comprehensive analysis of the clusters. The DXL sample offers a unique way to investigate the connections between the physical properties, the substructure content, the large-scale environment, and the mass assembly history of galaxy clusters observed at different stages of their dynamical history. To summarise, it is an ideal snapshot of the Universe that can be directly compared to the outcomes of N-body numerical simulations coupled to hydrodynamics, thus offering a great opportunity to study the physics driving cluster evolution.

Detailed X-ray analyses of the DXL clusters were performed by \cite{zhang04b,zhang05,zhang06} and \cite{finoguenov05}, providing results on the intra-cluster medium properties, the dynamical state of the clusters, the calibration of X-ray scaling relations, or active galactic nucleus feedback. \cite{braglia07} and \cite{braglia09} studied the galaxy content of two DXL clusters, Abell 2744 and RXCJ2308. Their results on the star formation activity as a function of environment suggest a link between the cluster assembly history and the properties of its galaxy population, with a notably enhanced activity found along the two filaments connected to A2744. \cite{pierini08} analysed the diffuse stellar emission around the brightest galaxies of three DXL clusters, finding different possible origins for the properties of this emission, as well as a probable link with the dynamical state of the clusters. In particular, the merging cluster A2744 presents a significantly bluer intra-cluster light around its central brightest cluster galaxies, most likely due to the shredding of star-forming low-metallicity dwarf galaxies. \cite{ziparo12} conducted a detailed analysis of the structure and dynamical state of A1300, with a methodology similar to that presented in this paper. This cluster has complex central dynamics, and it is embedded in a rich large-scale environment with filamentary structures.

In addition to the 13 clusters in the original DXL sample, three objects were added to cover a wider redshift range: two at redshift $z\sim0.45$, and RXCJ0225 at redshift $z\sim0.22$, which is analysed here for the first time. This paper also presents the first dynamical analysis of RXCJ0528. A detailed lensing analysis of RXCJ2308 was presented in \cite{newman13}, with brief results on its dynamics.

\subsection{Optical spectroscopy}
Multi-object spectroscopy (MOS) observations were carried out between 2003 and 2005 with the VIMOS instrument mounted at the Nasmyth focus B of VLT-UT3 {\it Melipal} at Paranal Observatory (ESO), Chile. When operated in MOS mode, VIMOS provides an array of four identical CCDs separated by a $2'$ gap, each with a field of view (FOV) of $7\,\times\,8\,\mathrm{arcmin^{2}}$ and a $0.205''$ pixel resolution.

The programme was designed to target galaxies from the cluster core up to well beyond $R_{200}$ (\citealt{zhang06} found an average $R_{500}\sim1.20$ Mpc for the DXL clusters, i.e. $R_{200}\sim1.7$ Mpc assuming a concentration $c_{200}=4$ typical for such massive objects). The observing strategy was the following: three pointings per cluster, extending along the major axis of the cluster shape as observed in X-rays, and overlapping in the centre to achieve a good sampling of the region of high galaxy density. Given the size of the VIMOS total FOV, the observations cover a roughly rectangular area of $9\times5$ Mpc$^2$ (see e.g. Fig. 1 in \citealt{braglia09}), with a continuous central region of radius $\sim2.5\,\mathrm{Mpc}$ at $z=0.3$. The selection of targets, which were detected on VIMOS pre-imaging, was performed only on the basis of their $I$-band luminosity to avoid any colour bias for the comparative analysis of passive and star-forming galaxies. The catalogues of targets produced with {\sc SExtractor} \citep{bertin96} were divided into bright and faint objects. They were observed with two different masks to optimise the allocation of the awarded time. Exposure times were calculated to reach a typical signal-to-noise ratio of 10 (5) for the bright (faint) targets, and divided in three exposures per mask.

Spectra were obtained with the low-resolution LR-Blue grism. It provides a spectral coverage from 3700 to 6700 $\AA$, has a spectral resolution of about 200 for $1''$ width slits, and does not suffer from fringing. Moreover, it allows up to four slits in the direction of dispersion, thus significantly increasing the number of targets per mask. Finally, redshifts up to $z\sim0.8$ can be reliably obtained with this grism and, for a galaxy at $z\sim0.3$, it covers important spectral features such as the $\mathrm{[OII]},\,\mathrm{[OIII]},\,\mathrm{H}_{\beta},\,\mathrm{H}_{\delta}$ emission lines, the $\mathrm{CaII_{H+K}}$ absorption lines, and the $4000\,\AA$ break. The data reduction was performed with the {\sc VIPGI} software \citep{scodeggio05}.

To estimate spectroscopic redshifts (hereafter $z_{\mathrm{spec}}$), we first used the EZ tool \citep{garilli10}. It relies on a decision tree based on the number and strength of detected emission lines, and also relies on a cross-correlation with the continuum and absorption lines. We ran EZ in blind mode, restricted to $z\in[0-2]$ and excluding star templates. We found that it was faster to remove stars a posteriori, rather than rerun EZ for the obvious mismatches. In the second step, where all spectra were reviewed by eye, we also made use of VIPGI for the manual detection and fit of spectral features in the case of probable misidentification. This step has proven to be necessary, in particular because strong residual sky lines were typically mistaken for the $\mathrm{[OII]}$ emission line. The EZ tool assigns different flags to its redshift estimates from 0 (not reliable) to 4 (highly reliable), and 9 for solutions based on a single strong emission line. While checking the spectra by eye, we readjusted the flags according to the visual identification of lines (especially for the solutions flagged with 9), and we finally kept objects with a flag higher than or equal to 2. The number of spectra and reliable redshifts are given in Table \ref{table:sample}. Since EZ does not provide redshift errors, we relied on repeated observations of the same object to estimate a typical uncertainty. We found an average value $\delta_{cz}\sim300\,\mathrm{km\,s^{-1}}$ with variations of $\sim50\,\mathrm{km\,s^{-1}}$ from cluster to cluster. Such a redshift uncertainty leads to overestimated velocity dispersions \citep{danese80}; all values quoted in the paper were corrected accordingly. For instance, a measured velocity dispersion $\sigma_{obs}=1000\,\mathrm{km\,s^{-1}}$ was corrected to $\sigma_v=\sqrt{\sigma_{obs}^2-\delta_{cz}^{2}/(1+z_c)^2}=973\,\mathrm{km\,s^{-1}}$ for a cluster with $z_c=0.3$.

\subsection{Optical imaging}
In addition to VIMOS spectroscopy, we used optical photometric data in the B, V, R, and I pass bands from the Wide Field Imager (WFI; \citealt{baade99}) mounted on the Cassegrain focus of the ESO/MPG 2.2 m telescope at La Silla, Chile. The WFI is a mosaic camera composed of $4\times2$ CCD chips, each made of $2048\times4096$ pixels with an angular resolution of $0.238"$/pixel. The total FOV is $34'\times33'$, which fully encompasses the region observed with VIMOS. The total exposure times are given in Table \ref{table:sample}.

The data reduction was performed with the THELI pipeline \citep{schirmer13}. It performs the basic pre-processing steps (bias subtraction, flat-fielding, background modelling and sky subtraction), and uses third-party software for the astrometry ({\sc Scamp}, \citealt{bertin06}) and the co-addition of mosaic observations ({\sc SWarp}, \citealt{bertin10}). The photometry was made with {\sc SExtractor} in dual mode with the detection in the R band. Stars, galaxies, and false detections were sorted according to their position in the magnitude/central flux diagram, their size with respect to that of the PSF, and their stellarity index (CLASS\_STAR parameter). Luminosities were estimated from the MAG\_BEST parameter, while colours were computed with MAG\_APER, measured in a fixed aperture of 3''.

\begin{figure}
\center
\includegraphics[width=6.5cm, angle=-90]{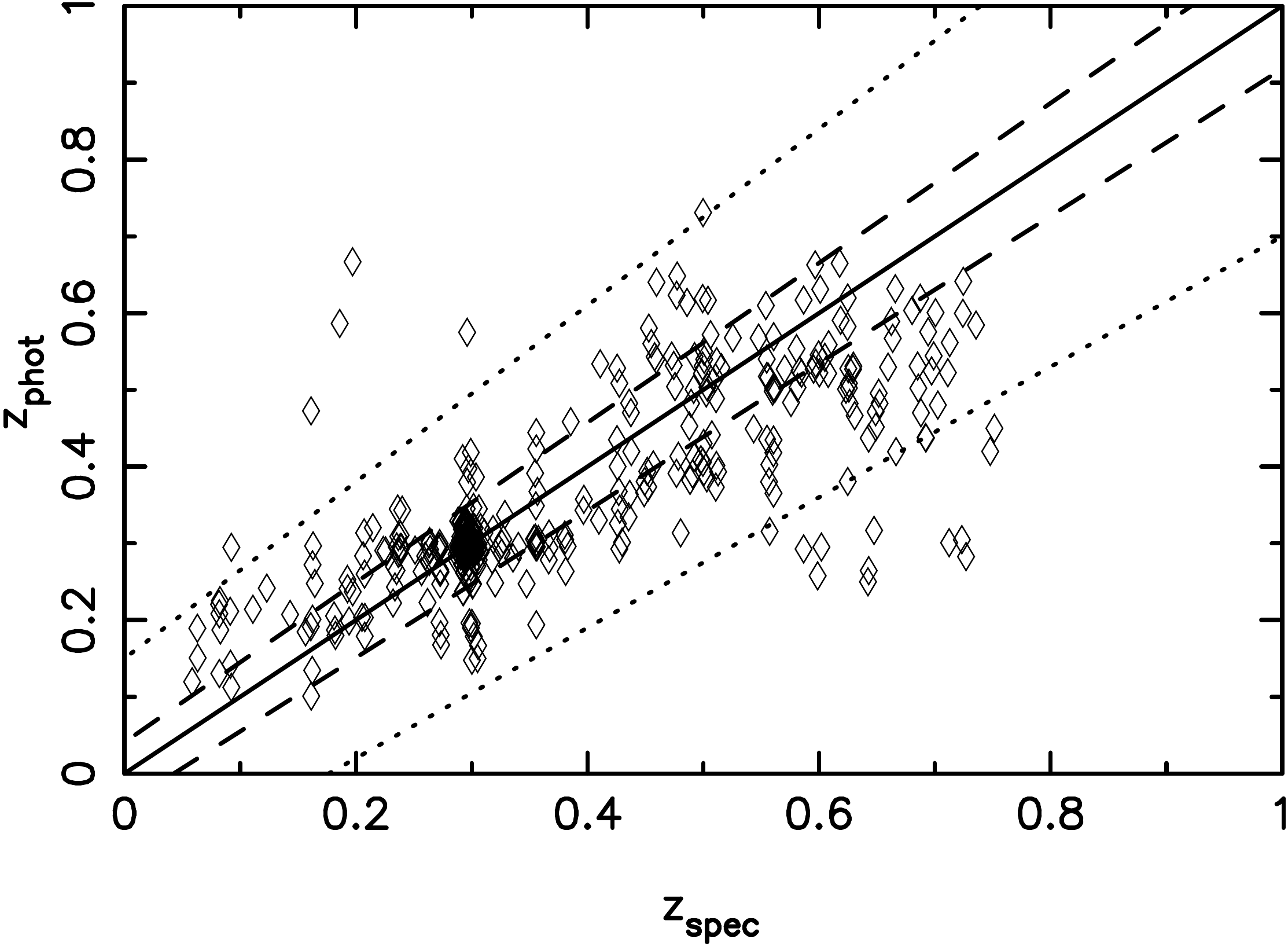}
\caption{Comparison between spectroscopic and photometric redshifts for RXCJ2308. The continuous line shows equality, dotted lines are for $z_{\mathrm{phot}}=z_{\mathrm{spec}}\pm0.15(1+z_{\mathrm{spec}})$, and dashed lines are for $z_{\mathrm{phot}}=z_{\mathrm{spec}}\pm\sigma_{z}(1+z_{\mathrm{spec}})$.}
\label{fig:zphot} 
\end{figure}

The WFI observations were used to compute photometric redshifts (hereafter $z_{\mathrm{phot}}$). Given the limited number of available bands, we employed the simple technique of the 'k-nearest neighbour' fitting (kNN; \citealt{altman92}). The basic idea of this method is that galaxies sharing similar observables should have a similar redshift. Therefore, the $z_{\mathrm{phot}}$ of a galaxy can be simply evaluated by averaging the $z_{\mathrm{spec}}$ of its closest galaxies in the parameter space, e.g. magnitudes or colours. The main advantage of this method is that it is self-contained, as external templates are not required. Moreover, using a training set (i.e. galaxies with a $z_{\mathrm{spec}}$) that is part of the target sample implies that accurate photometry is not mandatory. To be efficient, the kNN method needs a training set that covers the full observable space without being biased towards a specific galaxy population. The VIMOS target selection was made without any colour criterion, but within a limited magnitude range to observe mainly cluster members. Therefore, we expect the kNN algorithm to be more robust around the cluster redshift and for bright objects, but equally efficient for blue and red galaxies. We ran several tests to decide how the kNN algorithm should be employed. For each cluster, we divided the sample of $z_{\mathrm{spec}}$ into training and testing sets (60\% and 40\%, respectively). The testing sample is used to assess the quality of the $z_{\mathrm{phot}}$ according to the two usual quantities that are the fraction of catastrophic errors $\eta=|z_{\mathrm{phot}}-z_{\mathrm{spec}}|/(1+z_{\mathrm{spec}})>0.15$, and the redshift accuracy $\sigma_{z}=1.48\times \mathrm{med}[|z_{\mathrm{phot}}-z_{\mathrm{spec}}|/(1+z_{\mathrm{spec}})]$ \citep{ilbert06}. We investigated various combinations for the number of neighbours, the observables, the metric, and the weighting scheme. For each configuration, we repeated the measurements on 100 randomly selected training/testing sets to get a sense of the statistical fluctuations for $\eta$ and $\sigma_{z}$. Based on these two accuracy criteria, we chose the following procedure: ten neighbours, squared Euclidian distance measured in colour space, and weights equal to the inverse of said distance. Figure \ref{fig:zphot} presents the results for RXCJ2308, for which we obtained $(\eta,\sigma_{z})\sim(0.04,0.04)$. For RXCJ0225 and RXCJ0528 we obtained $(0.12,0.07)$ and $(0.09,0.05)$, respectively. The larger fraction of catastrophic errors for RXCJ0225 is due to the smaller number of spectroscopic redshifts available to sample the colour space (see Table \ref{table:sample}). We also estimated the redshift accuracy for the subsample of red-sequence galaxies, since most of our results rely on this population. We obtained $(\eta,\sigma_{z})\sim(0.03,0.06)$ for RXCJ0225, $(0.01,0.03)$ for RXCJ0528, and $(0.01,0.02)$ for RXCJ2308. These values are very good; therefore we can be confident that our photometric redshifts are correct, in particular for the red-sequence cluster members.
 
\subsection{X-ray observations}
To study the X-ray morphology of the clusters we used two combined Chandra observations for the RXCJ0225 (ObsID 15110,17476) and XMM-Newton observations for RXCJ0528 (ObsID 0042340801) and for RXCJ2308 (ObsID 0205330501).

The XMM-Newton observations for all three detectors were flare-cleaned, and out-of-time events were statistically subtracted from the pn data. Point sources and other sources unrelated to the galaxy clusters were removed. We removed the background contribution using the blank sky observation provided by \cite{read03}. The images from all three detectors were combined and the corresponding exposure maps were added with an appropriate weighting to match an effective pn exposure. The exposure-corrected and background-subtracted combined flux images are used in this paper.

We used all available archival Chandra ACIS-I observations for RXCJ0225. Standard data analysis was performed via CIAO 4.8 with calibration database (CALDB) 4.8.1. More details on the data reduction procedure are found in \cite{chon12c}.

\begin{table*}
\centering 
\begin{threeparttable}
\caption{General properties of the clusters and data sets.}
\label{table:sample}
\begin{tabular}{l c c c c c c c c c c}
\hline\hline\noalign{\smallskip}
Cluster & RA & Dec & z & \multicolumn{4}{c}{$\mathrm{T_{exp}\,(min)}$} & $m^*$ & $\mathrm{N_{spectra}}$ & $\mathrm{N_{z}}$\\
   & (J2000) & (J2000) & & B & V & R & I & (mag) & & \\
\noalign{\smallskip}\hline\noalign{\smallskip}
RXCJ0225.9-4154 & 02:25:54.6 & -41:54:35.3 & 0.2195 & 90 & 30 & 60 & 60 & 18.3 & 964 & 539\\
RXCJ0528.9-3927 & 05:28:52.8 & -39:28:17.7 & 0.2839 & 150 & 90 & 110 & - & 18.9 & 1114 & 628\\
RXCJ2308.3-0211 & 23:08:22.2 & -02:11:27.5 & 0.2966 &  50 & 45 & 45 & - & 18.9 & 1319 & 733\\
\noalign{\smallskip}\hline
\end{tabular}
    \begin{tablenotes}
      \small
      \item Columns: (1) Cluster name. (2-3) Equatorial coordinates of the X-ray peak. (4) Redshift, prior to our new estimates. (5-8) WFI exposure times in the B, V, R, and I bands. (9) R-band magnitude $m^*$, estimated from \cite{zenteno11}. (10-11) Number of spectra and reliable redshifts obtained from the VIMOS observations. 
    \end{tablenotes}
  \end{threeparttable}
\end{table*}

\section{Photometric analysis}
The first step of our analysis consists of selecting the cluster galaxies from either spectroscopic or photometric redshifts. The catalogues of cluster members are then used to construct surface density and luminosity maps, as well as ellipticity profiles. We also partition the cluster members into two broad categories to look for a possible segregation in their spatial distribution. 

\subsection{Selection of cluster members}

Among the various methods used to select cluster members from spectroscopic data (e.g. \citealt{wojtak07}), we opted for an iterative $3\sigma$ clipping combined with an iterative radial binning in the projected-phase space (hereafter PPS). This method extends the approach originally proposed by \cite{yahil77} by accounting for the radial variations in the velocity dispersion. The initial sample of cluster members was defined as the galaxies with an absolute rest-frame velocity difference $|\delta_v|\le4000\,\mathrm{km\,s^{-1}}$ (with respect to the cluster redshift given in Table \ref{table:sample}), from which we derived the initial $\sigma_P$. At each new iteration, we increased the number of radial bins by one unit, and computed their velocity dispersion using the galaxies selected in the previous iteration. We started with wide bins whose estimated $\sigma_P$ are robust against interlopers, and then moved towards a better estimate of the velocity dispersion profile. At each iteration, galaxies were allowed to re-enter the sample. The procedure was stopped either when the bins reached a limiting size of 300 kpc or when they contained a minimum of 30 galaxies. Velocity dispersions were estimated with the robust biweight scale estimator of \cite{beers90} (see e.g. \citealt{ruel14} for its unbiased version), and we adopted a $2.7\sigma_P$ rejection criterion, as advocated by \cite{MBM10}. At each iteration, we re-estimated the cluster redshift (used to determine rest-frame velocities) with the biweight location estimator of \cite{beers90}\footnote{Throughout this paper, mean redshifts and velocity dispersions are computed with the biweight location and scale estimators of \cite{beers90}. Statistical uncertainties are derived from bootstrapping.}. The cluster centre, needed for the radial binning, was chosen as the highest density peak in the galaxy surface density maps constructed from the galaxies with $|\delta_v|\le4000\,\mathrm{km\,s^{-1}}$.

To select the cluster member candidates from photometric redshifts, we proceeded as follows. First, we estimated the photometric redshift of the clusters by looking at the $z_{\mathrm{phot}}$ distribution of the spectroscopically confirmed cluster members. Due to the weighting scheme of the kNN algorithm (inverse distance in colour space), a galaxy with a $z_{\mathrm{spec}}$ has a nearly identical $z_{\mathrm{phot}}$, since its closest spectroscopic neighbour is the galaxy itself. Therefore, we again used a training/testing approach to derive the following quantities. We obtained $z_{\mathrm{c,phot}}=0.23\pm0.04$, $z_{\mathrm{c,phot}}=0.29\pm0.02$, and $z_{\mathrm{c,phot}}=0.30\pm0.01$ for RXCJ0225, RXCJ0528, and RXCJ2308. These values were used for the first selection, i.e. we only kept galaxies having $|z_{\mathrm{phot}}-z_{\mathrm{c,phot}}|<3\sigma_{\mathrm{z_{c,phot}}}$. To increase the purity of the catalogues, we then removed galaxies having $\sigma_{\mathrm{z,spec}}>0.1$, where $\sigma_{\mathrm{z,spec}}$ is the dispersion in $z_{\mathrm{spec}}$ of the kNN ten nearest neighbours. We estimated that these selection criteria lead to a typical completeness of $\sim75\pm5\%$ and a purity of $\sim60\pm5\%$ for the photometric sample, which increases by $\sim5\%-10\%$ for the combined catalogue, after the addition of the spectroscopically confirmed members. For the red-sequence galaxies (see below), the completeness reaches $\sim90\%$ and the purity is above $80\%$. 
 
\begin{figure}
\center
\includegraphics[width=6.5cm, angle=-90]{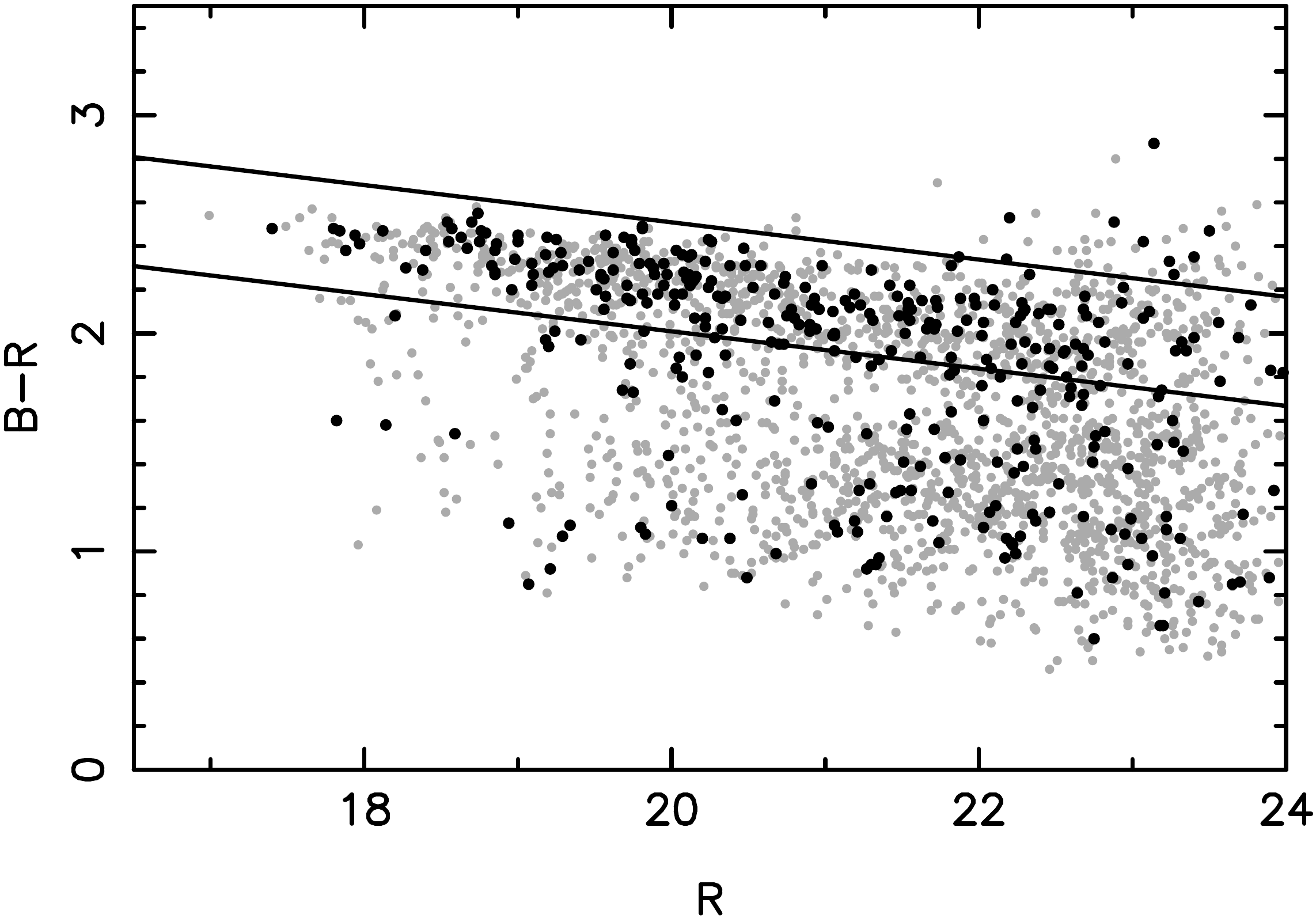}
\caption{Colour-magnitude diagram for RXCJ0225. Black circles are galaxies within 1 Mpc of the cluster centre (we note the presence of a brighter red galaxy, located outside this central region). The continuous lines show the red-sequence width $\pm3\sigma_{\mathrm{RS}}$.}
\label{fig:RS} 
\end{figure}

For each cluster, the spectroscopic and photometric catalogues were finally merged, giving priority to the spectroscopic classification when possible. From these combined catalogues, we fitted the clusters' red sequence in the (B-R)-R diagram using a $2\sigma$ clipping method (e.g. \citealt{stott09}; see the example in Figure \ref{fig:RS} for RXCJ0225). The locus and scatter, $\sigma_{\mathrm{RS}}$, of the red sequence were used to divide the catalogues into two broad populations: the red-sequence galaxies, i.e. those with a (B-R) colour within $3\sigma_{\mathrm{RS}}$, and the blue members. Additionally, the combined catalogues were cut to a limiting magnitude $m_{\mathrm{R}}\le m^{*}+3$ in order to reduce a residual contamination by faint background galaxies. The number of cluster members is given in Table \ref{table:members}. For the three clusters the red/blue fraction for the spectroscopic members is, interestingly, roughly equal to one, reflecting that the VIMOS target selection, designed to be unbiased towards a specific population of galaxies, worked reasonably well.

\begin{table}
\centering
\begin{threeparttable}
\caption{Summary of the cluster member selection.}
\label{table:members}
\begin{tabular}{l c c c c c c}
\hline\hline\noalign{\smallskip}
Cluster & \multicolumn{3}{c}{$\mathrm{N_{spec}}$} &  \multicolumn{3}{c}{$\mathrm{N_{spec}+N_{phot}}$}\\
   & all & red & blue & all & red & blue\\
\noalign{\smallskip}\hline\noalign{\smallskip}
RXCJ0225 & 228 & 120 & 108 & 1558 & 765 & 793\\
RXCJ0528 & 219 & 112 & 107 & 1126 & 321 & 805\\
RXCJ2308 & 307 & 173 & 134 & 1276 & 635 & 641\\
\noalign{\smallskip}\hline
\end{tabular}
    \begin{tablenotes}
      \small
      \item Columns: (1) Cluster name. (2-4) Number of spectroscopically confirmed cluster members, and their repartition into the red and blue populations. (5-7) Same as columns (2-4), but for the merged catalogues, and limited to $m_{\mathrm{R}}\le m^{*}+3$.
    \end{tablenotes}
  \end{threeparttable}
\end{table}

\subsection{Optical morphology}

To investigate the cluster structure, we constructed luminosity and surface density maps based on the combined catalogues. Starting from a grid of pixels of width 50 kpc and covering the entire WFI FOV, we measured $r_5$, the radius enclosing the fifth closest galaxy from the centre of a given pixel. The surface density was then defined as $\Sigma_N=5/(\pi r_5^2)$, and the corresponding luminosity was obtained by summing the luminosity of the individual galaxies within $r_5$. The distributions of pixels at large distance from the cluster centre were used to estimate the background levels $\Sigma_{bckg}$ and their dispersion $\sigma_{bckg}$. For the red (blue) population, we obtained a galaxy surface density $\Sigma_{bckg}=0.35\pm0.20\,(0.64\pm0.26)\,\mathrm{arcmin^{-2}}$ for RXCJ0225, $\Sigma_{bckg}=0.09\pm0.05\,(0.58\pm0.31)\,\mathrm{arcmin^{-2}}$ for RXCJ0528, and $\Sigma_{bckg}=0.16\pm0.08\,(0.47\pm0.20)\,\mathrm{arcmin^{-2}}$ for RXCJ2308.

To better quantify the clusters morphology, we computed integrated ellipticity profiles using the moment approach (e.g. \citealt{carter80}). We also measured centroid shifts with respect to the highest density peak, since they are a good indicator of the cluster substructures (e.g. \citealt{evrard93,mohr95,plionis02}). We also compared the orientation of the BCG to that of the large-scale morphology of the cluster. A correlation between the two has been observed in several studies (e.g. \citealt{lambas88,panko09,niederste-ostholt10,soucail15}) and can be interpreted as either a collimated infall of galaxies from filaments (e.g. \citealt{dubinsky98}) or due to tidal interactions (e.g. \citealt{faltenbacher08}).

\section{Dynamical analysis}

Several statistical tests based on velocity information have been devised to quantify the level of substructure in galaxy clusters, with varying sensitivity depending on the physical configuration \citep{pinkney96}. Here we focused on (i) measuring departures from Gaussianity in the velocity distribution, (ii) identifying gradients and discontinuities in the velocity and velocity dispersion profiles, and (iii) looking for deviations between the local and global velocity distributions.

\subsection{Velocity dispersion, virial mass and virial radius}

The mass of a cluster can be derived by applying the virial theorem for an isolated non-rotating spherical system (e.g. \citealt{limber60,heisler85,merritt88}):
\begin{equation}
\label{eq:M_V}
M_V=\frac{3\pi}{G}\sigma_{P}^{2}R_{PH}.
\end{equation}
The projected (line of sight) velocity dispersion $\sigma_P$ and the projected harmonic mean radius $R_{PH}$ are given by
\begin{eqnarray}
\label{eq:RPH}
\sigma_P=\sqrt{\frac{\sum_i{v_{rf,i}^2}}{N-1}},\\
R_{PH}=\frac{N(N-1)}{2\sum_{i>j}{R_{ij}^{-1}}},
\end{eqnarray}
with N the number of galaxies, $v_{rf}$ the rest-frame line-of-sight velocity, and $R_{ij}$ the angular-diameter distance between galaxy pairs (in practice, we used the robust biweight estimator to evaluate $\sigma_P$). The $M_V$ estimator also assumes that galaxies have the same spatial and velocity distribution as dark matter particles, and that all galaxies have the same mass. The latter approximation can be justified by the lack of observational evidence of a strong luminosity/mass segregation in the galaxy population (e.g. \citealt{adami98b,biviano02}). However, dark matter haloes are well represented by a cuspy NFW profile, whereas cluster members typically have a cored King-like spatial distribution. Moreover, numerical simulations have shown that a velocity bias exists between galaxies and dark matter particles (e.g. \citealt{berlind03,biviano06,munari13}), and many studies have concluded that clusters are not spherical (e.g. \citealt{limousin13}). Despite all these assumptions, the virial estimator is widely used due to its simple application.
 
Another consideration must be made before estimating the mass. The usual form of the scalar virial theorem $2E_k+E_P=0$ is only valid for isolated systems. However, galaxy clusters are embedded in dense environments, continuously accreting matter from their surroundings, far beyond their actual virial radius $R_V$. \cite{cupani08} estimated from numerical simulations that the turnaround radius $R_t$, i.e. the distance above which the Hubble flow prevents an infall of matter, is $R_t\sim3.5R_V$. Therefore, the virialised region of a cluster cannot be considered as an isolated system, which requires a modification of the virial theorem (e.g. \citealt{binney87,carlberg96}). Neglecting the dynamical pressure due to the radial infall of matter leads to virial masses typically overestimated by $\sim15-20\%$ \citep{carlberg97,girardi98}. In the case where the spectroscopic survey does not fully cover the virialised region, a larger correction should be used since the velocity dispersion in general decreases with radius. The limiting case is found for a singular isothermal sphere, whose constant velocity dispersion leads to a $50\%$ overestimation at all radii \citep{carlberg96}. As mentioned previously, our spectroscopic observations cover a continuous circular area of radius $\sim2.5$ Mpc at a redshift $z=0.3$ (regions at larger radius were only observed along the major axis of the cluster shape as observed in X-rays). Defining the virial radius as the distance encompassing an overdensity $\Delta_v(z)$ with respect to the critical density $\rho_c(z)=3H^2(z)/(8\pi G)$
\begin{equation}
\label{eq:R_V}
R_V=\left(\frac{2GM_V}{\Delta_v(z)H^2(z)}\right)^{1/3},
\end{equation}
and using the approximation given by \cite{bryan98} to estimate $\Delta_v(z=0.3)\sim125$ in a $\Lambda$CDM cosmology, we find that a cluster of mass $M_V\le 1.5\times10^{15}\,\mathrm{M_{\odot}}$ (i.e. with $R_V\le2.5$ Mpc) is fully covered by our observations. Even though the clusters analysed here were selected based on their high X-ray luminosity, their virial radius should not be much larger than the VIMOS FOV. Therefore, the $20\%$ correction factor will be sufficient for our purpose.

To evaluate the mass, we first started by estimating $\sigma_P$ and $R_{PH}$ within a circular aperture of radius 2.5 Mpc to avoid possible structures that are not part of the virialised region of the cluster. These values were used to estimate $R_V$, which is obtained by combining equations (\ref{eq:M_V}) and (\ref{eq:R_V}):
\begin{equation}
R_V=\left(\frac{6\pi}{\Delta_v(z)H^2(z)}\sigma_P^2R_{PH}\right)^{1/3}.
\end{equation}
Since we want to derive the mass contained within the virial radius, we recomputed $\sigma_P$ and $R_{PH}$ using the galaxies within this first estimate of $R_V$. The procedure was repeated until convergence on $R_V$. At each iteration, the harmonic radius was obtained from the combined photometric and spectroscopic catalogues. We followed the same approach to compute the dynamical $M_{200}$ (and corresponding radius $R_{200}$), i.e. replacing $\Delta_v=200$. We again used a $20\%$ correction factor to account for the surface pressure term, even though the aperture of radius $R_{200}$ is smaller. Alternatively, one can assume that the cluster has a NFW density profile, and simply convert $(M_V,R_V)$ into $(M_{200},R_{200})$; this approach requires knowledge of the concentration parameter, which was obtained from the mass-concentration relation of \cite{dutton14}. These two methods lead to equivalent mass estimates within their error bars, so in the following $M_{200}$ refers to the mass estimated from the virial theorem applied within $\Delta_v=200$. For comparison, we also estimated $M_{\sigma}=M_{200}(\sigma_P)$ from the scaling relation of \cite{biviano06}, under the assumption $\sigma_v=\sqrt3\sigma_P$. Application of the virial theorem gives radii $R_V\approx3.10$, 2.61, and 3.22 Mpc, and $R_{200}\approx$2.35, 2.10, and 2.56 Mpc (see Table \ref{table:Mvir}) for RXCJ0225, RXCJ0528, and RXCJ2308, respectively. These values are close to the size $R\approx2.5$ Mpc of the circular aperture containing the continuous VIMOS coverage; therefore, the $20\%$ correction on the virial masses proves to be a reasonable assumption.

Since our photometric redshifts are more accurate for the red galaxies, we expect a lower contamination by foreground/background interlopers for this galaxy population, hence providing a better estimate of the harmonic radius. Furthermore, the blue population should also contain a larger fraction of infalling galaxies located outside the virialised region of the cluster, hence biasing the estimate of its velocity dispersion. However, elliptical galaxies, in particular the massive ones, are subject to dynamical friction, which reduces their velocity dispersion (e.g. \citealt{merritt85}). \cite{biviano06} investigated the efficiency of the $M_V$ and $M_{\sigma}$ mass estimators with N-body numerical simulations. They found that interlopers cause an overestimate of the harmonic radius $R_{PH}$, and an underestimate of the velocity dispersion, the first effect being stronger than the second. Since we use the combined photometric and spectroscopic catalogue to estimate $R_{PH}$, we can suppose that its value is the main source of uncertainty in $M_V$ (a larger fraction of field galaxies than that of the spectroscopic catalogue). They conclude that $M_V$ typically overestimates the true mass by $\sim10\%$, whereas $M_{\sigma}$, which does not rely on $R_{PH}$, underestimates it by $\sim15\%$. Both estimators have an average $\sim35\%$ scatter (see also \citealt{saro13}). They also found that, unlike $M_{\sigma}$, the virial estimator $M_V$ is significantly improved when applied to the elliptical galaxies only, due to a smaller contamination by interlopers.

Our results, presented in Table \ref{table:Mvir}, are well explained by the above remarks. The masses $M_{200}$ are larger than the scaling masses $M_{\sigma}$, but there is a very good agreement between the $M_{200}$ estimated with the red galaxies and the $M_{\sigma}$ obtained with the full population. On the other hand, the $M_{200}$ obtained with the full population are significantly higher than the $M_{\sigma}$ for the red galaxies, in particular for RXCJ0225 (factor $\sim2.4$) and RXCJ0528 (factor $\sim3.3$). The virial theorem applied to the red galaxies should provide the most accurate masses and radii estimates within the density contrast $\Delta_v$ or $\Delta=200$ (thus $M_{200}$ and $R_{200}$ should not be confused with virial mass and virial radius). 

\begin{table*}
\centering
\begin{threeparttable}
\caption{Virial masses and radii before the substructure analysis.}
\label{table:Mvir}
\begin{tabular}{l c c c c c c c c c}
\hline\hline\noalign{\smallskip}
Cluster & pop. & $z_{spec}$ & $\sigma_{tot}$ & $\sigma_P$ & $M_V$ & $R_V$ & $M_{200}$ & $R_{200}$ & $M_{\sigma}$\\
   &  &  & ($\mathrm{km\,s^{-1}}$) & $(\mathrm{km\,s^{-1}})$ & ($\mathrm{10^{15}\,M_{\odot}}$) & (Mpc) & ($\mathrm{10^{15}\,M_{\odot}}$) & (Mpc) & ($\mathrm{10^{15}\,M_{\odot}}$)\\
\noalign{\smallskip}\hline\noalign{\smallskip}
RXCJ0225 & all & $0.2189\pm0.0003$ & $934_{-48}^{+34}$ & $962_{-53}^{+24}$ & $2.52_{-0.44}^{+0.24}$ & $3.10_{-0.10}^{+0.19}$ & $1.86_{-0.20}^{+0.26}$ & $2.35_{-0.09}^{+0.10}$ & $0.96_{-0.14}^{+0.12}$\\[3pt]
& red & $0.2195\pm0.0003$ & $857_{-51}^{+53}$ & $882_{-54}^{+46}$ & $1.57_{-0.23}^{+0.25}$ & $2.66_{-0.14}^{+0.13}$ & $1.22_{-0.21}^{+0.33}$ & $2.06_{-0.12}^{+0.16}$ & $0.76_{-0.15}^{+0.14}$\\[3pt]
\hline\noalign{\smallskip}
RXCJ0528 & all & $0.2837\pm0.0003$ & $874_{-39}^{+45}$ & $915_{-52}^{+31}$ & $1.64_{-0.22}^{+0.22}$ & $2.61_{-0.13}^{+0.11}$ & $1.40_{-0.26}^{+0.29}$ & $2.10_{-0.14}^{+0.14}$ & $0.78_{-0.10}^{+0.12}$\\[3pt]
& red & $0.2832\pm0.0004$ & $777_{-59}^{+62}$ & $776_{-81}^{+69}$ & $0.82_{-0.18}^{+0.25}$ & $2.08_{-0.17}^{+0.18}$ & $0.65_{-0.11}^{+0.22}$ & $1.73_{-0.16}^{+0.10}$ & $0.47_{-0.11}^{+0.16}$\\[3pt]
\hline\noalign{\smallskip}
RXCJ2308 & all & $0.2968\pm0.0003$ & $1101_{-38}^{+49}$ & $1150_{-41}^{+38}$ & $3.20_{-0.40}^{+0.35}$ & $3.22_{-0.14}^{+0.12}$ & $2.42_{-0.31}^{+0.44}$ & $2.56_{-0.11}^{+0.12}$ & $1.69_{-0.19}^{+0.24}$\\[3pt]
& red & $0.2967\pm0.0004$ & $1082_{-58}^{+79}$ & $1133_{-81}^{+67}$ & $2.22_{-0.45}^{+0.39}$ & $2.87_{-0.22}^{+0.15}$ & $1.89_{-0.35}^{+0.31}$ & $2.28_{-0.14}^{+0.12}$ & $1.68_{-0.25}^{+0.28}$\\[3pt]
\noalign{\smallskip}\hline
\end{tabular}
    \begin{tablenotes}
      \small
      \item Columns: (1) Cluster name. (2) Galaxy population. (3) Spectroscopic redshift. (4) Line-of-sight velocity dispersion within the entire FOV. (5) Line-of-sight velocity dispersion within the virial radius. (6) Cluster mass estimated from the virial theorem. (7) Virial radius estimated from Eq. \ref{eq:R_V}. (8) Cluster mass estimated from the virial theorem, within an overdensity $\Delta=200$. (9) Cluster radius, estimated from Eq. \ref{eq:R_V} for $\Delta=200$. (10) Cluster mass estimated from the scaling relation of \cite{biviano06}, i.e. $M_{\sigma}=M(\sigma_{P,200})$ with $\sigma_{P,200}$ the line-of-sight velocity dispersion estimated within $R_{200}$.
    \end{tablenotes}
  \end{threeparttable}
\end{table*}

\subsection{Velocity distribution}

To test for the departures from Gaussianity in the velocity distribution, we used the Kolmogorov-Smirnov (KS) test. When the parameters of the test distribution are inferred from the data, it is not possible to refer to the usual critical P-values to test the null hypothesis. Therefore, we proceeded as follows. First we measured the D-value between the data and the best-fit model, i.e. the maximum distance between their cumulative distribution function. Then we generated $10^4$ random velocity distributions from the Gaussian best fit to the data, with the same number of data points. For each realisation, we measured the D-value with respect to its Gaussian best fit. Finally, we estimated the significance of non-Gaussianity as the proportion of realisations having a D-value smaller than that obtained for the data.

Although it is a straightforward indicator, the KS test is mostly sensitive near the median of the distribution, and it is not robust to the presence of outliers. Furthermore, it does not provide information regarding the way the velocity distribution differs from a Gaussian. This is a strong limitation since we want to identify the physical mechanisms responsible from non-Gaussianity, e.g. infall of galaxies following radial orbits producing a peaked distribution, or presence of substructures with different velocities leading to a multimodal distribution. Therefore, we applied the method presented in \cite{zabludoff93}, which approximates the velocity distribution $N(v)$ by a series of Gauss-Hermite functions:
\begin{equation}
N(v)=\sum_{i=0}^{N_h}h_iH_i(x)\frac{e^{-x^2/2}}{\sqrt{2\pi S^2}},
\end{equation}
with 
\begin{equation}
x=\frac{(v-V)}{S}.
\end{equation}
The $H_i$ are the orthogonal Hermite polynomials (e.g. \citealt{vandermarel93}), and the projection coefficients $h_i$ are given by
\begin{equation}
h_i=\frac{2\sqrt{\pi}}{N}\sum_{j=1}^{N}H_i(x_j)\frac{e^{-x_j^2/2}}{\sqrt{2\pi}}.
\end{equation}
In practice, the series is truncated at $N_h=4$. The location and scale $(V,S)$ are free parameters. They are chosen so that the lowest order of the series, $H_0$, describes the best-fit Gaussian to the velocity distribution, which is obtained for $h_1=h_2=0$. Starting with $(V=0,S=\sigma_p)$, we performed the Gauss-Hermite decomposition (hereafter GH), iteratively changing $(V,S)$ until the criteria on $(h_1,h_2)$ were met. The $h_3$ and $h_4$ terms describe the asymmetric deviations ($h_3>0$ for an excess of positive velocities) and symmetric deviations ($h_4>0$ for a peaked distribution) from a Gaussian distribution, respectively. They are similar to the usual skewness and kurtosis, but are less sensitive to outliers. To interpret their magnitude, we ran the GH decomposition on $10^4$ random Gaussian distributions of parameters $(V,S)$ with as many data points as in the observed velocity distribution. The significance of $h_3$ and $h_4$ was then evaluated as the proportion of realisations with smaller coefficients (in absolute value). We applied the KS and GH tests on the full FOV, and within the central 1.5 Mpc, in order to focus on the dynamics of the main body.

When the velocity distribution deviates significantly from a Gaussian, it is possible to attempt a multicomponent fit to separate possible substructures from the main body of the cluster. The common approach makes use of the Kaye's mixture model algorithm (KMM; e.g. \citealt{ashman94}), which is a typical iterative expectation-maximisation algorithm for the modelling of a mixture of Gaussian distributions. It requires an initial guess for the location, scale, and mixing fraction of each component. The main freedom, hence uncertainty, of the algorithm is the number of components $g$ required to adequately describe the observed distribution. To determine the optimal number of Gaussians, one can compare the likelihood of a $g'$-mode model to that of a $g$-mode model. When the components have different scales, the significance of the likelihood ratio has to be calibrated with a Monte Carlo approach, i.e. generating a large number of $g$-mode models, applying the KMM algorithm with $g$ and $g'$ components, and estimating the corresponding likelihood ratios. The significance of the improvement in using $g'>g$ components is then given by the proportion of Monte Carlo realisations having a likelihood ratio smaller than that obtained for the data. The KMM algorithm was applied when the statistical tests suggested a non-Gaussian distribution; we verified that its outputs are weakly dependent on the initial guess values.

\subsection{Projected-phase space}

To obtain a better picture of the cluster dynamics, we measured integrated velocity and velocity dispersion profiles (iVP and iVDP) and differential velocity and velocity dispersion profiles (VP and VDP) (e.g. \citealt{denhartog96}). The profiles were centred on the central peak of the galaxy surface density map rather than on the BCG since the latter may be offset from the centre of the gravitational potential well. We also estimated smooth differential profiles with the LOWESS technique \citep{gebhardt94} to help visualise the local variations associated with substructures. Additionally, VDPs are an ideal way to investigate a possible dynamical segregation between the two populations of early- and late-type galaxies. The latter typically fall into the cluster for the first time, following radial orbits, and present a decreasing VDP. The early-type population is already virialized, hence following isotropic orbits producing a flatter VDP. As a consequence, early-type galaxies are expected to have a smaller velocity dispersion than late-type galaxies. This segregation has been observed in numerous studies, e.g. \cite{biviano92,colless96,adami98b,biviano04}. However, contradictory results have also been found (e.g. \citealt{rines05,rines13,girardi15}), hence the question remains open. 

\subsection{Combining velocity and sky coordinates}

The last series of tests we ran combine spatial and velocity information. They are based on the assumption that local departures from the overall dynamics can be attributed to substructures. Several implementations of this idea have been proposed. The original $\Delta$-statistics method developed by \cite{dressler88} uses the first two moments of the velocity distribution. For each galaxy, the local velocity $<v>_{\mathrm{loc}}$ and projected velocity dispersion $\sigma_{\mathrm{loc}}$, estimated with the $n_{NN}$ nearest neighbours, are used to quantify the deviation
\begin{equation}
\delta_i^2=\frac{n_{NN}+1}{\sigma_P^2}\left[(<v>_{\mathrm{loc},i}-<v>)^2+(\sigma_{\mathrm{loc},i}-\sigma_P)^2\right],
\end{equation}
where $<v>$ and $\sigma_P$ are the global values. The statistics are then obtained by summing the individual $\delta_i$. As pointed out by \cite{pinkney96}, gradients in the VDP can produce false positive detections of substructures. Therefore, following \cite{girardi15}, we used a radial-dependent $\sigma_P(R)$ as the `global' value against which $\sigma_{\mathrm{loc}}$ is compared. In practice, the velocity dispersion profile was fitted with a simple power law. One limitation of the $\Delta$-statistics method is that it mixes departures in velocity together with those in dispersion. Therefore, we used two additional statistics based on $\delta_{i,V}^2=[(n_{NN}+1)/\sigma_P^2]\times(<v>_{\mathrm{loc},i}-<v>)^2$ and $\delta_{i,S}^2=[(n_{NN}+1)/\sigma_P^2]\times(\sigma_{\mathrm{loc},i}-\sigma_P)^2$ (e.g. \citealt{girardi97,barrena11}). As for the $\Delta$-test, the global $\Delta_V$ and $\Delta_S$ values are obtained by summing the $\delta_{i,V}$ and $\delta_{i,S}$ of each galaxy. The number of neighbours $n_{NN}$ is somewhat arbitrary, but using $n_{NN}=\sqrt N$ has the advantage of being more sensitive to significant substructures and less sensitive to Poisson noise (e.g. \citealt{silverman86}).

The significance of the $\Delta$ values were estimated from $10^4$ random realisations of the galaxy distribution, where positions were fixed and velocities shuffled in order to erase any correlation between velocity and location. The $\delta_i$ values of the shuffled distributions were also used to define a criterion for selecting galaxies within substructures, i.e. the deviation threshold above which the local dynamics is significantly different than that of the main cluster. Its value is a rather arbitrary choice, and a compromise has to be made between the completeness and reliability of the selected galaxies (e.g. \citealt{biviano02}). We adopted the 95th percentile of the cumulated shuffled distributions as a threshold. For the combined test, it corresponds to a limit $\delta_{V+S}\sim2.2$, which is very similar to that chosen by \cite{biviano02}.

\section{Structure analysis from X-ray observations}

To investigate the cluster structure from X-ray observations, we fitted their surface brightness by a spherical $\beta$-model
\begin{equation}
S(r)=S_0\left(1+\frac{r^2}{r_c^2}\right)^{-3\beta+1/2}+b,
\end{equation}
where $S_0$ is the central brightness, $r_c$ the core radius, $\beta$ the shape parameter, and $b$ a residual background emission, assumed to be constant across the FOV; the centre was set on the X-ray emission peak. Such a simple model might be a poor representation of the actual surface brightness. However, we are only interested in the identification of substructures, thus it is sufficient to describe the smooth cluster emission. Once the best-fit parameters were obtained, we generated a signal-to-noise residual map following the prescription of \cite{neumann97} (see e.g. \citealt{neumann03,guennou14}). The surface brightness maps were smoothed with a Gaussian kernel of width 4'' for Chandra and 8'' for XMM-Newton prior to fitting the $\beta$-model. The noise of the residual map corresponds to the original signal (cluster emission plus background) smoothed with a kernel size $\sigma'=\sigma/\sqrt{2}$ (see e.g. Appendix in \citealt{neumann97}).

\section{Discussion}

For each cluster, we defined the most interesting regions, and we estimated some of their physical parameters. From the optical maps, we selected the most prominent galaxy overdensities, and we delimited the corresponding structures as ellipses englobing the $5\sigma_{bckg}$ isopleths in the surface density map of red-sequence galaxies. Within these regions, we located the brightest galaxy, and estimated the richness $N_{\mathrm{RS}}$ and optical luminosity $L_{\mathrm{RS}}$ (limited to the red sequence). They were corrected from the contribution of field galaxies, whose density was estimated for each cluster within the regions labelled R1 and R2 in Figures \ref{fig:reg_0225}, \ref{fig:reg_0528}, and \ref{fig:reg_2308}. Specifically, these regions were selected because they were devoid of significant galaxy overdensities $\Sigma_N\ge\Sigma_{bckg}+5\sigma_{bckg}$, defined with respect to the background levels obtained in Section 3.2. Assuming linear scalings $M\propto N$ and $M\propto L$, richnesses and luminosities can be used to approximate the relative mass of substructures with respect to the cluster. We also estimated the fraction $f_{\mathrm{RS}}$ of red-sequence galaxies contained in each region. The completeness and purity of the red galaxies are better than those of the blue ones (see Section 3.1) so we do not expect $f_{\mathrm{RS}}$ to be very accurate. Nonetheless, large values indicate a dense region in an advanced evolutionary state (e.g. \citealt{treu03,boselli06,huertas09}).

From the $\Delta$-tests, we defined additional structures as the regions showing a significant departure from the global dynamics. To separate them from the cluster main body, we ran the KMM algorithm. In contrast to the implementation presented in Section 4.2, we combined here velocity and sky position to estimate the membership probabilities, which are thus expressed as 3D multivariate Gaussian distributions. It is clear that the spatial distribution of galaxies within a cluster does not follow a Gaussian. However, the multivariate case provides a fast and easy way to partition the galaxies into substructures, whose rest-frame velocity and velocity dispersion are then readily estimated. The initial KMM guess parameters were evaluated from the galaxies belonging to substructures according to the $\Delta$-tests. For the substructures without a KMM partition, but spatially well separated from other components, we estimated their dynamical parameters using the spectroscopic members located within the corresponding region, as defined from the optical maps. In these cases, the results should be taken with caution, since a residual overlap with the main body cannot be entirely excluded.\\

For the substructures with dynamical information, we estimated the probability of their being bound to the cluster. The Newtonian criterion for gravitational binding of a two-body system, $E_k+E_P\le0$, can be expressed as \citep{beers82}
\begin{equation}
\delta_{V}R_{P}\le2GM_T\cos{\alpha}\sin^2{\alpha},
\end{equation}
where $\delta_{V}$ is the line-of-sight velocity difference between the two objects, $R_P$ their projected separation, $M_T$ the sum of their masses, and $\alpha$ the angle between the plane of the sky and the line joining their centres. Given $R_P$, $\delta_v$ and the total mass $M_T$, we can find the range of angles $\alpha_i\le\alpha\le\alpha_s$ for which the criterion is satisfied. The probability that the system is bound is then simply evaluated as $\int_{\alpha_i}^{\alpha_s}\cos{\alpha}d\alpha$, i.e. the fractional solid angle covered by $\alpha$.

The Newtonian two-body dynamical analysis can be extended by considering a degenerate elliptical Keplerian orbit \citep{beers82}. Assuming that there is no angular momentum and that the masses are constant, concentrated into a point in their respective centres, and with an initial zero separation, it is possible to obtain parametric solutions for the evolution of time, velocity difference, and radial separation as a function of a development angle $0<\chi<2\pi$ (eccentric anomaly). These solutions are the cycloid equations, and are equivalent to those of the spherical top-hat model of structure formation in an Einstein-de Sitter universe. The two-body model neglects the impact of a cosmological constant, which acts as a repulsive force proportional to the distance. On the other hand, the model assumes that no mass is present between the two systems. This is likely incorrect for clusters with significant substructures since it has been shown that they are preferentially found in overdense regions such as superclusters (e.g. \citealt{plionis02}). In this case, the local expansion rate is equivalent to that of a closed $\Lambda$CDM Universe. Estimating the impact of these two competing effects requires tintegrating the Friedmann equations, which is beyond the scope of this study. Thus, it should be noted that the results presented below are only approximate, and mostly used to discriminate between different general configurations.

The system of equations describing the evolution of the two bodies can be closed by making the further assumption that they are moving apart or coming together for the first time, i.e. by setting $t_0=0$ and $t=t(z)$ the age of the Universe at the redshift of the system. This approximation is most certainly valid for systems with a large projected separation, whereas substructures close to the cluster centre have a higher probability of being observed after their first pericentric passage. In this case, $t$ should be set to the time spent since core-crossing, which can be estimated from the merger configuration (e.g. \citealt{barrena09,girardi10}). Of the possible solutions, calculated as $M(\alpha,\chi)$ from the inputs $R_P$, $\delta_{V}$, and $t$, there are two bound incoming (i.e. collapsing, $\chi>\pi$), one bound outgoing (i.e. still expanding, $\chi<\pi$), and one unbound outgoing solution. Their relative probabilities are obtained, as above, from the range of $\alpha$ for which the mass criterion $M(\alpha)$ matches the estimated mass $M_T$, i.e. within $M(\alpha_i)=M_T-\sigma_M$ and $M(\alpha_s)=M_T+\sigma_M$; we note that a system can meet the Newtonian binding criterion without having a bound solution according to the two-body model. To determine the mass of each body, we adopted the following approach. First, we estimated the velocity dispersion from the (blue and red) galaxies associated with the corresponding KMM partition, which were then converted into $M_{200}$ using the scaling relation of \cite{biviano06}. Using the mass-concentration relationship of \cite{dutton14}, we obtained $c_{200}$, which was converted into $c_{vir}$, to give finally $M_{V}=M_{200}\times(\Delta_v/200)\times(c_{vir}/c_{200})^3$. Mass uncertainties were obtained by error propagation, leading to an average $\delta M/M=3\times\delta\sigma/\sigma\sim60\%$. For the mass of the main body, we applied the virial estimator as described previously (red galaxies only), but using only the main KMM partition to derive $\sigma_P$, and after cutting out annular sectors englobing the different substructures to estimate the harmonic radius.

\subsection{RXCJ0225}

\subsubsection{Optical analysis}

\begin{figure}
\center
\includegraphics[width=8.5cm, angle=-90]{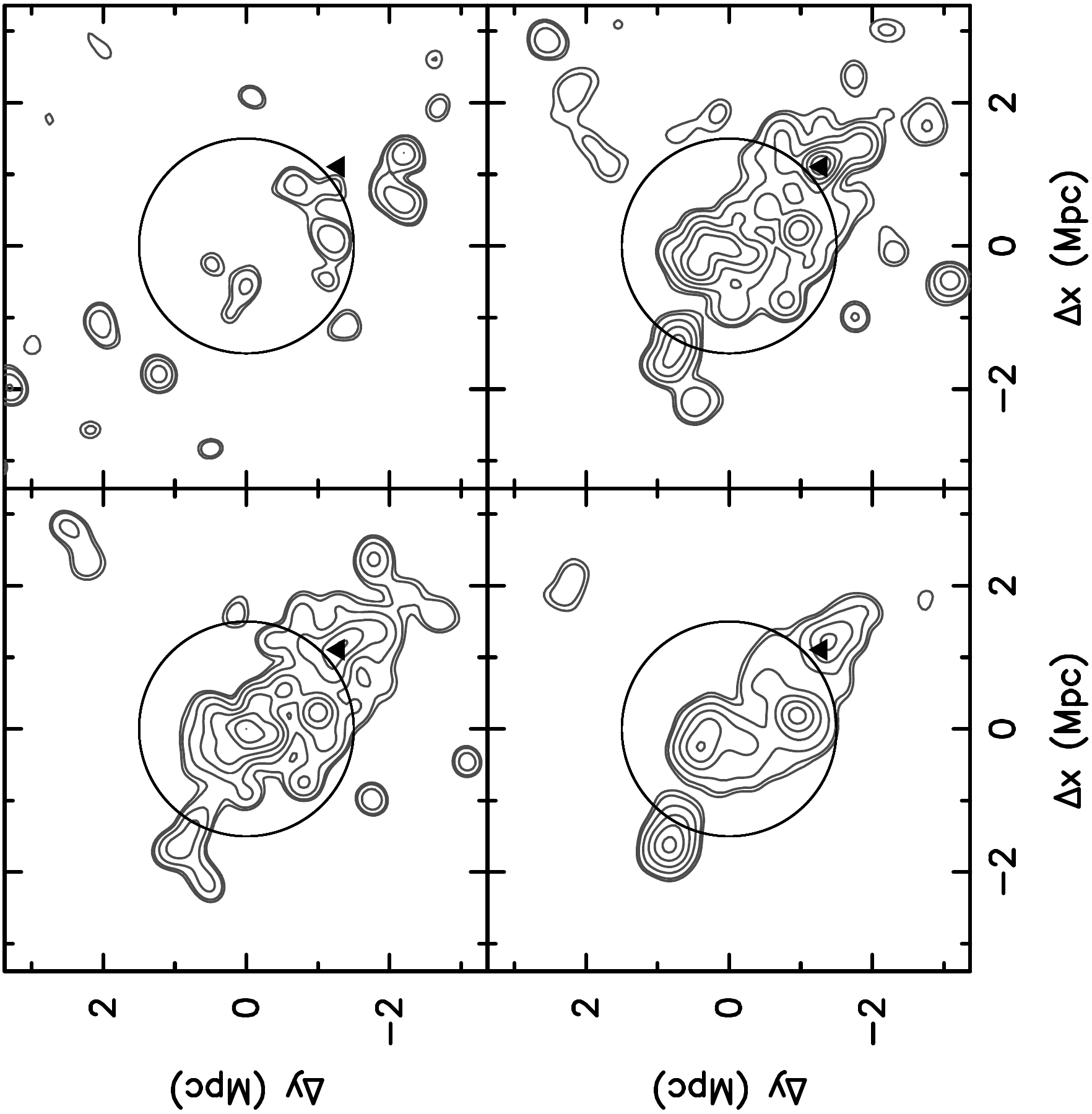}
\caption{Galaxy surface density maps for the red (top left), blue (top right), and bright red (bottom left) galaxy populations of RXCJ0225. The bottom right panel shows the luminosity density map for the red galaxies. In each panel, the triangle marks the position of the central BCG. The circle has a radius of 1.5 Mpc, and is centred on the highest density peak of red galaxies. Contours start at 5(3)$\sigma$ for the red (blue) galaxies, and follow a square-root scale.}
\label{fig:0225_Nmap} 
\end{figure}

The presence of several overdensities of blue galaxies distributed along a NE-SW axis (Fig. \ref{fig:0225_Nmap}) suggests that RXCJ0225 is embedded in a filamentary structure. The distribution of red galaxies shares the same large-scale orientation. Several substructures of similar density and extent are found in the central region, in particular for the bright $m<m^*+1$ red galaxies. They are located within four well-resolved overdensities, which are also aligned along the same NE-SW axis. As mentioned above (Fig. \ref{fig:RS}), the central BCG of RXCJ0225 is not the overall brightest red-sequence galaxy. It is actually located in the SW overdensity, which corresponds to the highest density peak in the luminosity map. 

\begin{figure}
\center
\includegraphics[width=6.5cm, angle=-90]{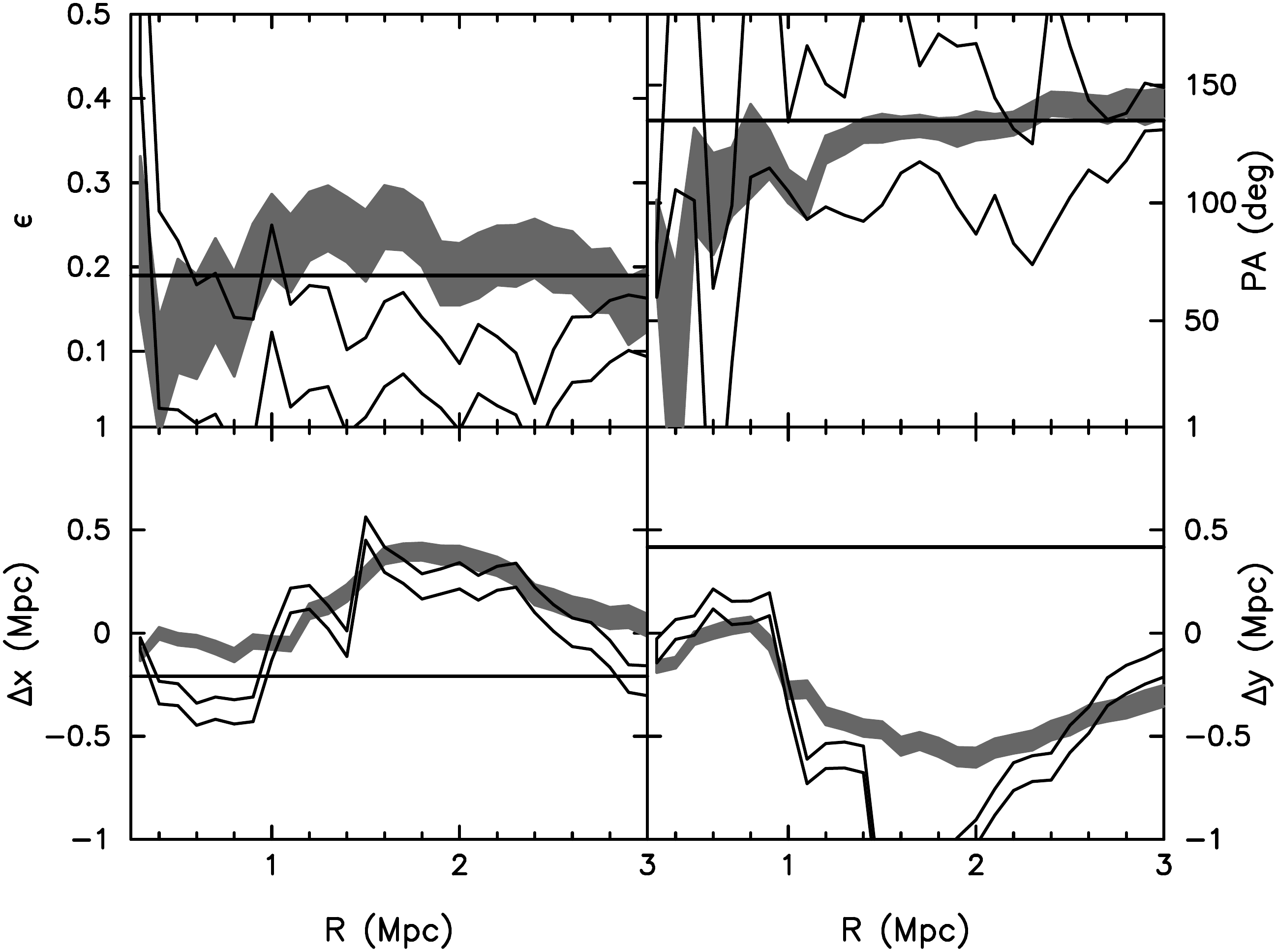}
\caption{RXCJ0225: ellipticity (top left), position angle (top right, anticlockwise from the E-W axis), and centroid shift (bottom left and right) as a function of radius (size of the circular aperture). The $1\sigma$ error around the profiles for the red-sequence galaxies is traced by the shaded regions, while the emptyregions are for the blue population. In each panel, the horizontal line marks the corresponding value of the central BCG.}
\label{fig:0225_ell} 
\end{figure}

The ellipticity profile of RXCJ0225 (Fig. \ref{fig:0225_ell}) reflects the elongation observed in the optical maps, with a good match for the position angle between the red and blue populations. The distribution of red galaxies has a rather large ellipticity $e\sim0.25$ within the central 1-2 Mpc. The main feature of these profiles is the significant centroid shift around $R\sim2$ Mpc due to the SW galaxy overdensity, which implies that this galaxy clump has a galaxy content similar to that of the main body. The central BCG is clearly offset from the highest density peak, and from the centroid of the large-scale galaxy distribution, which confirms that RXCJ0225 has a complex morphology at all scales. Interestingly, we see that the shape of the BCG matches that of the large-scale morphology of the cluster. Given the narrow and elongated galaxy distribution observed NE and SW from the core, we can suppose here that the shape of the BCG results from a collimated infall of material onto the cluster.

\subsubsection{Dynamical analysis}

\begin{figure}
\center
\includegraphics[width=6.5cm, angle=-90]{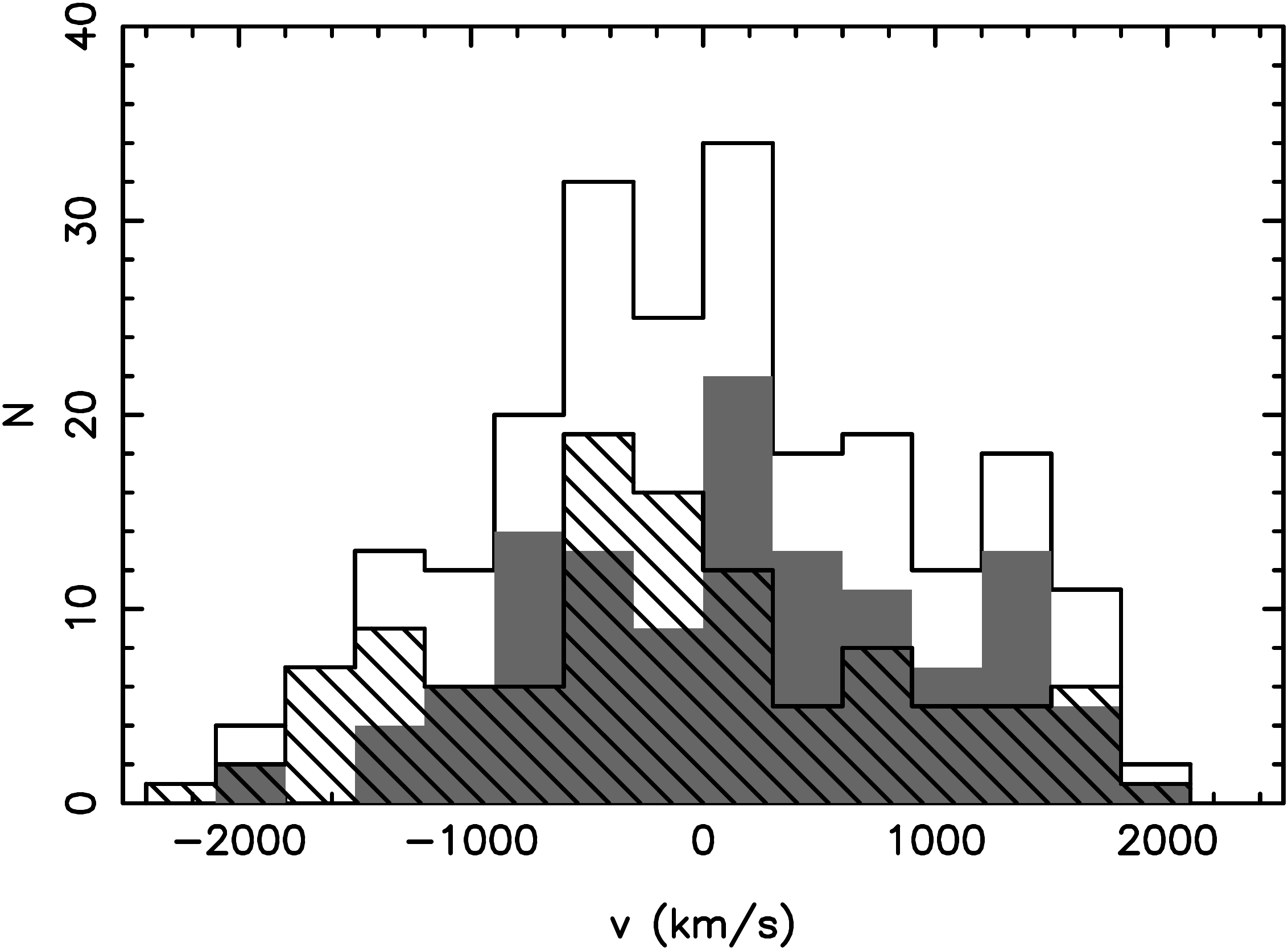}
\caption{Velocity distribution of RXCJ0225. The empty histogram represents the distribution for all galaxies. The filled histogram is for the red population, and the hatched one for the blue population.}
\label{fig:nz_0225} 
\end{figure}

According to the KS test, the velocity distribution of RXCJ0225 (Fig. \ref{fig:nz_0225}) does not differ significantly from a Gaussian, even though an excess of galaxies with $v\sim+1300\,\mathrm{km\,s^{-1}}$ is clearly seen. This is confirmed by the GH test, which returns a positive value of $h_3$ with a significance probability $p=0.8$. We also find that the velocity distribution of red galaxies has a negative $h_4$ component ($p=0.9$), resulting from a symmetric excess of high positive and negative galaxies. We applied a two-sided KS test to compare the distributions of red and blue galaxies, and we obtained a probability $p=0.99$ that they are different; the probability increases to $p=0.996$ when the distributions are limited to within 1.5 Mpc. This difference can be attributed to their average velocities (see e.g. the redshifts given in Table \ref{table:Mvir}), and to the excess of blue galaxies with $v\sim-1500\,\mathrm{km\,s^{-1}}$. These results motivated us to run the KMM algorithm with a three-mode model. The probability that this model provides a better fit than a single Gaussian is $p=0.73$, which is to small to be conclusive. However, the location of the galaxies assigned to each KMM partition reveals that the NE galaxy clump is mainly populated by high-velocity galaxies.

\begin{figure}
\center
\includegraphics[width=7cm, angle=-90]{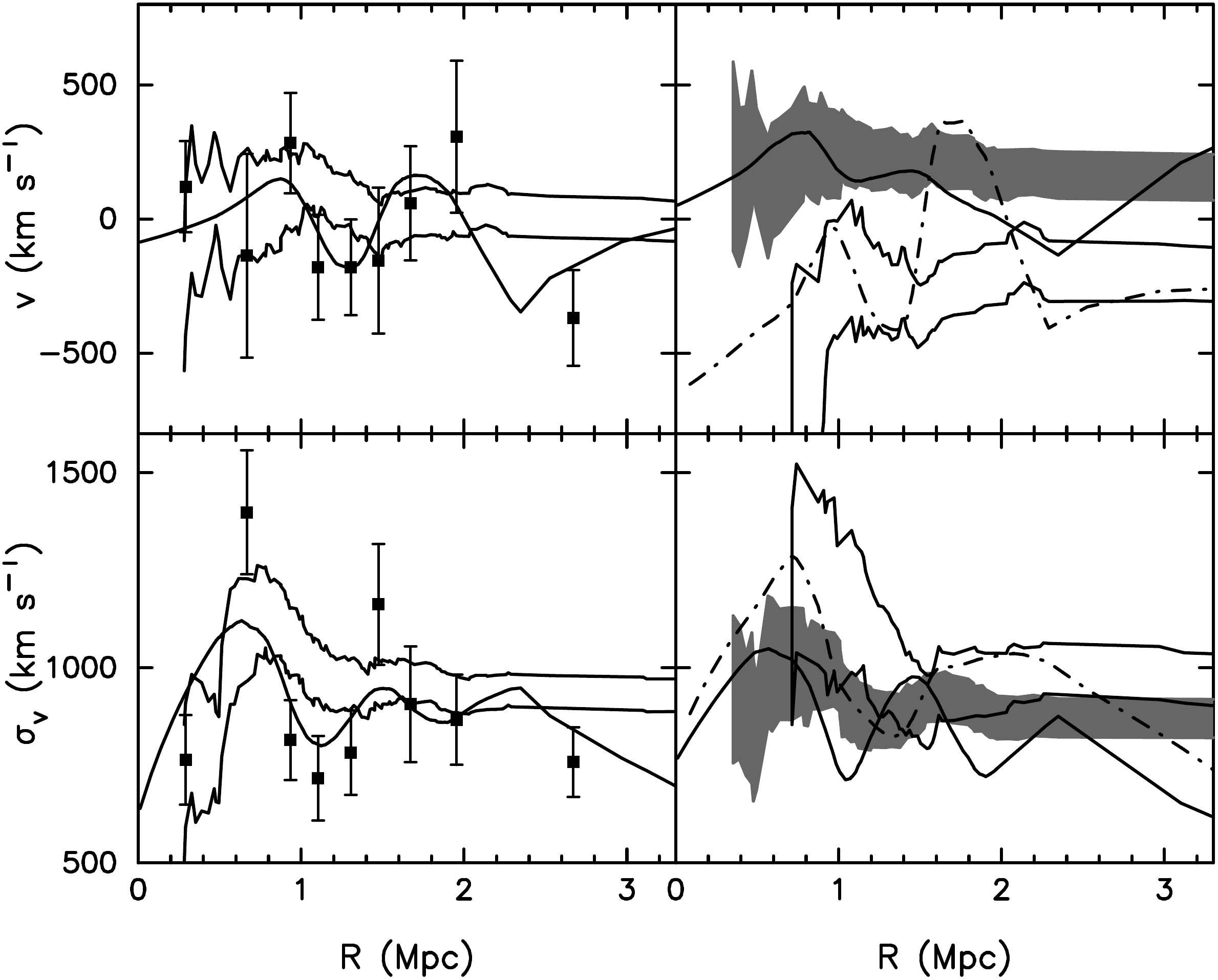}
\caption{RXCJ0225 velocity and velocity dispersion profiles. Top left panel: $1\sigma$ uncertainty on the iVP (uneven continuous lines), VP (squares with $1\sigma$ error bars), and its LOWESS version (smooth continuous line). Bottom left panel: same as the top left panel, but for the iVDP and VDP. Top right panel: iVP for the red (filled region) and blue (empty region) populations, and the smoothed LOWESS VP (solid line for the red galaxies, dot-dashed for the blue ones). Bottom right panel: same as top right panel, but for the iVDP and VDP.}
\label{fig:vprof_0225} 
\end{figure}

The VPs and VDPs of RXCJ0225 (Fig. \ref{fig:vprof_0225}) exhibit three main features. First of all, we see that the excess of blue galaxies with high negative velocities is mainly located in the central region. This results in a velocity difference $\sim400\,\mathrm{km\,s^{-1}}$ between the two galaxy populations within $R_{200}$. Second, the shape of the VDP for the full population, and to a lesser extent, those of the red and blue galaxies rise from low values to nearly $1500\,\mathrm{km\,s^{-1}}$ at $R\sim0.7$ Mpc, and then decrease at larger radii. Inverted VDPs can have different origins, e.g. dynamical friction inducing isotropic orbits, cuspy density profile, or significant differences between the mass and galaxy distributions (e.g. \citealt{denhartog96}, and references therein). Another possibility comes from the mixing of structures with different rest-frame velocities, which is a very likely solution given the results obtained previously. Finally, we see that the iVDPs of the red and blue populations show a marginal agreement at large radius: RXCJ0225 seems to confirm that late-type galaxies are characterised by a larger velocity dispersion.

\begin{figure}
\center
\includegraphics[width=6.5cm, angle=-90]{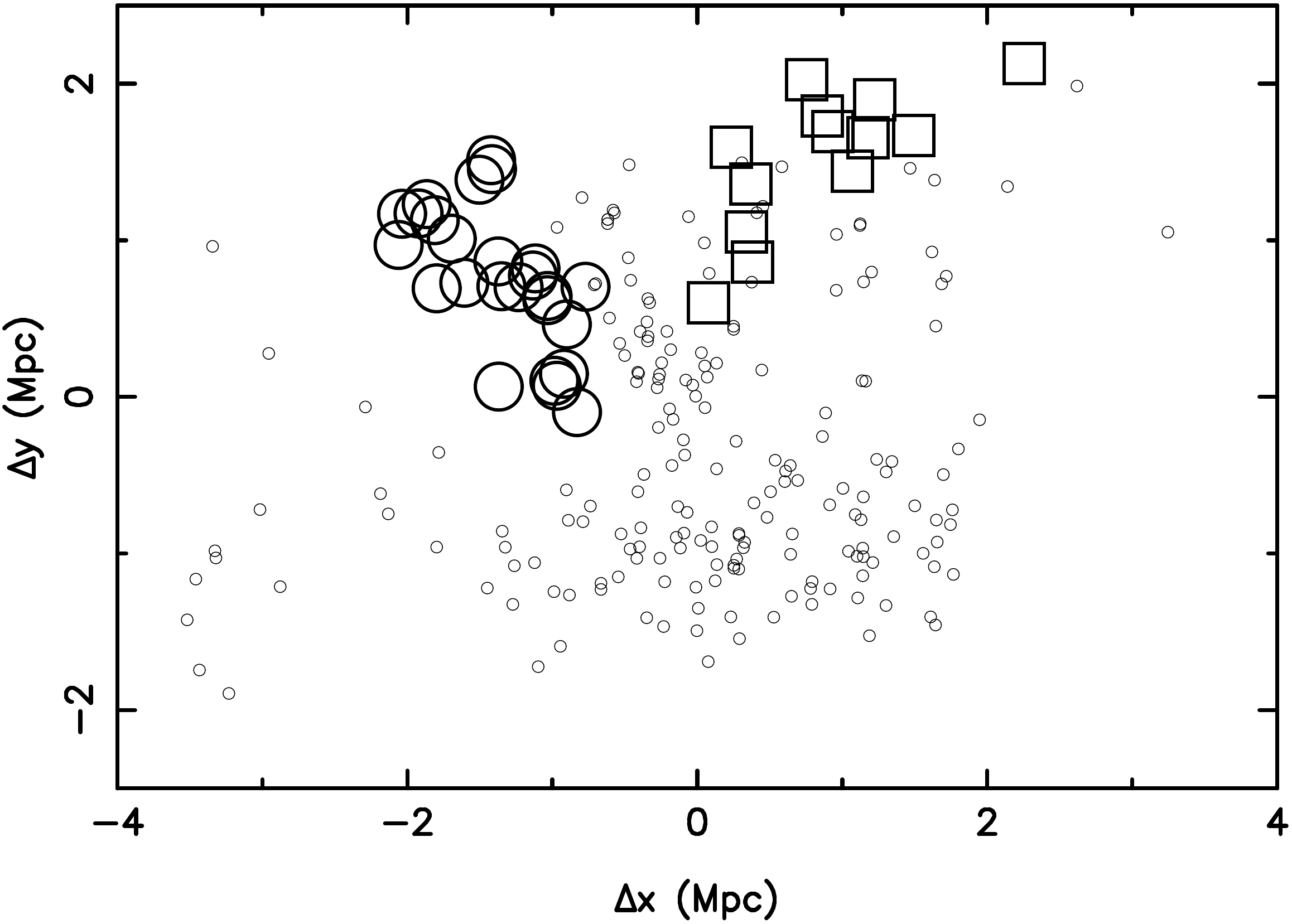}\\
\includegraphics[width=6.5cm, angle=-90]{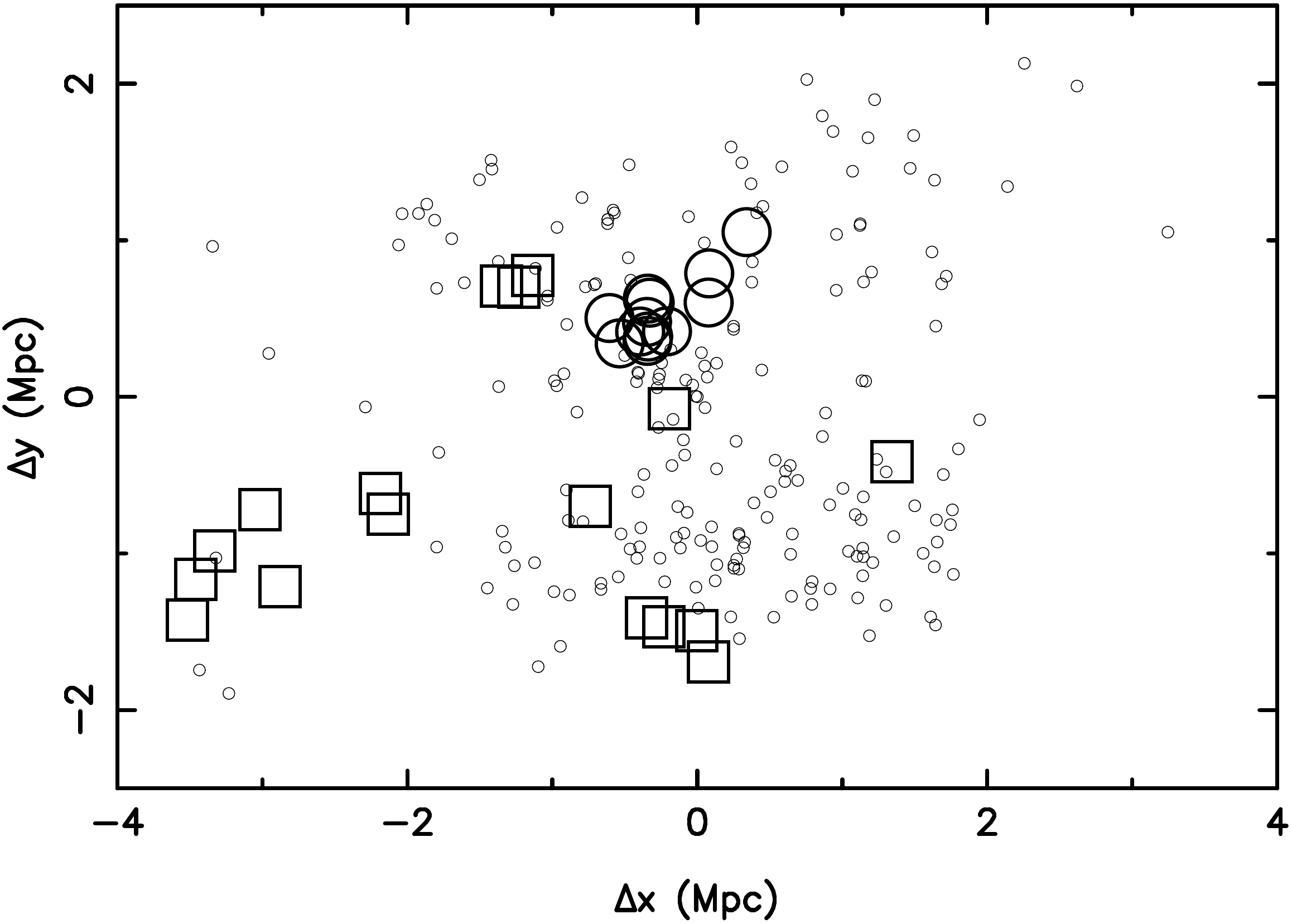}
\caption{Results of the $\Delta_V$ (top panel) and $\Delta_S$ (lower panel) tests for RXCJ0225. Large circles (positive rest-frame velocity or larger velocity dispersion) and squares (negative rest-frame velocity or smaller velocity dispersion) circles show the positions of galaxies for which the local velocity distribution is significantly different from the global value. Small circles show galaxies for which no significant deviation is found.}
\label{fig:0225_DS} 
\end{figure}

The Dressler \& Shectman tests highlight the complex dynamics of the cluster. The $\Delta_V$ statistic finds a probability $p=0.996$ that the cluster contains substructures, and it identifies 38 galaxies having a local velocity significantly different from the overall value. The $\Delta_S$-test returns $p=0.99$ with 29 galaxies associated with cold/hot groups, and the combined $\Delta_{V+S}$-test gives $p=0.998$ with 38 galaxies whose local dynamics differ from the average. By combining the results of the three tests, we find that $27\%$ of the cluster members are part of substructures, $47\%$ of which are red galaxies. These values depend on the selection threshold, but they are nonetheless a good indicator of the dynamical structure of the cluster. As we can see in Figure \ref{fig:0225_DS}, there are three regions of interest: the SE quadrant, which is populated by a cold group of $\sim10$ galaxies; the NW quadrant, which contains $\sim15$ galaxies with negative velocities; and the NE part of the cluster, which contains the most prominent substructure, made of a group of $\sim25$ high-velocity galaxies. The blue galaxies with high negative velocities detected previously do not show up in the $\Delta_V$-test. A more careful inspection of their position reveals that four of them are within the central 0.5 Mpc, with velocities $v\sim-2000\,\mathrm{km\,s^{-1}}$. It is difficult to determine whether these galaxies are foreground interlopers or really part of the cluster. Nonetheless, their presence explains why the $\Delta_S$-test finds a compact hot spot of $\sim10$ galaxies near the cluster centre.

\subsubsection{X-ray analysis}

\begin{figure}
\center
\includegraphics[width=9cm]{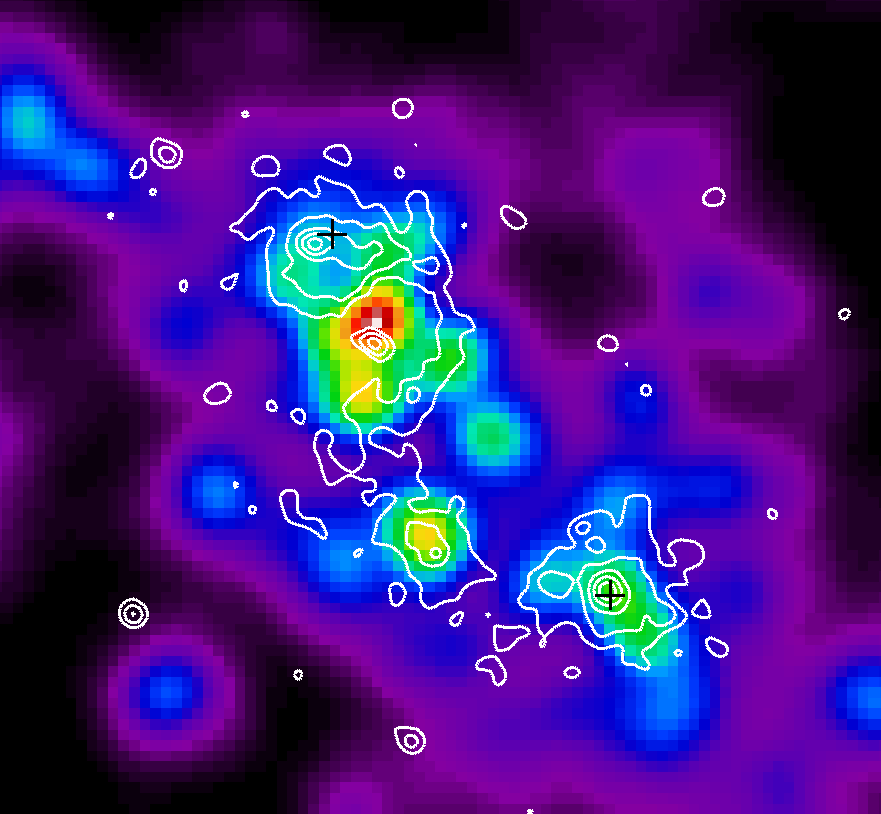}
\caption{X-ray residual emission of RXCJ0225 shown in white contours (starting at $1\sigma$, and increasing by 1 unit), overlaid on the surface density map of red-sequence galaxies. The two crosses are located on the BCG of the two main galaxy clumps. They are approximately $10'$ apart, i.e. $\sim2.1$ Mpc at the cluster redshift.}
\label{fig:0225_X} 
\end{figure}

The X-ray emission has a bimodal morphology within the main galaxy clump, as is seen from the significant peak found next to the BCG (Fig. \ref{fig:0225_X}). The cluster centre, which we defined as the peak of the galaxy surface density, also has a residual X-ray emission. Such a highly disturbed gas distribution indicates a young dynamical state, which is also supported by the large separation between the BCG and the cluster centre. We see a clear diffuse emission associated with the SW galaxy clump, matching perfectly the position of its BCG. Interestingly, there is also some residual emission in between these two main galaxy overdensities, in particular at the position of a third clump. The residual map, smoothed with a Gaussian kernel of width $8''$, shows a continuous contour at the $1\sigma$ level that connects this small clump to the main cluster. Moreover, the overall agreement between the morphology of the X-ray surface brightness and the galaxy surface density suggests that this extended emission is real rather than noise. Recently \cite{eckert15} reported the X-ray observation of filaments around the massive cluster A2744, finding a mass fraction $\sim5-10\%$ associated with baryonic gas. The X-ray emission observed along the filamentary structure connected to RXCJ0225 makes it a very interesting case to confirm their findings and to study the gas properties in this low-density region. A more accurate description of the gas properties within and outside RXCJ0225 will be presented in Chon et al. (in preparation).

\subsubsection{Substructure analysis}

\begin{figure}
\center
\includegraphics[width=8cm, angle=-90]{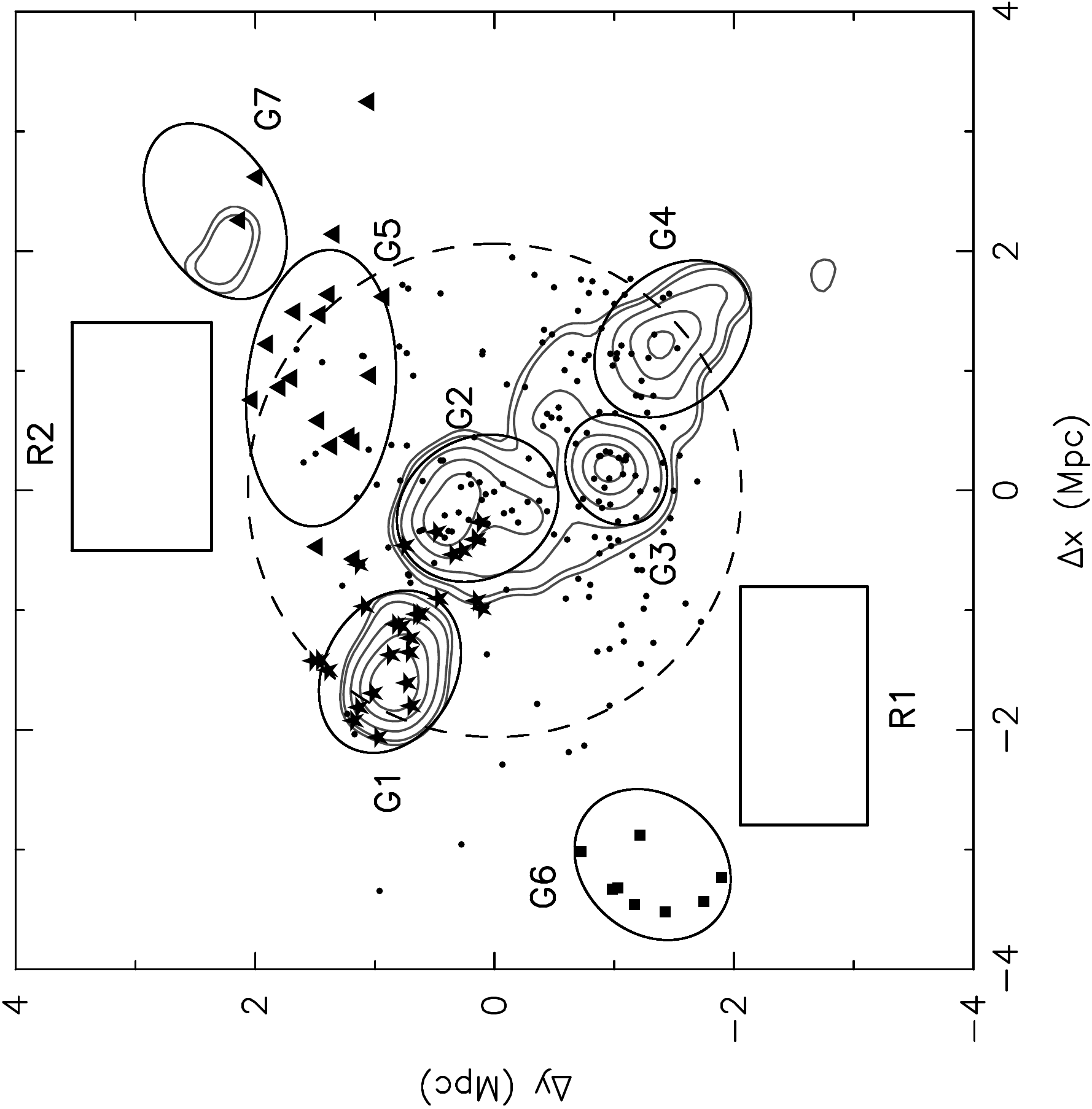}
\caption{Regions of interest for RXCJ0225. Regions R1 and R2, which do not contain any significant galaxy overdensity, were used to estimate the average surface density of field galaxies. The latter is used to correct the richness of each substructure. The elliptical regions labelled G1-G7 are the substructure candidates identified by the galaxy overdensities in the optical maps, and/or from the $\Delta$-tests. The contours trace the surface density of the bright red-sequence galaxies (starting at $5\sigma$ and omitting the two innermost levels, for clarity). Symbols show the location of the spectroscopic members associated with the different KMM partitions. The dashed circle has a radius of $R_{200}\sim2$ Mpc.}
\label{fig:reg_0225} 
\end{figure}

We combined the results of the analyses described above to select and study the substructure candidates of RXCJ0225 (Fig. \ref{fig:reg_0225}). We first defined the four central regions labelled G1-G4, following the NE-SW axis of accretion. Region G2 has a radius $R=\sqrt(ab)\sim650\,\mathrm{kpc}\sim0.3\,\mathrm{R_{200}}$ (where $a$ and $b$ are the semi-major and semi-minor axes of the ellipse G2), thus it only traces the cluster core, and its galaxy content is not representative of the cluster total mass. Interestingly, it has two different centres, depending on whether one looks at the galaxy surface density or luminosity. While the former matches the X-ray emission peak, the latter is located on the BCG, and it has a clear residual X-ray emission (its peak has a S/N$\sim8$; see Fi.g \ref{fig:0225_X}); they are $\sim500$ kpc apart. Therefore, G2 has already undergone the merger of a smaller substructure, and it will significantly increase its mass with the future merging of the surroundings structures. Region G4 contains the overall BCG, has a high fraction of red galaxies $f_{\mathrm{RS}}>0.8$, and its galaxy content suggests a mass $\sim0.7$ smaller than the main clump of region G2. Region G3, which is located between G2 and G4, has a smaller $f_{\mathrm{RS}}$, a fainter BCG, and contains roughly half the number of  galaxies that G4 has. Thus it is likely a small galaxy group, which has a counterpart in the X-ray residual map.

Regions G3 and G4 do not show up on the $\Delta$-tests, so we did not attempt to isolate them with the KMM algorithm. However, using the spectroscopic members located at their position, we estimated their rest-frame velocity and velocity dispersion. Region G3 is nearly at rest compared to the main clump, and its velocity dispersion suggests a mass half that of the main clump. However, owing to its proximity to the main clump, its velocity dispersion may be overestimated (its galaxy content suggest a lower mass, $\sim1/3$ that of G2, thus even smaller when compared to the full cluster). The two-body model finds a very high probability that they are bound. The best solution is incoming, with an angle $\alpha<5\degree$ with the plane of the sky, which makes it difficult to estimate the true infall velocity or the expected time before collision. Region G4 is barely covered by the VIMOS observations, hence its dynamical properties have large uncertainties. Moreover, its VDP may not be flat; adding galaxies located at larger distance from its centre may thus lead to a smaller velocity dispersion. Nonetheless, the dynamical analysis confirms that G4 is a massive object. Like G3, the rest-frame velocity of G4 is compatible with zero, thus the two-body model favours a bound-incoming solution nearly in the plane of the sky. 

Region G1 presents a receding velocity $\delta_v\sim+1000\,\mathrm{km\,s^{-1}}$, and is clearly associated with a KMM partition (stars in Fig. \ref{fig:reg_0225}). According to the two-body model, the most likely solution for G1 is bound-incoming with $\alpha\sim48\degree$, corresponding to an infall velocity of $\sim1600\,\mathrm{km\,s^{-1}}$ at a distance $R\sim2.2$ Mpc from the cluster centre, i.e. around $R_V$. We estimate that G1 will be accreted within the next $\sim0.8$ Gyr. Since the position and elongation of G1 closely matches the orientation of the NE structure, we can suppose that the latter is connected to the cluster from the front side.

In the NW quadrant, a smaller clump was also detected in the optical maps (bottom right panel in Fig. \ref{fig:0225_Nmap}), which we labelled G7, and whose optical properties indicate that it is a galaxy group ($f_{\mathrm{RS}}>0.8$). The lack of spectroscopic redshifts around G7 does not allow us to firmly conclude on its membership to RXCJ0225. However, the $\Delta$-tests suggested the presence of another structure between G7 and the main cluster (top panel in Fig. \ref{fig:0225_DS}). According to the KMM results, we defined the corresponding region G5 (triangles in Fig. \ref{fig:reg_0225}). It is less compact but still presents the typical characteristics of a coherent object, i.e. $f_{\mathrm{RS}}>0.7$ and a rather bright galaxy ($\sim0.65$ mag fainter than G2's BCG, but $\sim0.3$ mag brighter than G7's). Its dynamical properties correspond to a low-mass, high-velocity (negative, hence falling from behind) object with a probability of $\sim60\%$ of being bound to RXCJ0225. The two-body model favours a bound-incoming solution characterised by an angle $\alpha\sim47\degree$, corresponding to an infall velocity of $\sim1400\,\mathrm{km\,s^{-1}}$ at a distance $R\sim2.1$ Mpc from the centre, and $t_{coll}\sim0.9$ Gyr.

The last region of interest (SE quadrant, labelled G6, squares in Fig. \ref{fig:reg_0225}) was only detected based on its specific dynamics (bottom panel in Fig. \ref{fig:0225_DS}). It barely stands out from the background and its galaxy content is dominated by blue members. Because of the low velocity difference with the main body, it has a high probability of being bound to it. The two-body model favours a bound-incoming solution with $\alpha\sim20\degree$, $v\sim1100\,\mathrm{km\,s^{-1}}$, $R\sim3$ Mpc, and $t_{coll}\sim1.5$ Gyr.

\subsubsection{Summary}

To summarise, RXCJ0225 has a complex multimodal morphology. It is made of three overdensities of bright red-sequence galaxies (G2, G3, and G4), which are also detected via their X-ray emission; another substructure, mostly populated by faint members, is also observed further SW (see Fig. \ref{fig:0225_Nmap}, top left panel). They are distributed along the main axis of accretion, extending SW over at least $\sim4$ Mpc, and most likely very close to the plane of the sky. The relatively small magnitude gaps between the BCG of these galaxy clumps ($\sim0.4$ mag) add more evidence to the possibility that we are observing RXCJ0225 in an early phase of its dynamical history (e.g. \citealt{smith10}). The cluster is embedded in a rich large-scale environment: the distribution of red-sequence galaxies covers a continuous narrow band extending over $\sim6$ Mpc (as traced by the $3\sigma$ contour of the galaxy surface density), from the SW corner to the substructure G1. We found additional evidence of a large-scale filamentary-like structure with a chain of five overdensities of blue galaxies extending in the NE direction (top right panel in Fig. \ref{fig:0225_Nmap}). Hence the total size of the structure is 8 Mpc at the cluster redshift, but it could be even larger, since it reaches the limit of the WFI FOV in the SW and NE corners. Based on the galaxy content of the different structures, and using the velocity dispersion of those associated with a KMM partition, we estimate that RXCJ0225 will accrete $\sim15-25\%$ of its current mass during the next Gyr from the NE and NW regions (G1, G5, and G7). The two main clumps along the SW part of the large-scale structure (G3 and G4) will further increase the total mass by a factor of $\sim1.5-2$. Overall, RXCJ0225 appears to be a growing massive cluster, as traced by the large amount of substructures located within (or close to) its virial radius.

This cluster also highlights the need for a multiwavelength approach to obtain a precise picture of its properties. On the one hand, the KMM partition of the main body has a velocity dispersion $\sigma_P\sim850\,\mathrm{km\,s^{-1}}$ within $R_{200}$. This corresponds to a mass $M_{\sigma}$ that is $\sim1.4$ times smaller than that obtained in Section 4.1, i.e. prior to removing the substructures. On the other hand, the dynamical analysis alone does not allow for the detection of G3 and G4, thus the estimator $M_{\sigma}$ misses a significant fraction of the cluster total mass. We can make a similar comparison for the virial estimator $M_{200}$. We assume that the main clump is G2, and that G3 has not been accreted yet. Hence we exclude the full SW region that covers G3 and G4. We also cut out the regions containing G1, G5, and G7. Finally, we also excluded the galaxies within G3 and G4 from the main KMM partition. In doing so, we obtained a mass $M_{200}=0.84_{-0.15}^{+0.13}\times\mathrm{10^{15}\,M_{\odot}}$, i.e. $\sim1.5$ times smaller than the value derived assuming a single component. Assuming a NFW profile, we find $R_{500}\sim1.15$ Mpc, which is the projected separation between G2 and G3. Thus, this small substructure currently sits in the outskirts of the cluster. Adding the crude estimates for G1, G3, and G4, we obtain a total mass of $\sim1.5\times\mathrm{10^{15}\,M_{\odot}}$ within the virial radius, $\sim50\%$ of which is contained within substructures.

\subsection{RXCJ0528}

\subsubsection{Optical analysis}

\begin{figure}
\center
\includegraphics[width=8.5cm, angle=-90]{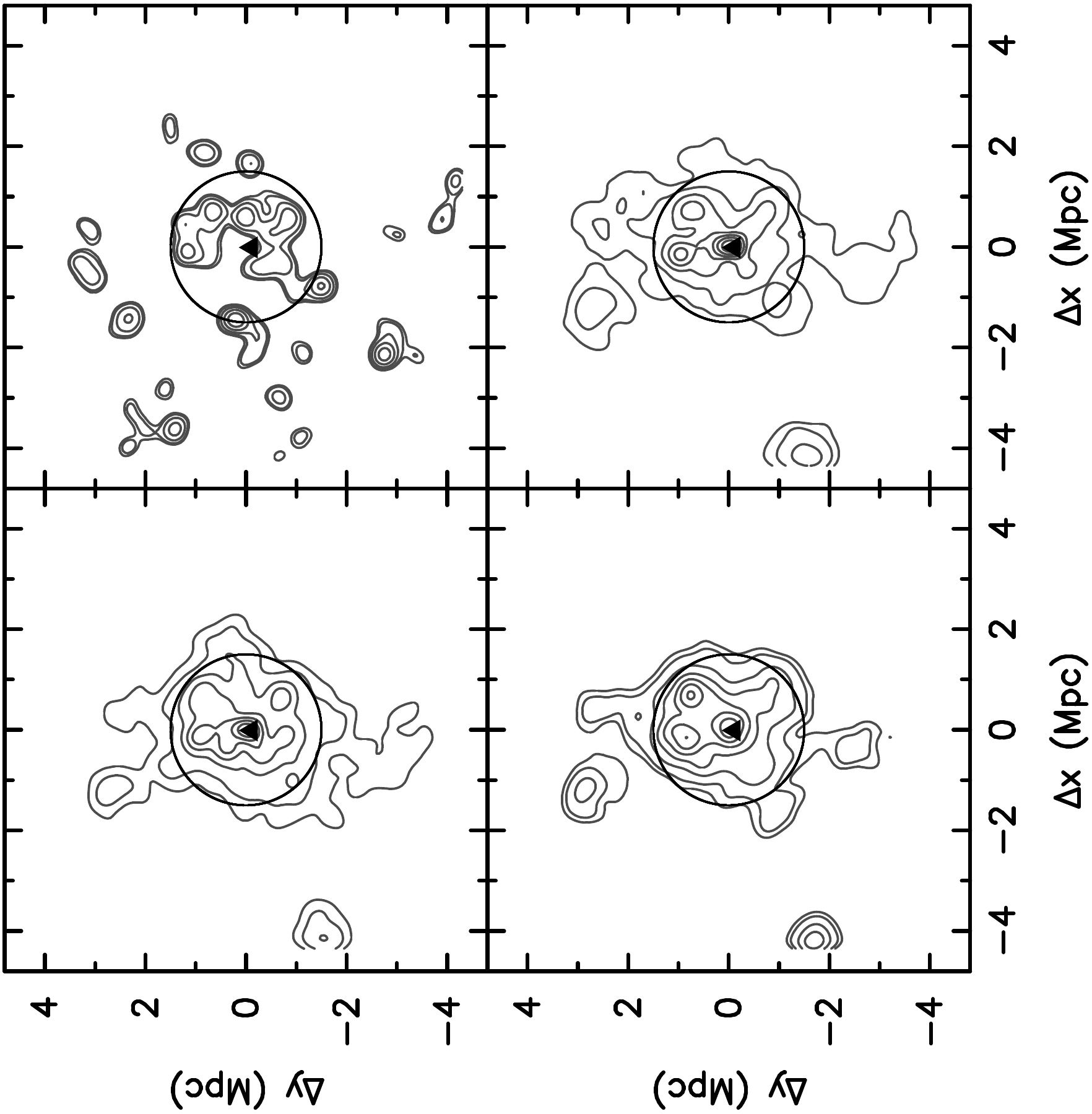}
\caption{Same as Fig. \ref{fig:0225_Nmap}, but for RXCJ0528.}
\label{fig:0528_Nmap} 
\end{figure}

The large-scale galaxy distribution of RXCJ0528 is characterised by a well-defined galaxy clump (Fig. \ref{fig:0528_Nmap}). The overdensities of the blue galaxies do not show strong evidence for a filamentary structure; the blue galaxies are simply more scattered around the cluster than the red galaxies. It should be noted that the total fraction of red galaxies is significantly smaller than those of the other clusters (see Table \ref{table:members}). This is partly explained by its smaller mass (Table \ref{table:Mvir}), but also by the lack of massive and evolved structures in its surroundings: only two major secondary clumps are actually detected in the luminosity and surface density maps, N and SE from the main body (more pronounced for the bright members). Furthermore, a foreground structure was detected in PPS at only $\sim-4500\,\mathrm{km\,s^{-1}}$ in the cluster rest-frame, i.e. at a radial distance $R=c\delta z/H(z)\sim70$ Mpc. Owing to the larger uncertainty in the photometric redshifts of the blue galaxies, it is not surprising that part of these galaxies were included in the combined catalogue, hence reducing the fraction of red members. In fact, the foreground structure could be associated with the overdensities of blue galaxies found $\sim1.5$ Mpc east of the BCG (top right panel in Fig. \ref{fig:0528_Nmap}). Alternatively, the larger fraction of blue galaxies can be also partly attributed to a filamentary structure close to the line of sight.

\begin{figure}
\center
\includegraphics[width=6.5cm, angle=-90]{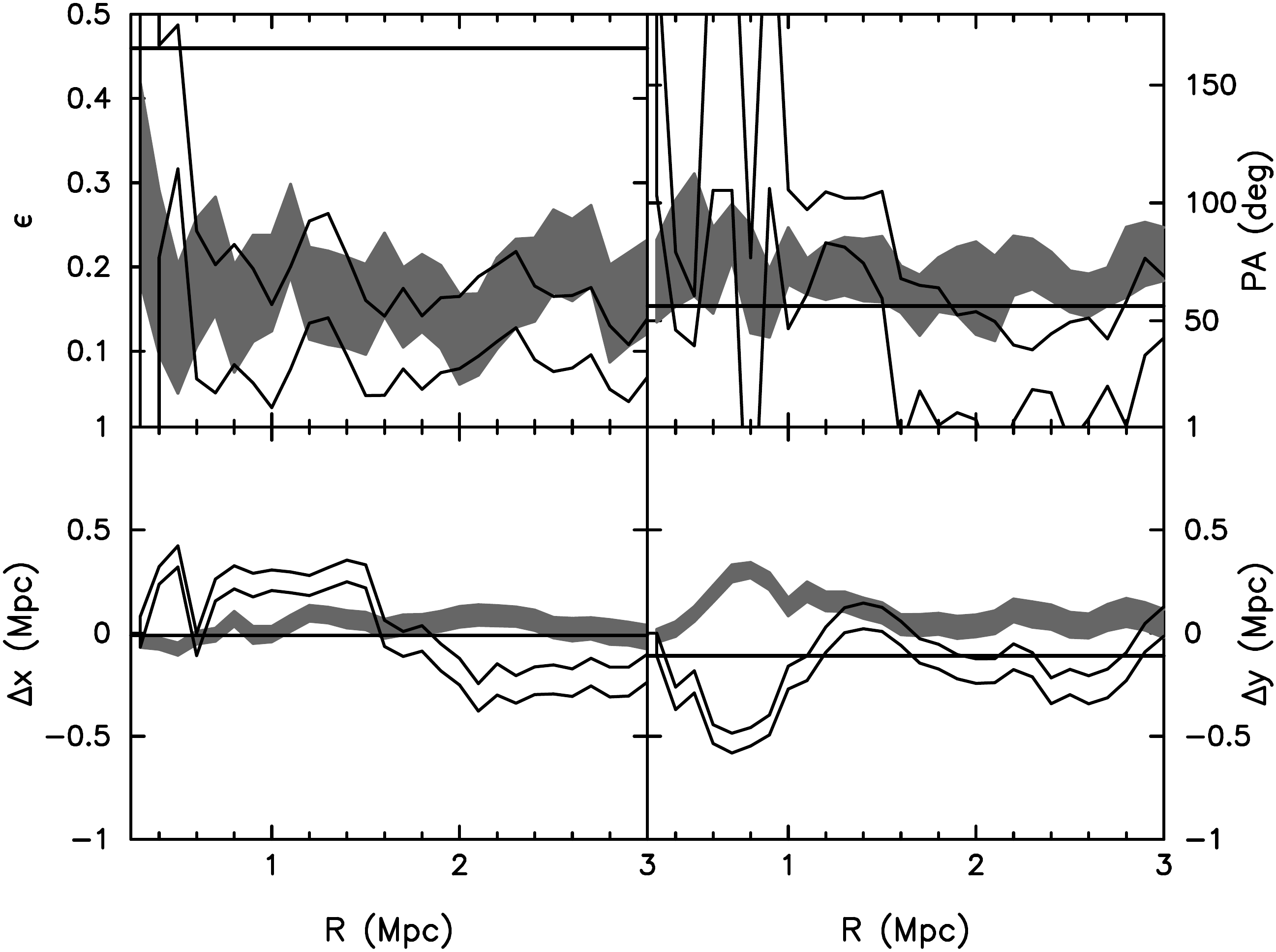}
\caption{The same as Fig. \ref{fig:0225_ell}, but for RXCJ0528.}
\label{fig:0528_ell} 
\end{figure}

The ellipticity profile of RXCJ0528 indicates a mild $e\le0.2$ N-S elongation (Fig. \ref{fig:0528_ell}). It is roughly constant at all radii, indicating the absence of major substructures. Only a small centroid shift is detected at $R\sim0.7$ Mpc for the red population. It corresponds to the two small galaxy overdensities found in the surface density of bright red-sequence galaxies. The BCG of RXJ0528 is located at its centre, which is typical for a cluster that has not undergone any recent major mergers. However, a second bright galaxy only $\sim0.35$ mag fainter is found $\sim200$ kpc in the north, which suggests that the cluster core has not yet reached equilibrium. The orientation of the BCG matches that of the distribution of galaxies within $R_{200}$; in addition to the mild cluster ellipticity, the two galaxy overdensities, and the second bright galaxy, this points towards an accretion history along the N-S axis.

\subsubsection{Dynamical analysis}

\begin{figure}
\center
\includegraphics[width=6.5cm, angle=-90]{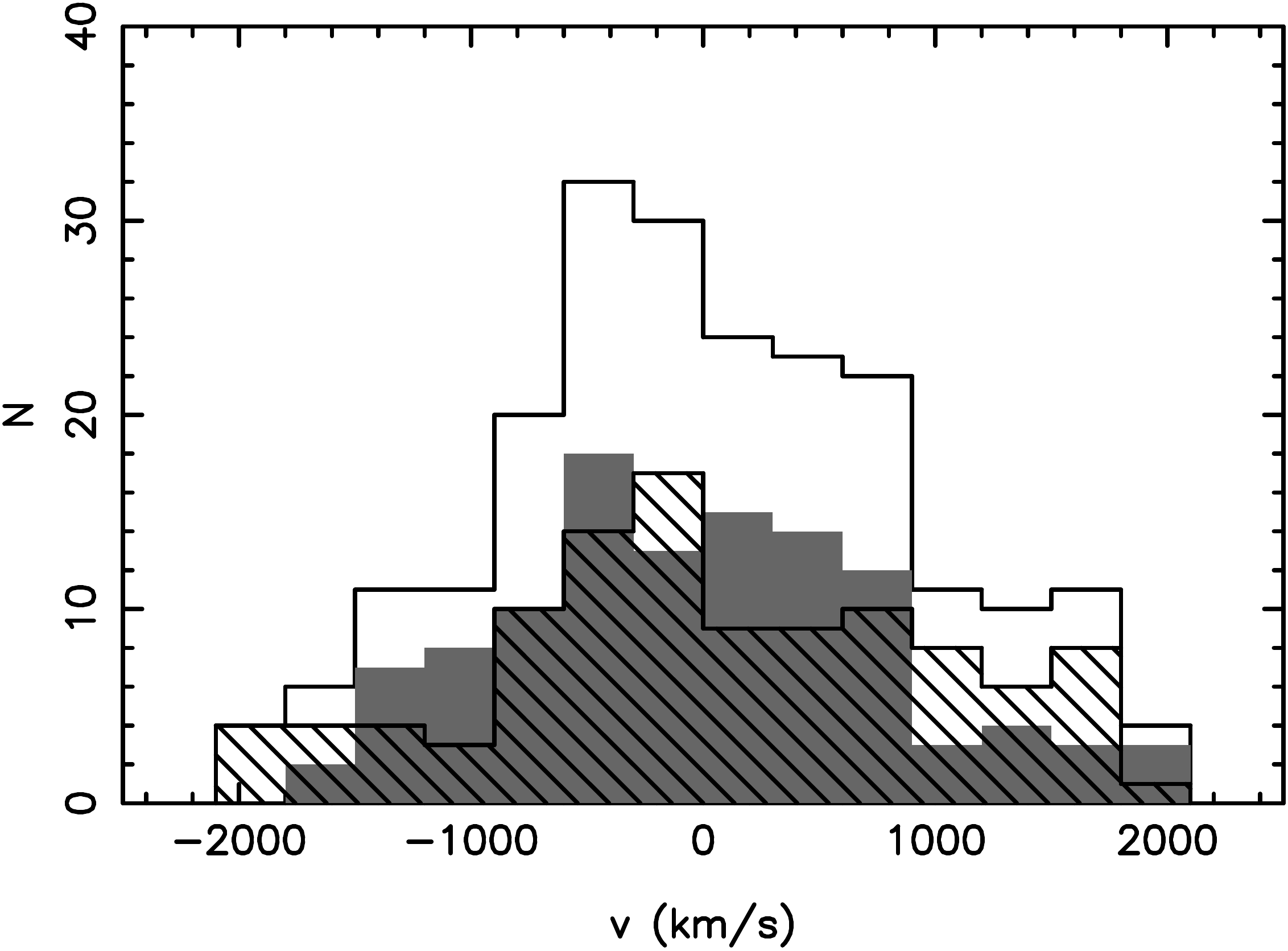}
\caption{Same as Fig. \ref{fig:nz_0225}, but for RXCJ0528. The distributions of red and blue galaxies are not statistically different according to the KS test.}
\label{fig:nz_0528} 
\end{figure}

According to the KS test, the velocity distribution of RXCJ0528 (Fig. \ref{fig:nz_0528}) deviates from a Gaussian with a probability $p=0.87$. It is characterised by an excess of galaxies at $v\sim+1500\,\mathrm{km\,s^{-1}}$. This is confirmed by the GH test, which returns a positive $h_3$ with a probability $p=0.8$ (0.88 when limiting the distribution within the central 1.5 Mpc). A two-sided KS test does not find a significant difference between the two galaxy populations, even though the high-velocity galaxies are mostly blue members. We used the KMM algorithm to fit the velocity distribution with two components, limited to the galaxies within 2 Mpc, since the cluster is well located within this region. As an initial guess, we used the combination N(v)=$0.7\times\mathcal{N}(0,700)+0.3\times\mathcal{N}(1500,400)$, where $\mathcal{N}(\mu_v,\sigma_v)$ describes the parameters of a Gaussian distribution. The best fit corresponds to the combination N(v)$=0.85\times\mathcal{N}(-165,746)+0.15\times\mathcal{N}(1360,328)$, with a probability of improvement $p=0.85$; the high-velocity KMM partition contains 17/26 blue galaxies.

\begin{figure}
\center
\includegraphics[width=7cm, angle=-90]{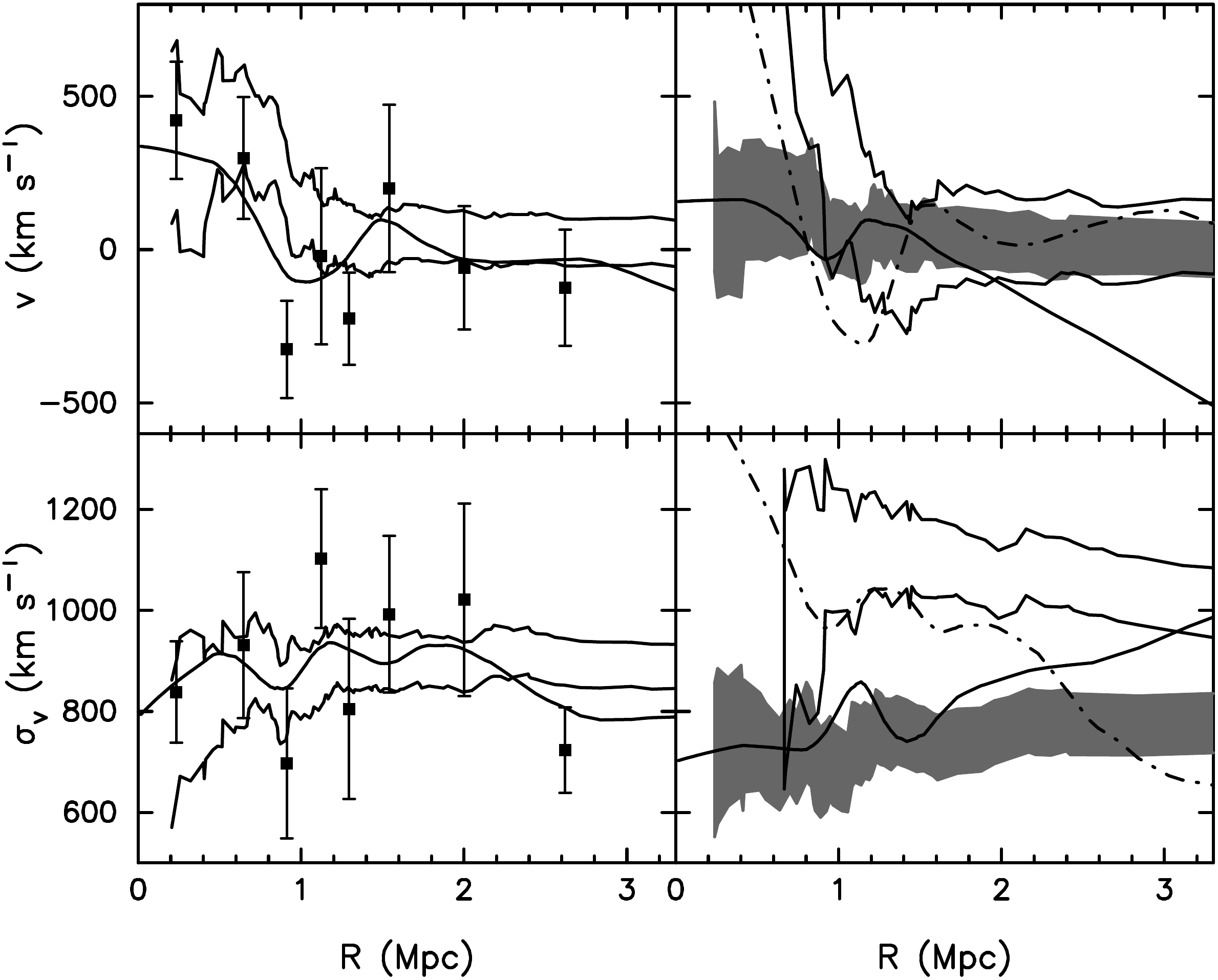}
\caption{Same as Fig. \ref{fig:vprof_0225} but for RXCJ0528.}
\label{fig:vprof_0528} 
\end{figure}

The VP of blue galaxies indicates that this high-velocity component is located in the central region, as seen by its steep gradient within $\sim1.2$ Mpc (Fig. \ref{fig:vprof_0528}). The VP of the red population is flat in the central region, and decreases outside $R_{200}$, which produces an increase in its VDP. The VDP of the blue population presents a strong negative gradient at all radii, as expected from the typical anisotropy profile of late-type galaxies and from the mixing with the central high-velocity component. At large radius, the iVDP of the two populations are significantly different: within $R_{200}$, the velocity dispersion of the red population is $\sim400\,\mathrm{km\,s^{-1}}$ smaller. However, this difference is mainly driven by the central high-velocity blue galaxies, which may contain interlopers or high-velocity infalling galaxies. Therefore, the apparent contrast in velocity dispersion should be taken with caution. The VDP of the full population is rather flat at all radii, indicating a predominance of red galaxies, and the absence of major structures with a large rest-frame velocity.

\begin{figure}
\center
\includegraphics[width=6.5cm, angle=-90]{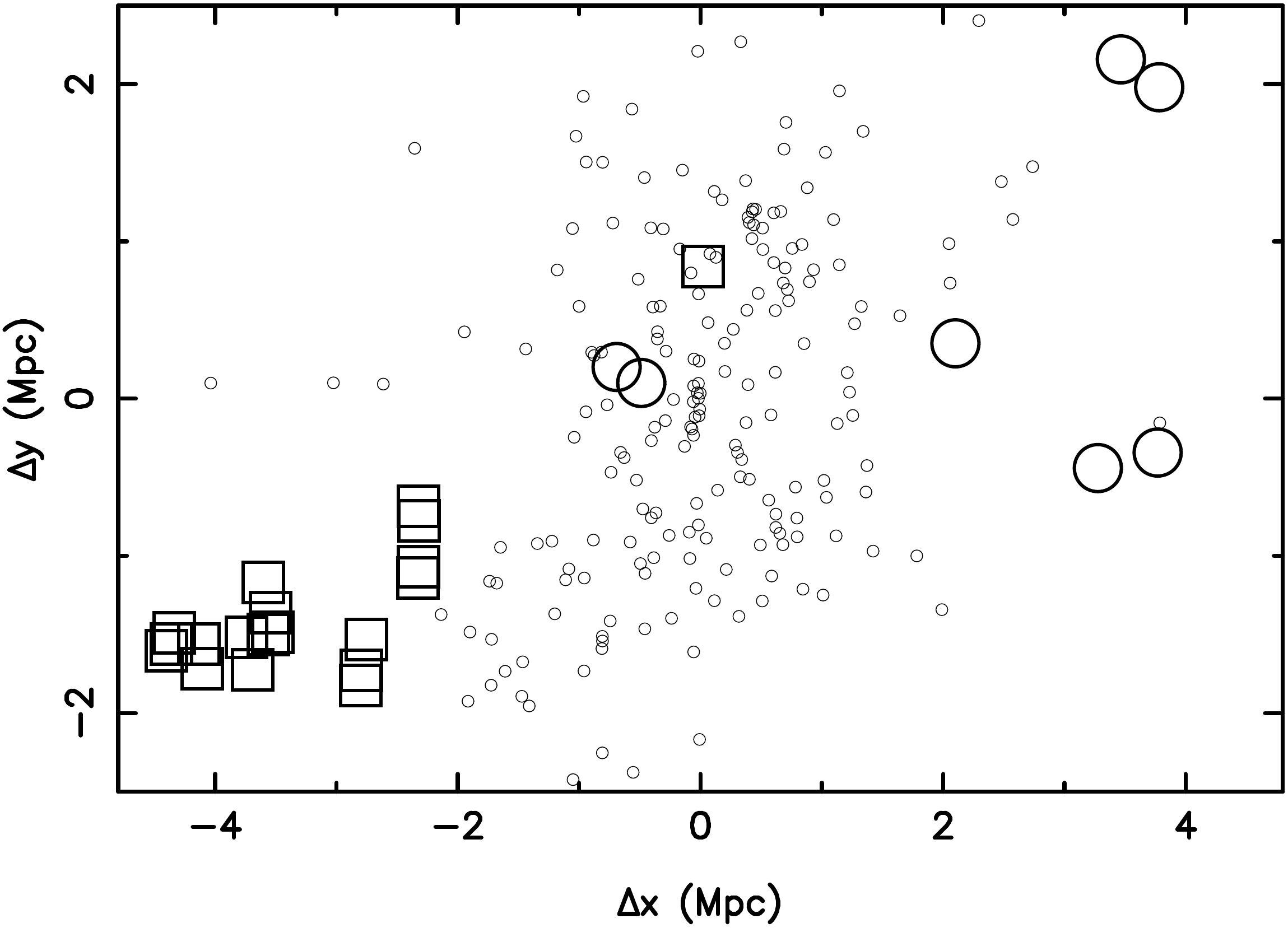}\\
\includegraphics[width=6.5cm, angle=-90]{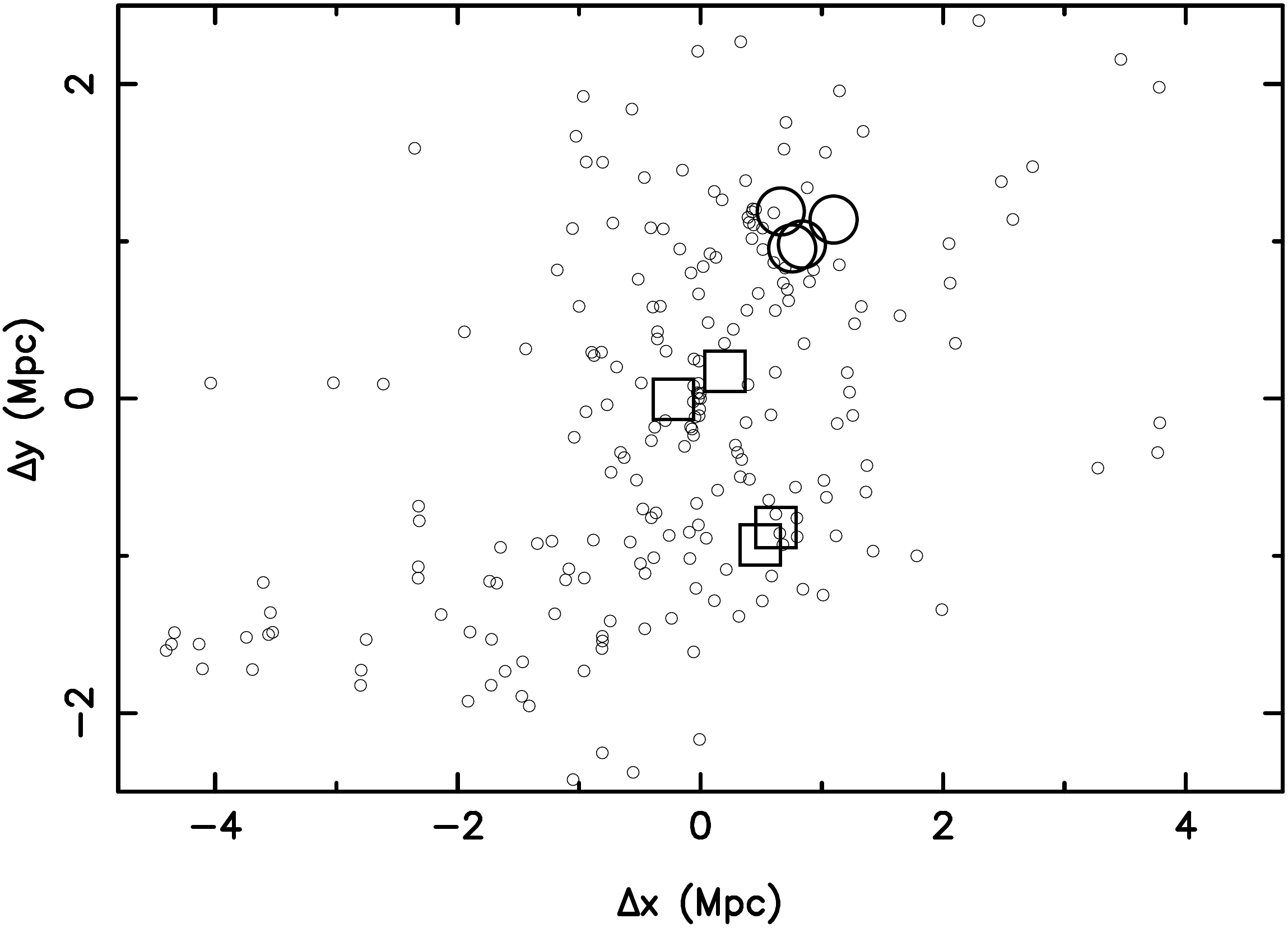}
\caption{Same as Fig. \ref{fig:0225_DS} but for RXCJ0528.}
\label{fig:0528_DS} 
\end{figure}

The $\Delta$-tests confirm the apparent quiet dynamical state of RXCJ0528 (Fig. \ref{fig:0528_DS}). The $\Delta_V$-test returns a probability of substructure $p=0.92$, with 26 galaxies having a local different velocity. However, inspection of their location reveals only one main structure. It is located in the SE quadrant, $\sim4.5$ Mpc from the cluster centre. It contains $\sim20$ galaxies, and is characterised by a local velocity $v\sim-800\,\mathrm{km\,s^{-1}}$, which explains the gradient in the VP of the red galaxies. The $\Delta_S$-test only finds eight galaxies associated with substructures, for a probability $p=0.79$. Half are part of a hot group, which most likely results from the presence of the high-velocity blue galaxies. In total, only a small fraction of RXCJ0528 spectroscopic members ($\sim16\%$) are part of substructures, most of which are located far beyond $R_{200}$. The high-velocity component does not show up as a compact group in the $\Delta_V$-test. Therefore, it is difficult to determine whether these galaxies are background interlopers, or are falling onto the cluster from a foreground filament close to the line of sight. The presence of several overdensities of blue galaxies distributed around the cluster centre may support the second scenario, although, as stated previously, part of them are confirmed foreground interlopers at a distance $\sim70$ Mpc. 

\subsubsection{X-ray analysis}

\begin{figure}
\center
\includegraphics[width=9cm]{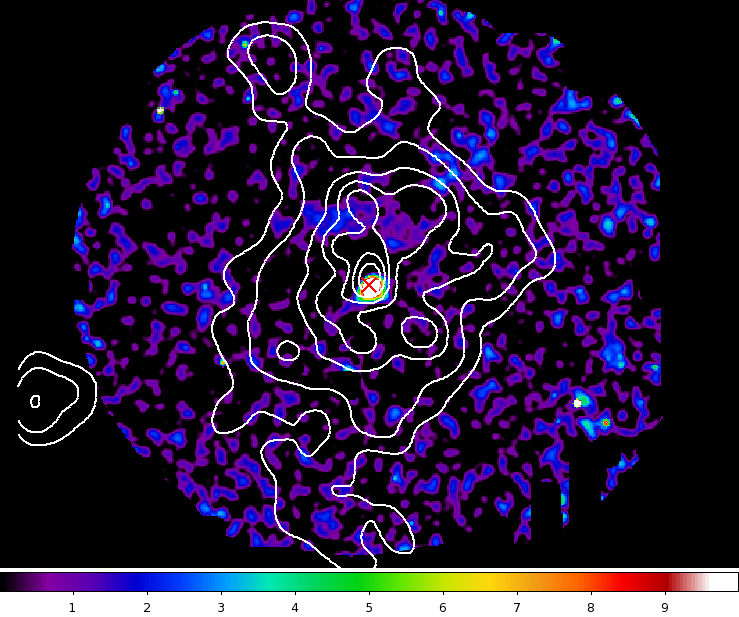}
\caption{Residual X-ray emission of RXCJ0528 (colour-coded in $\sigma$ levels). The white contours trace the surface density of the red-sequence galaxies. The red cross marks the position of the central BCG.}
\label{fig:0528_X} 
\end{figure}

The X-ray emission of RXCJ0528 is characterised by a slight centroid shift along the NE-SW axis. Therefore, to avoid the characteristic half-moon shape in the residual emission, we did not use the X-ray peak to centre the profile. Instead, we used the centre of the best-fit ellipse to the surface brightness isophote at a distance $\sim0.5\,\mathrm{R_{500}}$ from the emission peak. It is located $\sim100$ kpc NE from the BCG (above this radius the position of the isophotes does not change significantly). As seen in Figure \ref{fig:0528_X}, the residual emission presents only a small asymmetry due to the shifted position of the cool core with respect to the large-scale emission. No major substructures are found, which agrees with the picture of a quiet state as deduced from the dynamical analysis.

\subsubsection{Substructure analysis}

\begin{figure}
\center
\includegraphics[width=8cm, angle=-90]{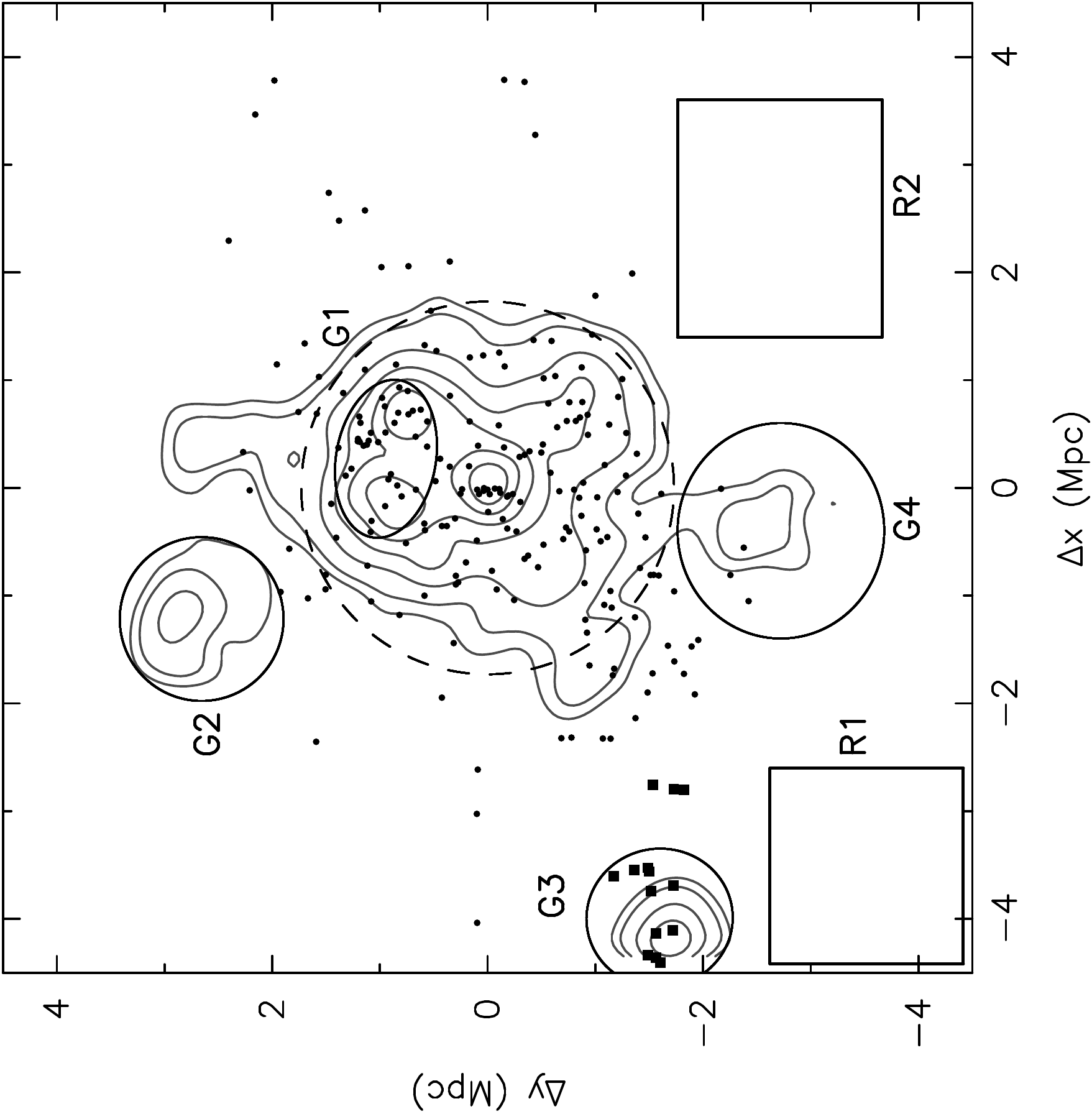}
\caption{Same as Fig. \ref{fig:reg_0225}, but for RXCJ0528: substructure candidates (regions G*), regions used to estimate the density of field galaxies (R1 and R2), and contours tracing the surface density of the bright red-sequence galaxies. Symbols show the location of the spectroscopic members associated with the different KMM partitions. The dashed circle has a radius of $R_{200}\sim1.7$ Mpc.}
\label{fig:reg_0528} 
\end{figure}

The main feature of RXCJ0528 is the presence of two galaxy clumps slightly north of the centre (bottom left panel in Fig. \ref{fig:0528_Nmap}). Therefore, we defined a first region encompassing them (G1, see Fig. \ref{fig:reg_0528}). Located $\sim1$ Mpc from the BCG, it contains $\sim25\%$ of the red-sequence galaxies found within $R_{200}$ and hosts a bright galaxy $\sim0.6$ mag fainter than the RXCJ0528 BCG. As noted previously, a second BCG located only $\sim200$ kpc from the cluster centre is another indication of a possible recent merger. The velocity of G1 is similar to that of the main body, so the merging axis must be very close to the plane of the sky. Owing to its proximity with the cluster centre, we expect an overlap with the main body. Hence we did not estimate its dynamical mass. However, its optical properties, which are most likely overestimated, suggest that it contains the remnants of two small galaxy groups. The overall N-S elongation is also seen at larger scales with two relatively small galaxy overdensities located beyond $R_{200}$, labelled G2 and G4. The southern one, G4, contains more galaxies, whereas G2 is more luminous (its BCG is $\sim0.5$ mag brighter), and has a larger fraction of red galaxies. This suggests that G2 is an evolved galaxy group, while G4 most likely traces a filamentary structure feeding the main cluster from the south. From their optical properties, we estimate a combined mass $\mathrm{M_{G2+G4}}\sim0.10-0.15\times M_{200}$. However, since no dynamical information is available for these two regions, we cannot conclude firmly about their connection to RXCJ0528.

The last significant structure was found $R_P\sim4.3$ Mpc east of from the cluster (region G3, squares in Fig. \ref{fig:reg_0528}). It sits near the edge of the WFI FOV, so its optical properties are most likely underestimated. It appears to be dense and dominated by bright red galaxies ($L\approx2N$). However, given its position and velocity, G3 has a very low probability of being connected to RXCJ0528 and must be instead a foreground galaxy group.

\subsubsection{Summary}

To summarise, RXCJ0528 presents mild evidence of a N-S axis of accretion close to the plane of the sky. North of a well-defined centre, two small galaxy clumps hosting a bright member indicate that the cluster is not yet fully relaxed. This is further supported by the presence of another bright galaxy very close to the centre, and by a centroid shift of $\sim100$ kpc between the position of the cool core and the large-scale X-ray emission. The central velocity distribution departs slight from Gaussianity due to a possible foreground filament feeding the cluster with high-velocity blue galaxies. This scenario would also explain the presence of numerous overdensities of blue galaxies, although projected interlopers cannot be entirely ruled out. We found that the two galaxy populations are characterised by a significantly different velocity dispersion. Owing to the lack of major substructures, we conclude that RXCJ0528 is a cluster that does not require precise modelling to estimate its mass, provided that red galaxies are used as dynamical tracers. 

\subsection{RXCJ2308}

\subsubsection{Optical analysis}

\begin{figure}
\center
\includegraphics[width=8.5cm, angle=-90]{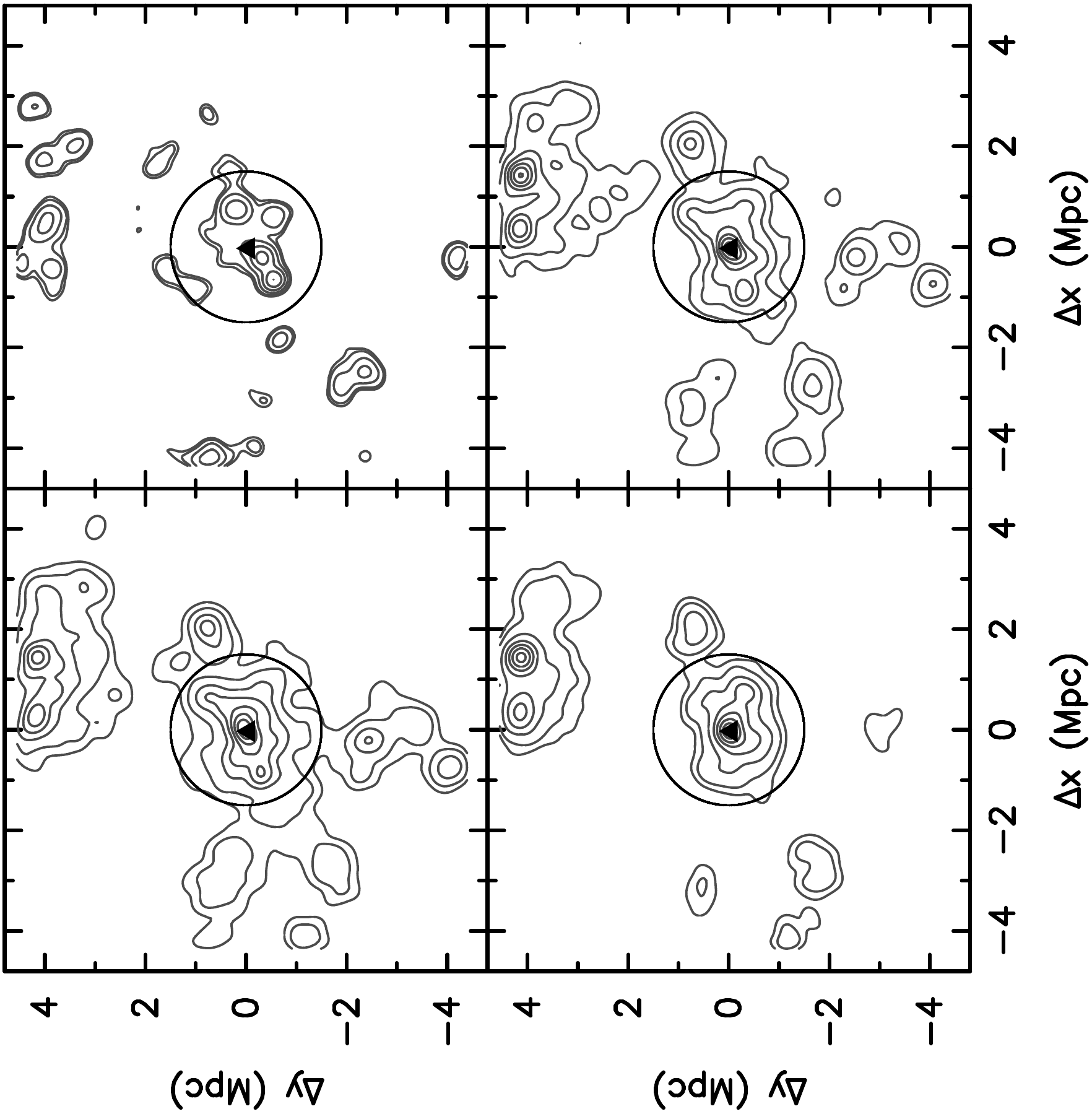}
\caption{Same as Fig. \ref{fig:0225_Nmap}, but for RXCJ2308.}
\label{fig:2308_Nmap} 
\end{figure}

RXCJ2308 is embedded in a very rich environment containing many secondary clumps (Fig. \ref{fig:2308_Nmap}). They are distributed along two main axes, roughly oriented N-S and E-W. In the northern part, a very large structure is detected at $\sim4.5$ Mpc. It is composed of two clumps whose combined extent is similar to that of the main cluster. At $\sim2$ Mpc west of the centre, a clear overdensity is observed, in particular in the distribution of bright members. Towards the south, two more concentrations are detected at $\sim3$ and $\sim4.5$ Mpc from the BCG, and to the east we find two other elongated structures above and below the E-W axis passing through the cluster centre. Most of these structures are also detected in the density map of blue galaxies. Furthermore, they are populated by bright red-sequence galaxies, which indicate that they are already evolved galaxy groups. Such a rich environment suggests that RXCJ2308 is part of a supercluster. 

\begin{figure}
\center
\includegraphics[width=6.5cm, angle=-90]{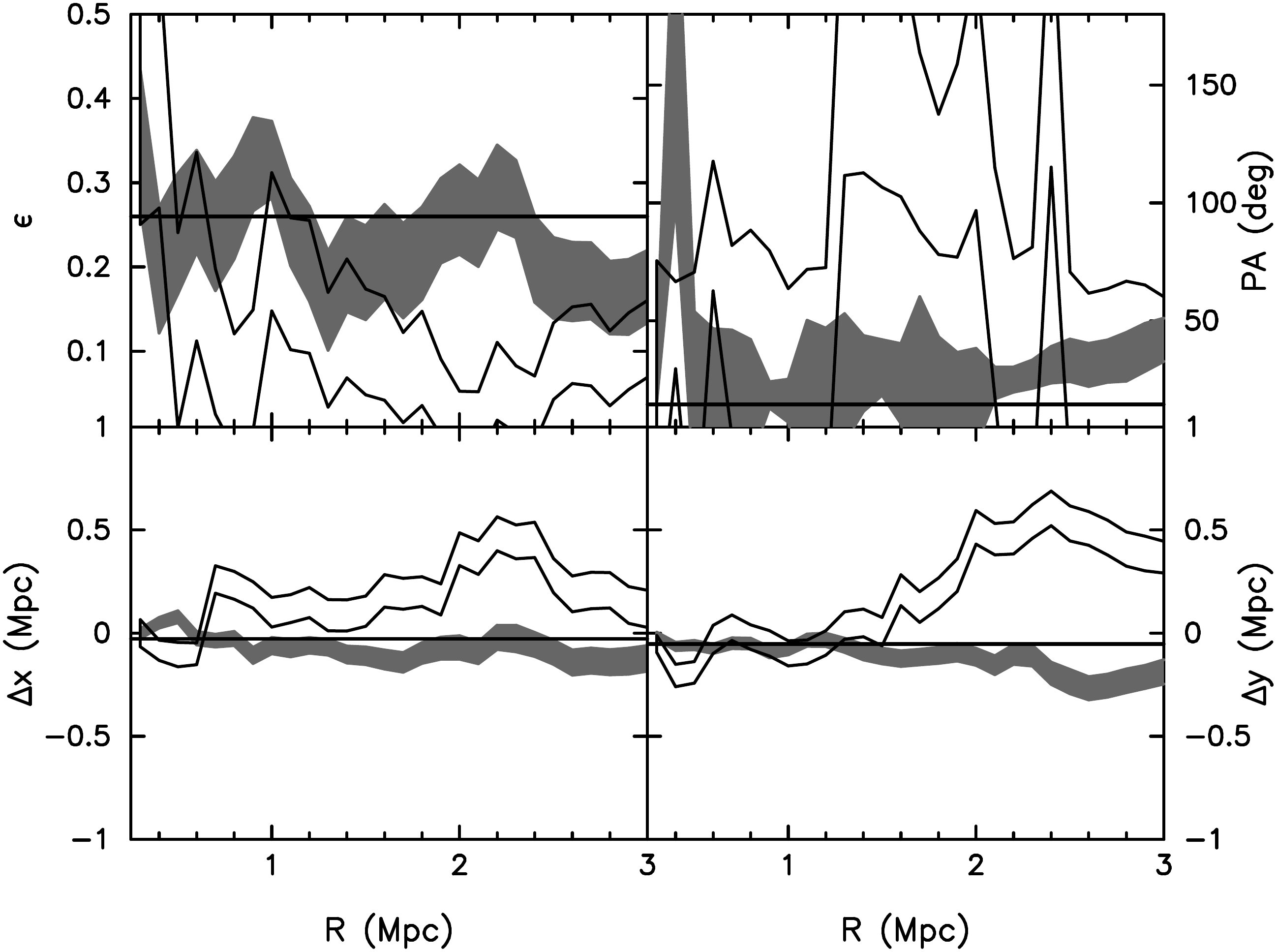}
\caption{The same as Fig. \ref{fig:0225_ell} but for RXCJ2308.}
\label{fig:2308_ell} 
\end{figure}

The main body of the cluster is rather elongated, $e\sim0.25$, along an E-W axis (Fig. \ref{fig:2308_ell}). Owing to the several galaxy clumps found at larger scale, the ellipticity profile shows strong radial variations. However, it is interesting to note that the centroid of the red population remains constant at all radii, which implies that RXCJ2308 sits in the core of its large-scale environment. The BCG is also found at the centre of the main body, sign of a quiet recent formation history.

\subsubsection{Dynamical analysis}

\begin{figure}
\center
\includegraphics[width=6.5cm, angle=-90]{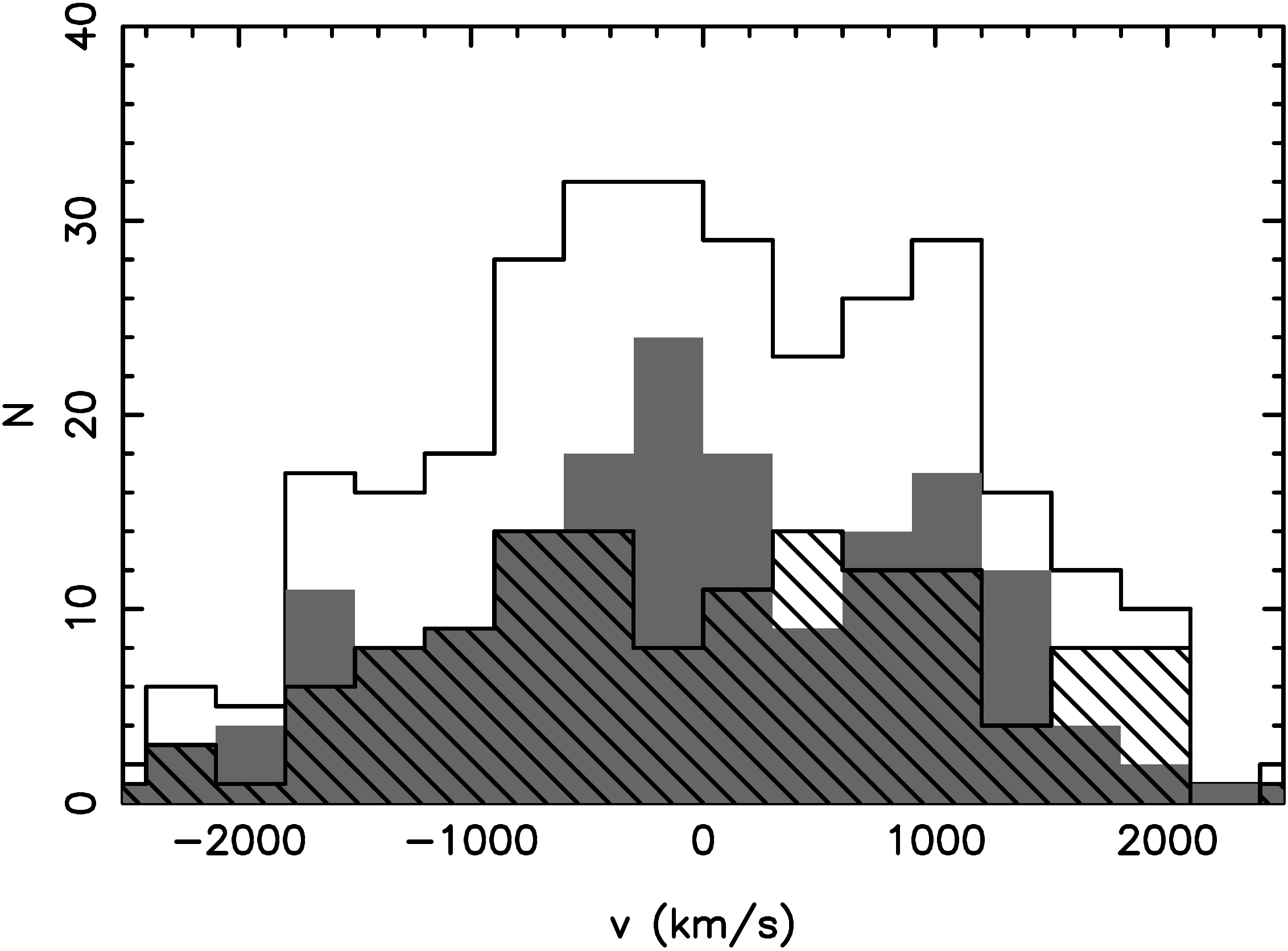}
\caption{Same as Fig. \ref{fig:nz_0225}, but for RXCJ2308. The distributions of red and blue galaxies are marginally different according to the KS test (p=0.72).}
\label{fig:nz_2308} 
\end{figure}

The velocity distribution of RXCJ2308 (Fig. \ref{fig:nz_2308}) deviates from a Gaussian with a probability $p=0.93$ according to the KS test ($p=0.96$ within 1.5 Mpc). The GH test returns a negative value for $h_4$ with $p=0.88$ ($p=0.96$ within 1.5 Mpc), due to a large number of galaxies at $\sim+1000\,\mathrm{km\,s^{-1}}$ and $\sim-1500\,\mathrm{km\,s^{-1}}$, giving the visual impression of a multimodal distribution. The test also finds a negative $h_3$ within 1.5 Mpc ($p=0.83$), consequence of an excess of galaxies with positive velocities. According to the two-sided KS test, the blue and red galaxies have a similar velocity distribution.

We first applied the KMM algorithm with a two-Gaussian model. We limited the analysis within 2 Mpc, since the departure from Gaussianity is greater in the central area. With an initial guess N(v)$=0.6\times\mathcal{N}(0,800)+0.4\times\mathcal{N}(1000,400)$, we obtained a probability of improvement $p=0.98$ for a best-fit mixture N(v)$=0.75\times\mathcal{N}(-731,894)+0.25\times\mathcal{N}(1183,621)$. Interestingly, the main component has a rest-frame velocity $v\sim-700\,\mathrm{km\,s^{-1}}$. This suggests that we underestimated the cluster redshift: it is likely that some of the structures found in its surroundings have a non-negligible rest-frame velocity that bias the overall redshift estimate. Alternatively, a third component with a high negative velocities could be responsible for this value. To check whether this scenario is a viable alternative, we ran the KMM algorithm for a three-Gaussian model. Starting with estimated values N(v)$=0.6\times\mathcal{N}(-250,600)+0.3\times\mathcal{N}(1200,600)+0.1\times\mathcal{N}(-1500,350)$, we obtained a best-fit mixture N(v)$=0.45\times\mathcal{N}(-539,547)+0.33\times\mathcal{N}(1043,653)+0.22\times\mathcal{N}(-1640,629)$, providing a probability of improvement $p=0.98$. Following the same approach, we estimated the improvement of using three Gaussians instead of two. We found $p=0.70$, a value that is not conclusive either way.  

\begin{figure}
\center
\includegraphics[width=7cm, angle=-90]{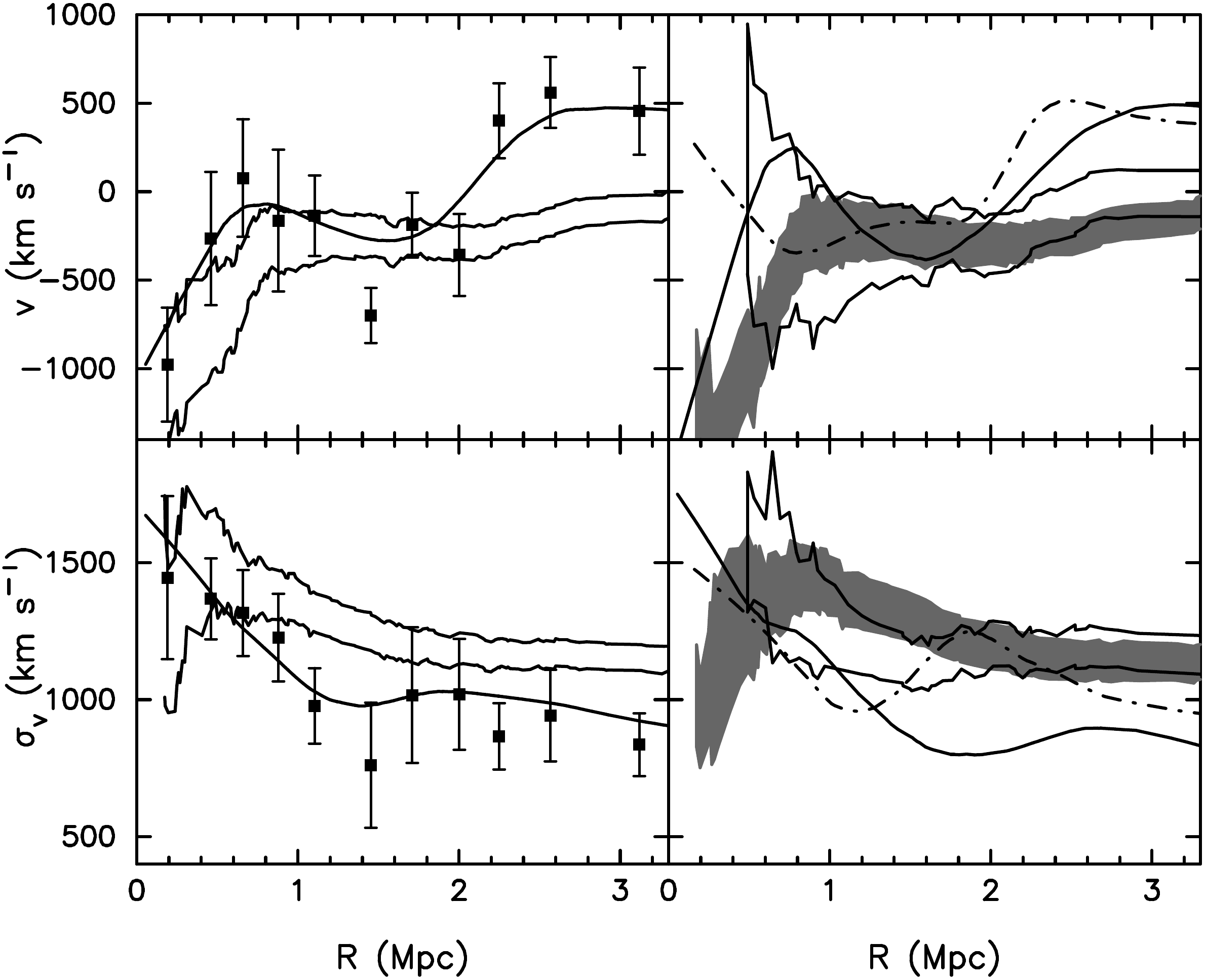}
\caption{Same as Fig. \ref{fig:vprof_0225}, but for RXCJ2308.}
\label{fig:vprof_2308} 
\end{figure}

The VPs and VDPs of RXCJ2308 (Fig. \ref{fig:vprof_2308}) allow us to better appreciate the complex dynamics of this cluster. The most striking feature is found in the VP: it starts at $\sim-1000\,\mathrm{km\,s^{-1}}$, reaches zero at $\sim0.8$ Mpc, remains flat up to $\sim2$ Mpc, and then peaks up again to reach $\sim+500\,\mathrm{km\,s^{-1}}$. The second gradient suggests the presence of a significant structure located at a great distance, which is responsible for an overestimation of the cluster redshift. However, the flat section of the VP within $\sim1-2$ Mpc has a velocity $v\sim-250\,\mathrm{km\,s^{-1}}$, which is not low enough to be the only explanation for the velocity $v\sim-700\,\mathrm{km\,s^{-1}}$ of the main KMM partition in the two-Gaussian model. The shape of the VP is mainly driven by the population of red galaxies, thus the central high negative velocities are not due to blue interlopers. The VDP shows a negative gradient, which is steeper in the central $\sim1.5$ Mpc. It can be explained by the combination of two effects: the typical decreasing VDP of late-type galaxies, and the mixing of the above-mentioned structures with different rest-frame velocities. The latter effect is clearly seen in the VDP profile of red galaxies, which are usually characterised by a flat profile. A significant peak is found in the blue VDP at $\sim1.8$ Mpc without a counterpart in that of the red population. This could be attributed to a clump mainly populated by late-type galaxies or to the azimuthal average of galaxies falling onto the cluster from different angles. It is interesting to note that the two galaxy populations have the same velocity dispersion integrated within $R_{200}$, despite the rich dynamical structure of the cluster.

\begin{figure}
\center
\includegraphics[width=6.5cm, angle=-90]{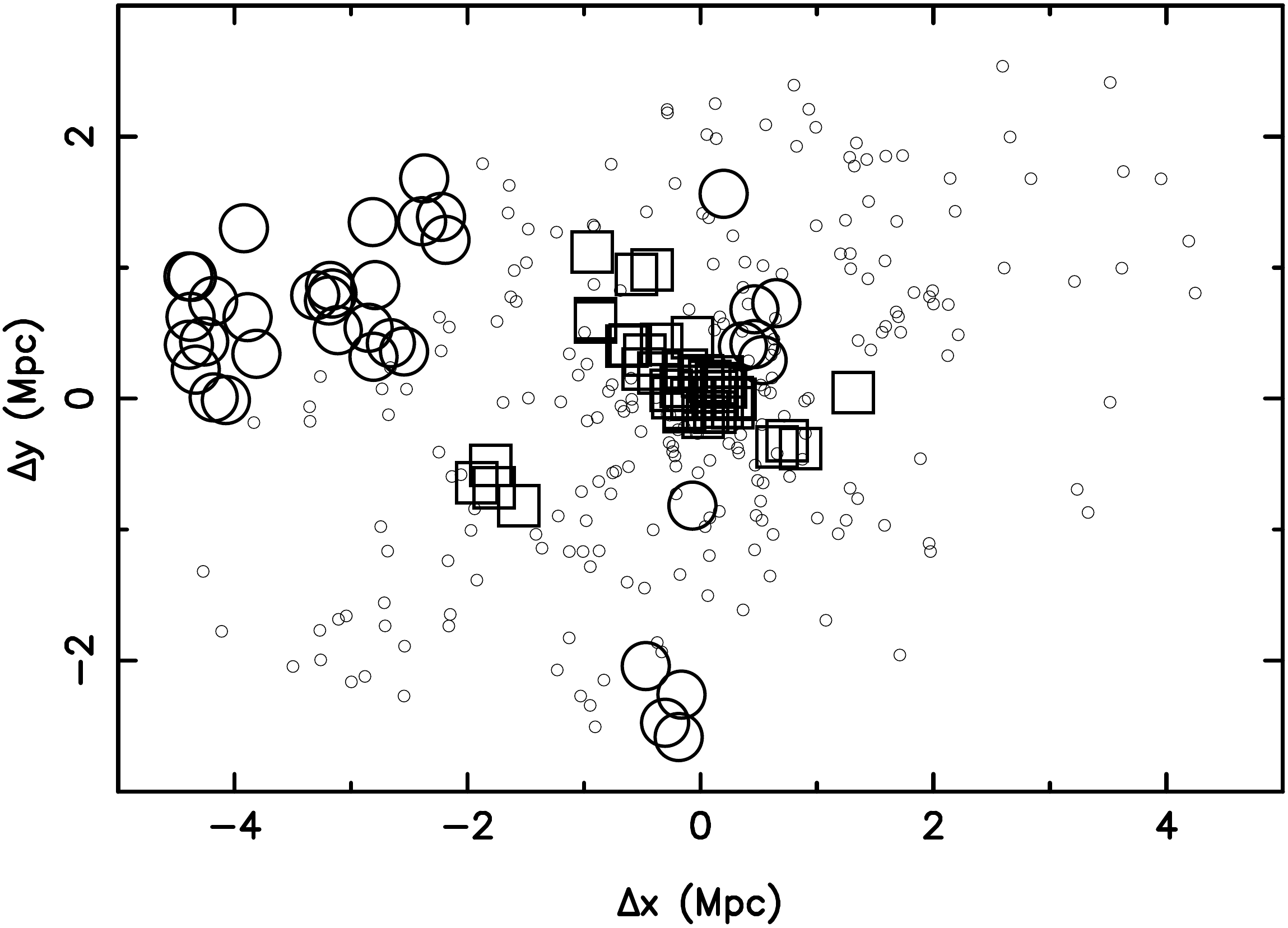}\\
\includegraphics[width=6.5cm, angle=-90]{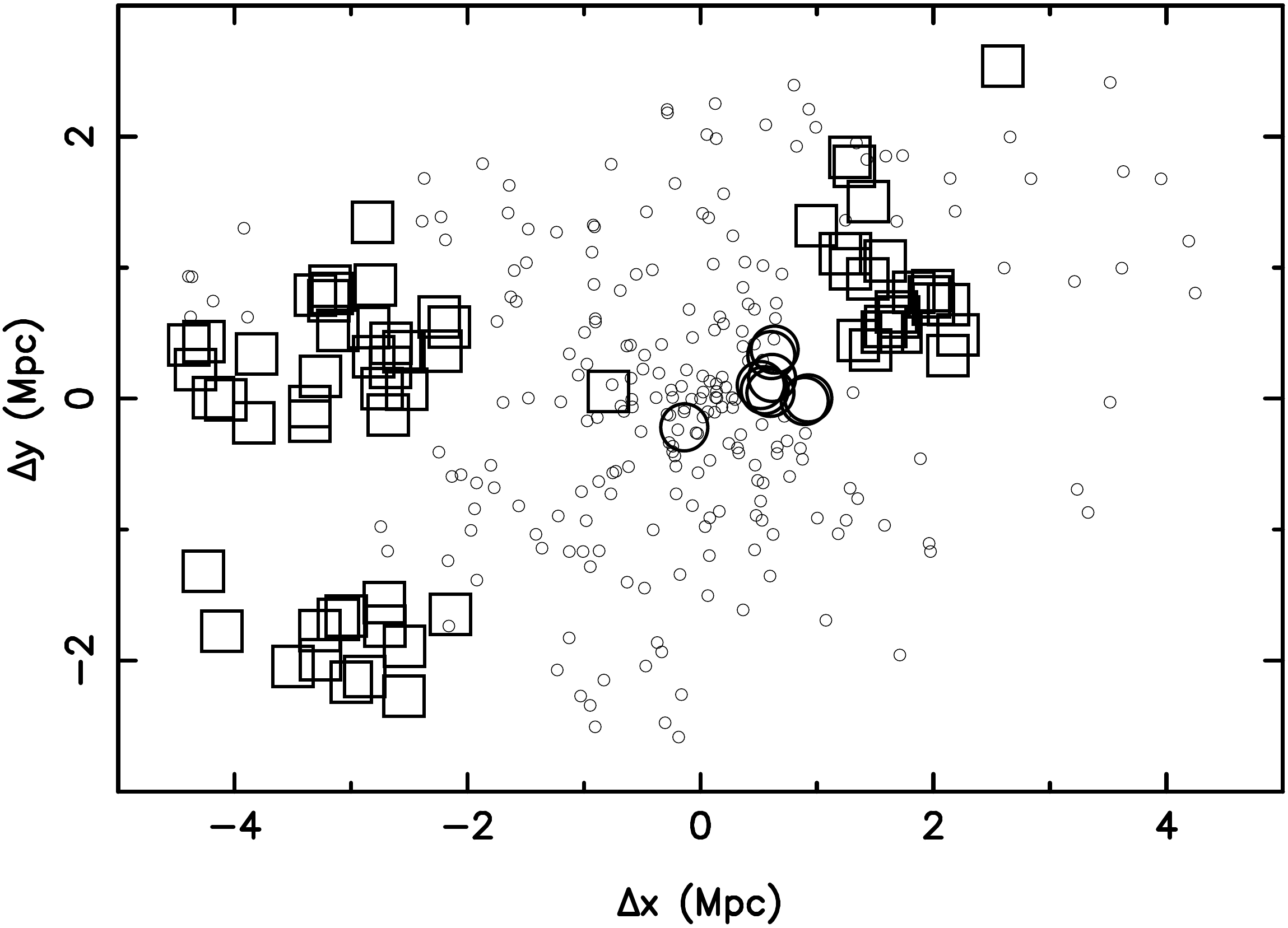}
\caption{Same as Fig. \ref{fig:0225_DS}, but for RXCJ2308.}
\label{fig:2308_DS} 
\end{figure}

As expected, the $\Delta$-tests find a high probability that RXCJ2308 has substructures. In fact, none of the shuffled realisations had $\Delta$ values as large as those found for the data. A large fraction ($\sim45\%$) of the spectroscopic members are associated with substructures (Fig. \ref{fig:2308_DS}); $\sim65\%$ of them are red galaxies, indicating that they are part of dense hence evolved and massive objects. Inspection of their location reveals several interesting regions. In the NE quadrant, we find a cold cluster of $\sim30$ galaxies, and with a local velocity $v\sim+800\,\mathrm{km\,s^{-1}}$; its position explains the second gradient seen in the VP. The $\Delta_S$-test detects two other cold groups: one of $\sim10$ galaxies in the SE quadrant, and a compact one of $\sim25$ galaxies NW of the cluster centre. Several other groups are detected by the $\Delta_V$-test, in particular at the cluster centre, which is dominated by galaxies whose local velocity is smaller than the average value.

\subsubsection{X-ray analysis}

\begin{figure}
\center
\includegraphics[width=9cm]{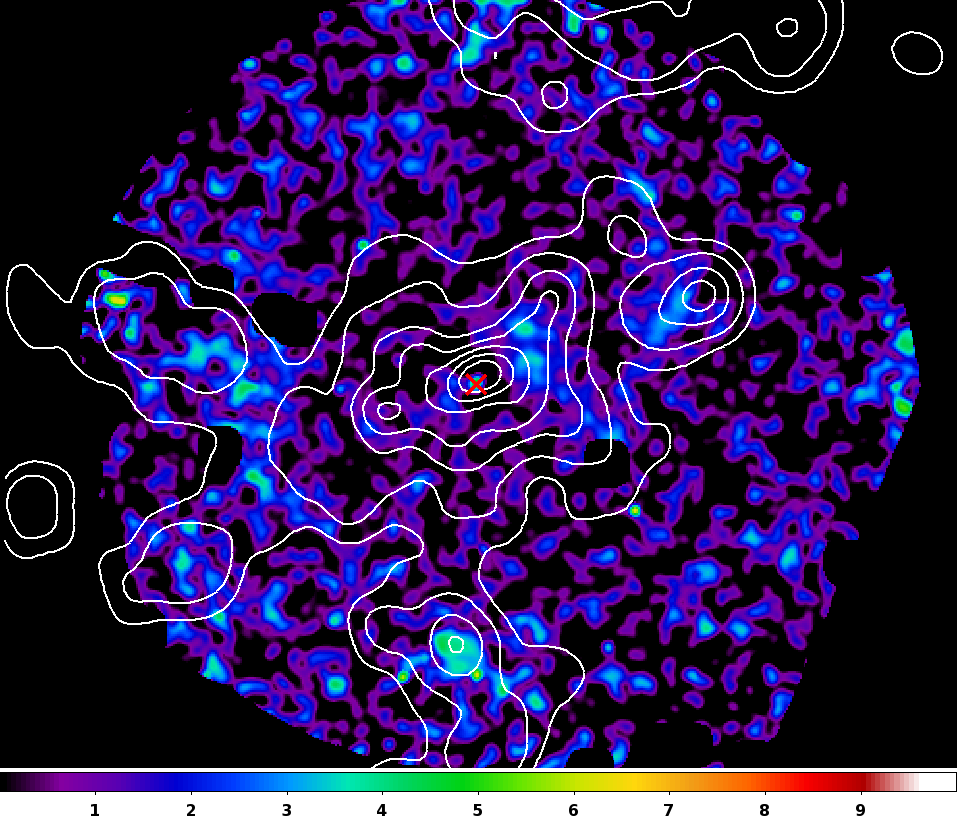}
\caption{Same as Fig. \ref{fig:0528_X}, but for RXCJ2308.}
\label{fig:2308_X} 
\end{figure}

\begin{figure}
\center
\includegraphics[width=9cm]{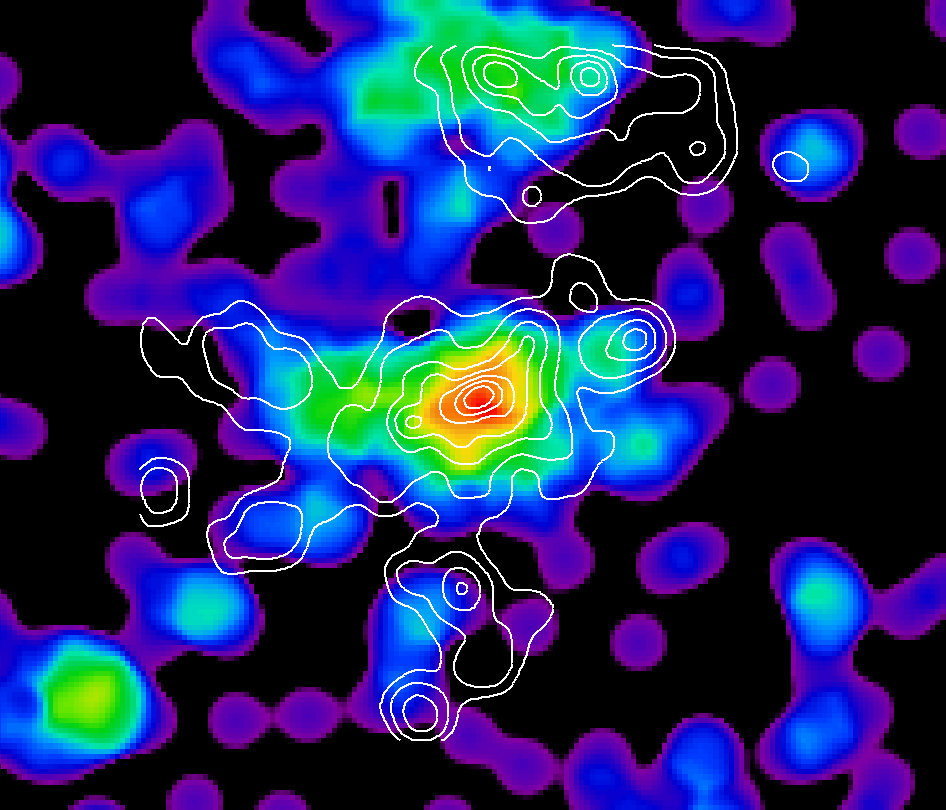}
\caption{ROSAT X-ray emission of RXCJ2308. White contours trace the surface density of red-sequence galaxies.}
\label{fig:2308_rosat} 
\end{figure}

The residual X-ray emission of RXCJ2308 (Fig. \ref{fig:2308_X}) presents several good matches with the density of red-sequence galaxies. The southern galaxy clump has a clear emission, indicating that it is a dense and massive group. The galaxy overdensity west of the cluster also has some associated X-ray emission, although less pronounced. The two extended galaxy distributions found on the east side of the cluster are also associated with some residual; however, they are more diffuse and close to the noise level, in particular the one in the south. Another excess of emission is found slightly west of the BCG, which is not surprising given the results of the $\Delta_V$-tests. The XMM-Newton FOV does not extend far enough in the north to cover the large structure found in the optical map. However, it shows up on the ROSAT image (Fig. \ref{fig:2308_rosat}), which also clearly shows the E-W extension matching the optical morphology. In fact, the northern object is the galaxy cluster RXCJ2308.3-0155, which is part of the REFLEX-II sample \citep{chon12}. Furthermore, \cite{chon13} associated the two clusters with a single supercluster, or, more exactly, to a `superstes-cluster', i.e. a matter overdensity that will eventually collapse and form a virialised structure \citep{chon15}. 

\subsubsection{Substructure analysis}

Given the results obtained for RXCJ2308, we defined several (sub)structure candidates (Fig. \ref{fig:reg_2308}). To the east of the centre are the two regions labelled G5 and G6. Region G5 is more diffuse, but it contains more galaxies, and has a more convincing X-ray counterpart (Fig. \ref{fig:2308_X}). Region G6, which encompasses two small overdensities, has a larger fraction of red galaxies, contains a bright BCG, and overall is more luminous. Their orientation/elongation match the central morphology of RXCJ2308, in particular G6. Furthermore, bridges of red-sequence galaxies connect them to the main clump (top left panel in Fig. \ref{fig:2308_Nmap}), thus G5 and G6 most likely trace filamentary structures feeding RXCJ2308. Both region are associated with a KMM partition (squares for G5 and stars for G6 in Fig. \ref{fig:reg_2308}). Region G5 has a large rest-frame line-of-sight velocity $\delta_v\sim+750\,\mathrm{km\,s^{-1}}$, hence only a probability of $\sim60\%$ of being bound to the cluster. Its velocity dispersion, as well as its galaxy content, suggests that it is a rich galaxy group with $\mathrm{M_{G5}}\sim0.2\times M_{200}$. The two-body model only finds a bound-outgoing solution at an angle $\alpha\sim75\degree$. Since the galaxy distribution suggests that G5 is connected to the cluster, we can suppose that we underestimated the total mass of the system, and look for the best bound-incoming solution. Increasing the total mass by $\sim25\%$ gives a collapsing solution with $\alpha\sim40\degree$, corresponding to an infall velocity $v\sim1100\,\mathrm{km\,s^{-1}}$ at a distance $R\sim4.3$ Mpc. With this configuration, G5 will pass the RXCJ2308 core in $t_{coll}\sim2$ Gyr. For G6, we estimate a smaller mass $\mathrm{M_{G6}}\sim0.1\times M_{200}$, but given its small velocity difference with RXCJ2308, they have a very high probability of being gravitationally bound. The best two-body solution is collapsing with $\alpha\sim22\degree$, $v\sim800\,\mathrm{km\,s^{-1}}$, $R\sim4.0$ Mpc, and $t_{coll}\sim2.5$ Gyr.

West of the cluster centre we defined the region G3, which is also associated with a KMM partition (triangles in Fig. \ref{fig:reg_2308}). This region has properties similar to those of G5 and G6, hence another likely galaxy group being accreted by RXCJ2308. A faint X-ray counterpart is detected at the same position (Figs. \ref{fig:2308_X} and \ref{fig:2308_rosat}), which confirms that it is a massive object. It is nearly at rest with respect to the main body and has a high probability of being bound to it. The estimated total mass of the two-body system spans all possible configurations. The best one is bound-incoming with $\alpha\sim5\degree$, $v\sim1700\,\mathrm{km\,s^{-1}}$, $R\sim2.0$ Mpc, and $t_{coll}\sim0.7$ Gyr, i.e. a high-velocity substructure already within $R_{200}$ and reaching the end of an orbit on the plane of the sky. Its optical properties give a mass $\mathrm{M_{G3}}\sim0.1\times M_{200}$. Its velocity dispersion is most likely biased low due to its small spatial and redshift separation with the main body, preventing the KMM algorithm from making a correct partition (which can be seen by the inclusion of several galaxies well beyond the optical overdensity).

In the southern part of the cluster, the distribution of galaxies suggests another elongated structure extending down to the edge of the WFI image (best seen in the top left panel of Fig. \ref{fig:2308_Nmap}). We defined accordingly the region G7. It encompasses a rather large amount of galaxies within two small overdensities, and it contains a BCG only $\sim0.6$ mag fainter than that of RXCJ2308. There is a clear X-ray emission matching the position of the northern clump (Fig. \ref{fig:2308_X}), hence it is definitely a massive galaxy group. The VIMOS observations barely cover this region; however, we found a KMM partition associated with the northern clump within the region G7 (stars in Fig. \ref{fig:reg_2308}). It has a probability of less than $50\%$ of being bound to RXCJ2308. However, due to the limited number of redshifts and the large error bar on its rest-frame velocity we cannot entirely rule out the possibility that G7 is connected to the main body. Similarly, the estimate of its velocity dispersion must be taken with caution, even though its value $\sigma_P\sim700\,\mathrm{km\,s^{-1}}$ is consistent with a massive galaxy group. As for G5, a bound-outgoing solution is probable (at an angle $\sim78\degree$), but the presence of a bridge of galaxies motivates us to assume again that we underestimated the total mass of the system. Furthermore, using the lower limit of the velocity difference, which raises the bound probability to $\sim65\%$, gives a collapsing orbit with $\alpha\sim40\degree$, $v\sim1100\,\mathrm{km\,s^{-1}}$, $R\sim4.3$ Mpc, and $t_{coll}\sim2$ Gyr. From its galaxy content, we estimate that the extended region G7 contains a mass $\mathrm{M_{G7}}\sim0.2\times M_{200}$.

We labelled G4 the northern galaxy cluster RXCJ2308.3-0155. It has a bright central galaxy ($\sim0.25$ mag fainter than the RXCJ2308 BCG), and is characterised by two distinct concentrations of red galaxies, labelled G41 and G42. \cite{chon12} obtained four spectroscopic redshifts within G41, from which we estimate a rest-frame line-of-sight velocity $|\delta_v|\sim1600\pm700\,\mathrm{km\,s^{-1}}$, and a velocity dispersion $\sigma_P\sim1000\,\mathrm{km\,s^{-1}}$, i.e. a mass similar to that of the RXCJ2308 main body. Owing to the velocity difference and large projected separation $R_P\sim3.8$ Mpc, the two systems have a very low probability of being gravitationally bound. However, the rest-frame velocity of the second cluster was only derived with four redshifts, and thus has a large uncertainty. Using the lower limit on the velocity difference, the probability rises to $\sim60\%$. Moreover, due to the superstes-cluster nature of the system, it is clear that the assumption of isolated point masses is wrong. Thus the probability of observing a bound system must be higher due to the additional matter located in between the two bodies. In this case, the two-body model favours a bound-outgoing solution at angle $\sim77\degree$, giving a physical separation $R\sim17$ Mpc. This solution agrees with the classification of the system as a superstes-cluster: the two clusters are still moving apart from each other, but they will eventually collapse. 

\begin{figure}
\center
\includegraphics[width=8cm, angle=-90]{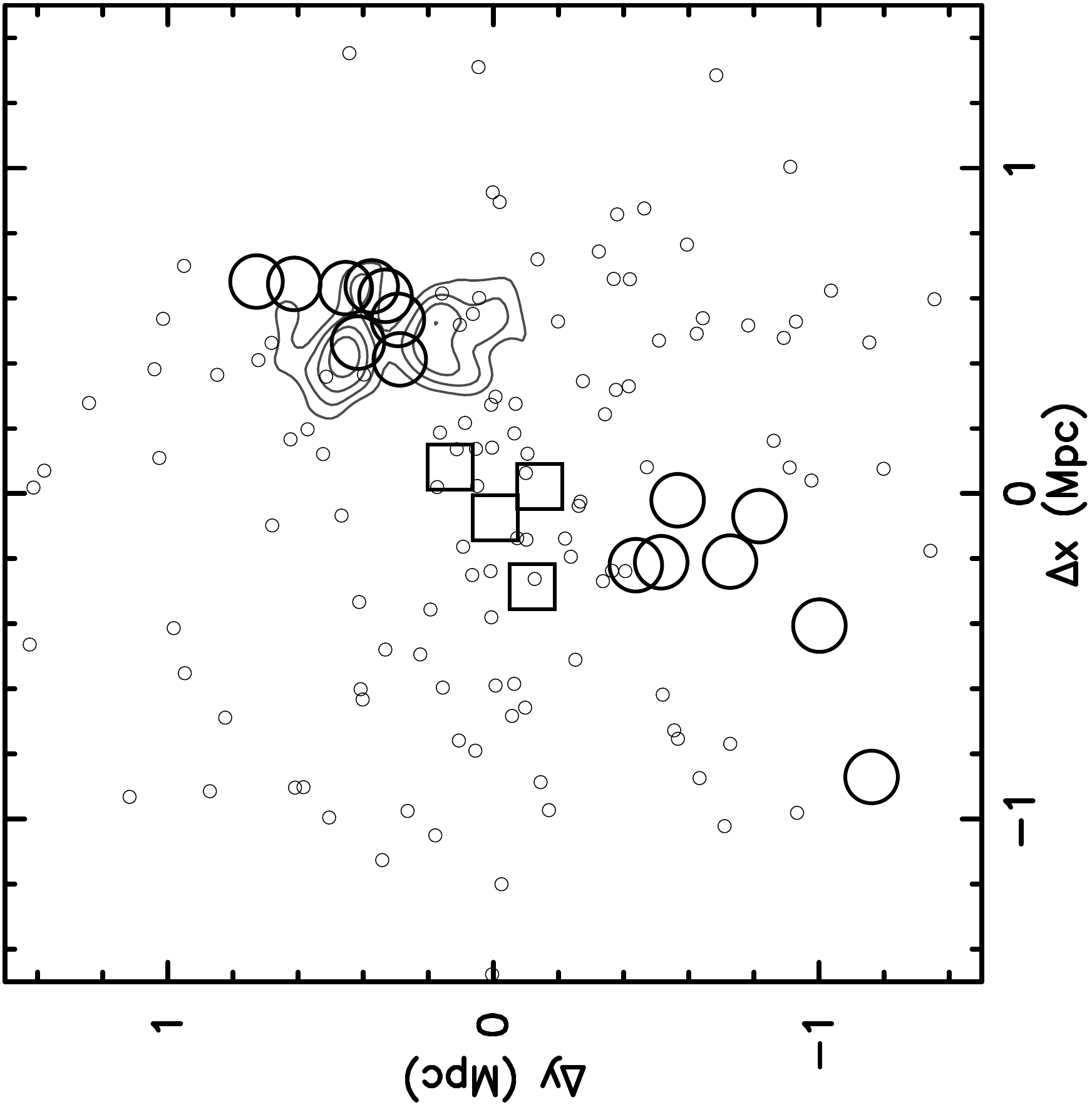}
\caption{Result of the $\Delta_V$-test, limited within the central 1.5 Mpc of RXCJ2308. The contours show the residual X-ray emission (starting at 2 $\sigma$, and increasing by 0.5).}
\label{fig:2308_DS_centre} 
\end{figure}

\begin{figure}
\center
\includegraphics[width=8cm, angle=-90]{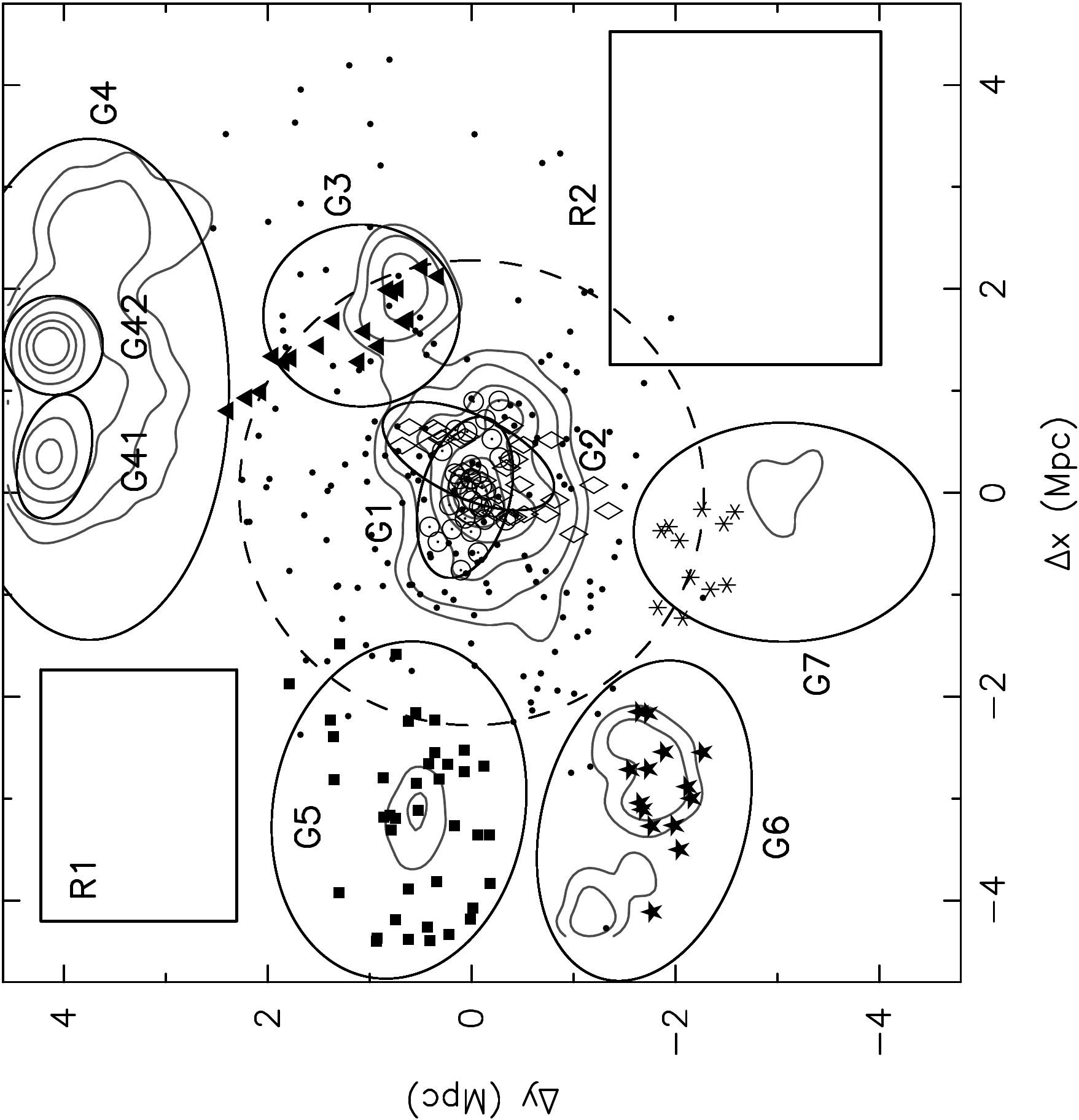}
\caption{Same as Fig. \ref{fig:reg_0225}, but for RXCJ2308: substructure candidates (regions G*), regions used to estimate the density of field galaxies (R1 and R2), and contours tracing the surface density of the bright red-sequence galaxies. Symbols show the location of the spectroscopic members associated with the different KMM partitions. The dashed circle has a radius of $R_{200}\sim2.3$ Mpc.}
\label{fig:reg_2308} 
\end{figure}

The central region of the cluster is more difficult to analyse. The dynamical tests indicate the presence of a high-velocity component, and perhaps a second one with a negative rest-frame velocity (top panel in Fig. \ref{fig:2308_DS}). According to the $\Delta_V$-test, the cluster centre is dominated by galaxies whose local rest-frame velocity is negative. As shown with the VPs (Fig. \ref{fig:vprof_2308}), the cluster redshift was overestimated due to the several receding systems found in its surroundings. Therefore, we applied again the $\Delta_V$-test, but limited within the central 1.5 Mpc. The results are presented in Figure \ref{fig:2308_DS_centre}. Clearly, our first application of the $\Delta_V$-test was biased by an overestimation of the cluster redshift since very few galaxies with a local negative rest-frame velocity are left; four of them are still present, located on the cluster centre. Interestingly, we also find a very good match between the X-ray residuals and a group of galaxies with high velocities. Finally, we find another group of high-velocity galaxies, $\sim0.6-0.8$ Mpc south of the centre.

At this point of the analysis, we can make a few more tests to assess the central dynamics. We started by cutting out the component associated with the X-ray residuals. The corresponding velocity distribution (limited within 1.5 Mpc) has a KS probability $p=0.95$ to differ from a Gaussian. We ran the KS test again after cutting out the second group of high-velocity galaxies as well, and we obtained $p=0.78$, i.e. a better agreement with a Gaussian. Thus, we are left with two possibilities: the velocity distribution of the remaining galaxies presents a slight departure from Gaussianity due to a non-virialised state, or to a third component close to the line of sight. The latter scenario is supported by the VP: the central velocity of the red galaxies is $\sim-1000\,\mathrm{km\,s^{-1}}$ (top left panel in Fig. \ref{fig:vprof_2308}), which still corresponds to $\sim-500\,\mathrm{km\,s^{-1}}$ after correcting the cluster redshift. Therefore, we applied the 3D KMM algorithm with three components: one for the cluster and two for the galaxies with negative and positive velocities, respectively. According to the new $\Delta_V$-test, the receding galaxies could be located in two different substructures. However, to avoid over-interpreting a dynamical configuration that appears to be very complex, we decided to merge them into a single structure. The corresponding KMM partition are labelled G1 and G2 (circles and diamonds in Fig. \ref{fig:reg_2308}). Since they are very close to the cluster centre, we did not estimate their optical properties. Furthermore, they have a higher probability of being observed post-merging, and for substructures deep in the cluster's gravitational potential well, dynamical friction, angular momentum, or tidal forces cannot be ignored. So we did not apply the two-body model for these two components. Additionally, the 3D KMM algorithm truncates the tails of the velocity distribution, hence it artificially reduces the velocity dispersion of the main body and overestimates that of the other partitions. Furthermore, using a Gaussian-shaped spatial distribution tends to make assignments regardless of redshift for the galaxies close to the centre of a given group. This is evident for G1 since nearly all the central galaxies are associated with this partition. Nonetheless, we find that G1 and G2 have a relatively small velocity dispersion, and that they have a rest-frame velocity difference of $\delta_v\sim2800\,\mathrm{km\,s^{-1}}$.

As a last remark, we should mention the completeness of the spectroscopic catalogue: it is $\sim50\%$ for the faint $m^*+1<m<m^*+3$ galaxies, and goes down to $\sim30\%$ for the bright ones. In other words, our results are subject to statistical fluctuations, in particular due to possible high-velocity interlopers. Since none of the tests that we used allows us to securely distinguish between a two-, three-, or even four-component model, we leave open the exact interpretation of the dynamical configuration of the cluster core. In any case, it is likely that the cluster centre has not yet reached virialisation due to a merger close to the line of sight. In addition to our findings, we mention the results of \cite{rossetti11}, who classified the cluster as non-cool core, as well as those of \cite{newman13}, whose strong-lensing analysis requires a second mass clump close to the centre. \cite{braglia09} noted a large magnitude gap between the BCG and the second brightest red-sequence member. However, a confirmed cluster member located $\sim250$ kpc from the BCG exhibits a small magnitude gap $\Delta m\sim0.3$ mag, which supports our findings for a non-relaxed core.

 \subsubsection{Summary}
 
To summarise, RXCJ2308 presents a complex structure from small to large scales. It has the characteristics of a superstes-cluster, with two main axes of accretion (N-S and E), along which we identified several massive and compact galaxy overdensities, including a confirmed galaxy cluster. The core of the cluster has a very complex configuration, possibly with three distinct high-velocity substructures, indicating a recent merger activity. We find that excluding the different components with a specific dynamics leads to a velocity dispersion significantly different from the value derived prior the substructure analysis. In fact, we find that the corresponding mass $M_{\sigma}$ is divided by a factor of $\sim2$. Consequently, we estimate that the main body will increase its mass by a factor of $\sim3$ from the accretion of the various structures found in its surroundings. We applied the procedure described previously to re-estimate the dynamical mass, i.e. cutting out annular sectors embedding substructures prior to estimating the harmonic radius, and using the red galaxies of the main KMM partition to estimate the velocity dispersion. We obtained a mass $M_{200}=0.72_{-0.21}^{+0.32}\times\mathrm{10^{15}\,M_{\odot}}$, which is $\sim2.5$ times smaller than the value found previously. This new mass is most likely underestimated because of the truncation of the velocity distribution by the KMM algorithm. Moreover, the high-velocity components in the central region increase the number of close pairs, which decreases the value of the harmonic radius. We can also consider that we misidentified the component G1. Adding the corresponding galaxies to the main body increases it velocity dispersion to $\sim1030\pm65\,\mathrm{km\,s^{-1}}$. The corresponding mass is $M_{200}=1.10_{-0.24}^{+0.21}\times\mathrm{10^{15}\,M_{\odot}}$, which is not significantly different from the value obtained after excluding G1. Either way, we find that cutting out the different structures leads to a smaller dynamical mass. Owing the uncertainties in characterising the core structure, it is difficult to obtain a robust estimate of the mass fraction contained in substructures. Assuming that we correctly identified G1, G2, and G3 (the other groups are too far away to be considered part of the cluster), we obtain a total mass $M\sim1.1\pm0.4\times\mathrm{10^{15}\,M_{\odot}}$, with a mass fraction of $\sim35\%$ within the substructures. Alternatively, merging G1 with the main body gives a slightly larger total mass $M\sim1.27\pm0.3\times\mathrm{10^{15}\,M_{\odot}}$, and a smaller mass fraction $\sim15\%$ within substructures.

As for RXCJ0228, we find that a combined analysis can lead to very different conclusions, as compared to a single-wavelength approach. The optical analysis, and to some extent the X-ray observations, identified the very rich environment of the cluster, but it underestimates the complexity of its core. The dynamical analysis reveals an overall bi-modality in the redshift distribution, as well as a multimodal core. However, when applied without spatial information, it does not allow the galaxies to be partitioned into the several distinct structures found in the cluster surroundings. Hence, it fails at correctly estimating the total mass that will be accreted by the main body. Our analysis also shows the limitations of the simple dynamical mass estimator in the case of a complex configuration. Therefore, we plan to apply the combined lensing and dynamical methodology introduced in \cite{verdugo16} to obtain a better picture of the mass distribution within the core of RXCJ2308.


\begin{table*}
\centering
\begin{threeparttable}
\caption{Photometric and dynamical properties of the substructures.}
\label{table:sub_prop}
\begin{tabular}{l c c c c c c c c c c}
\hline\hline\noalign{\smallskip}
Region & $N_{\mathrm{RS}}$ & $L_{\mathrm{RS}}$ & $f_{\mathrm{RS}}$ & $m_{\mathrm{BCG}}$ & $\mathrm{R_P}$ & $\mathrm{N_z}$ & $\delta_V$ & $\sigma_P$ & $\sigma_P^3$ & $\mathrm{P_{bound}}$\\
& $\overline{N_{200}}$ & $\overline{L_{200}}$ & & (mag) & $\overline{R_{200}}$ & & $(\mathrm{km\,s^{-1}})$ & $(\mathrm{km\,s^{-1}})$ & $\overline{\sigma_{200}^3}$ & \\
\noalign{\smallskip}\hline\noalign{\smallskip}
RXCJ0225-R200 & 1 & 1 & 0.80 & 16.99 & 0 & 197 & $41\pm80$ & $933\pm44$ & 1 & -\\
RXCJ0225-G1 & $0.11\pm0.01$ & $0.11\pm0.01$ & 0.73 & 17.74 & 0.72 & 28 & $1072\pm103$ & $424\pm61$ & $0.10\pm0.04$ & $\sim0.55$\\
RXCJ0225-G2$^{\dag}$ & $0.29\pm0.02$ & $0.25\pm0.01$ & 0.84 & 17.40 & 0.08 & - & - & - & - & -\\
RXCJ0225-G3$^{\dag}$ & $0.10\pm0.01$ & $0.10\pm0.01$ & 0.74 & 17.79 & 0.42 & 24 & $-90\pm172$ & $745\pm95$ & $0.45\pm0.19$ & $>0.95$\\
RXCJ0225-G4$^{\dag}$ & $0.19\pm0.02$ & $0.17\pm0.01$ & 0.83 & 16.99 & 0.80 & 18 & $-344\pm350$  & $1025\pm206$ & $1.20\pm0.70$ & $>0.95$\\
RXCJ0225-G5 & $0.04\pm0.01$ & $0.06\pm0.01$ & 0.76 & 17.64 & 0.69 & 19 & $-1099\pm114$ & $339\pm68$ & $0.05\pm0.03$ & $\sim0.60$\\
RXCJ0225-G6 & $0.02\pm0.01$ & $0.02\pm0.01$ & 0.31 & 18.32 & 1.39 & 8 & $291\pm298$ & $339\pm65$ & $0.05\pm0.03$ & $>0.80$\\
RXCJ0225-G7 & $0.08\pm0.01$ & $0.07\pm0.01$ & 0.81 & 17.96 & 1.35 & - & - & - & - & -\\
RXCJ0225-MB & - & - & - & - & 0.20 & 163 & $-63\pm83$ & $839\pm55$ & $0.73\pm0.18$ & 1\\
\noalign{\smallskip}\hline\noalign{\smallskip}
RXCJ0528-R200 & 1 & 1 & 0.68 & 17.03 & 0 & 152 & $49\pm86$ & $895\pm51$ & 1 & -\\
RXCJ0528-G1 & $0.23\pm0.01$ & $0.25\pm0.01$ & 0.65 & 17.67 & 0.57 & - & - & - & - & -\\
RXCJ0528-G2 & $0.04\pm0.01$ & $0.07\pm0.01$ & 0.30 & 17.98 & 1.72 & - & - & - & - & -\\
RXCJ0528-G3 & $0.05\pm0.01$ & $0.10\pm0.01$ & 0.69 & 17.41 & 2.54 & 14 & $-910\pm166$ & $325\pm82$ & $0.05\pm0.04$ & $<0.10$\\
RXCJ0528-G4 & $0.08\pm0.01$ & $0.08\pm0.01$ & 0.52 & 18.53 & 1.62 & - & - & - & - & -\\
RXCJ0528-MB & - & - & - & - & 0.05 & 152 & $49\pm86$ & $895\pm51$ & 1 & 1\\
\noalign{\smallskip}\hline\noalign{\smallskip}
RXCJ2308-R200 & 1 & 1 & 0.74 & 17.43 & 0 & 217 & $-236\pm84$ & $1164\pm52$ & 1 & -\\
RXCJ2308-G1 & - & - & - & - & 0.03 & 33 & $-1561\pm92$ & $572\pm123$ & $0.12\pm0.08$ & $>0.90$\\
RXCJ2308-G2$^{\dag}$ & - & - & - & - & 0.12 & 27 & $1330\pm121$ & $387\pm104$ & $0.04\pm0.03$ & $\sim0.75$\\
RXCJ2308-G3$^{\dag}$ & $0.10\pm0.01$ & $0.13\pm0.01$ & 0.62 & 18.46 & 0.95 & 18 &  $-300\pm89$ & $290\pm72$ & $0.02\pm0.01$ & $>0.90$\\
RXCJ2308-G4$^{\dag}$ & $0.78\pm0.03$ & $0.84\pm0.03$ & 0.78 & 17.69 & 1.80 & - & - & - & - & -\\
RXCJ2308-G41 & $0.17\pm0.01$ & $0.18\pm0.01$ & 0.73 & 17.69 & 1.90 & 4 & $1583\pm710$ & $1034\pm425$ & $0.70\pm0.87$ & $<0.60$\\
RXCJ2308-G42 & $0.18\pm0.01$ & $0.24\pm0.01$ & 0.89 & 17.86 & 2.01 & - & - & - & - & -\\
RXCJ2308-G5$^{\dag}$ & $0.17\pm0.02$ & $0.15\pm0.01$ & 0.52 & 18.70 & 1.47 & 39 & $743\pm114$ & $720\pm114$ & $0.24\pm0.12$ & $\sim0.60$\\
RXCJ2308-G6 & $0.12\pm0.01$ & $0.19\pm0.01$ & 0.58 & 18.01 & 1.68 & 14 & $162\pm140$ & $454\pm120$ & $0.06\pm0.05$ & $>0.85$\\
RXCJ2308-G7$^{\dag}$ & $0.18\pm0.01$ & $0.17\pm0.01$ & 0.75 & 17.97 & 1.43 & 7 & $845\pm340$ & $703\pm146$ & $0.22\pm0.14$ & $<0.65$\\
RXCJ2308-MB & - & - & - & - & 0.06 & 134 & $-235\pm90$ & $935\pm56$ & $0.52\pm0.12$ & 1\\
\noalign{\smallskip}\hline
\end{tabular}
    \begin{tablenotes}
      \small
      \item Columns: (1) Name of the region. For each cluster, the first line corresponds to the circle of radius $R_{200}$. The last line is the main body of the cluster, i.e. the principal KMM partition. (2) Background-subtracted number of red-sequence galaxies belonging to the region, normalised to the value obtained within $R_{200}$. (3) Associated R-band luminosity, corrected from an average background contribution and normalised by its value within $R_{200}$. (4) Fraction of red-sequence galaxies. (5) R-band magnitude of the brightest red-sequence galaxy. (6) Projected separation between the region centre and the cluster centre, in unit of $R_{200}$. (7) Number of spectroscopic members associated with the corresponding KMM partition, limited within $R_{200}$ for the R200 and MB regions. (8) Rest-frame velocity. (9,10) Line-of-sight velocity dispersion, and the third power of its ratio to the dispersion within $R_{200}$. The latter quantity can be used as a relative mass estimator, assuming a scaling $M\propto\sigma^3$. We note that the $\sigma_{200}$ are different from the values from Table \ref{table:Mvir}, since here we used the $R_{200}$ derived from the virial estimator applied to the red galaxies (see Section 4.1). (11) Probability that the structure is bound to the main body, according to the Newtonian criterion. Regions for which a significant X-ray counterpart was detected are indicated by a $^{\dag}$.
    \end{tablenotes}
  \end{threeparttable}
\end{table*}

\section{Conclusions}

In this paper, we presented the first combined morphological, X-ray, and dynamical analysis of three massive galaxy clusters. We used WFI optical imaging to compute photometric redshifts with the k-nearest neighbour fitting method, from which we made galaxy surface density and luminosity maps. These were used to constrain the morphology of the clusters, and to detect structures up to several Mpc in radius. From VIMOS optical spectroscopy, we performed the dynamical analysis of the clusters. An iterative $3\sigma$ clipping scheme with radial binning allowed us to secure the cluster members from which dynamical masses were estimated. We ran several statistical tests to assess the dynamical state of the clusters. By combining spatial and velocity information, we were able to partition the cluster members into different components, and to make the connection with the large-scale environment. X-ray counterparts were detected for several galaxy overdensities, confirming that they are massive objects rather than the result of fortuitous projections. For each cluster, we finally drew a global picture of their structure at all scales: 

\begin{itemize}
\item RXCJ0225 has a bi-modal core, indicating an active dynamical state. Two other massive components are found within its virial radius, aligned in a structure extending SW over $\sim4$ Mpc that is also detected in X-rays. The structure is further detected in the NE quadrant with a chain of five galaxy overdensities reaching the limit of the WFI FOV at $\sim4$ Mpc from the cluster centre. The most massive substructure has a mass comparable to that of the cluster core, hence RXCJ0225 will experience a major merger that will take place nearly on the plane of the sky. Overall, RXCJ0225 appears to be in an early stage of its accretion history, and we estimate that the mass of its core, i.e. G2, will more than double by the accretion of the substructures found within its virial radius. The peculiar configuration of this cluster makes it an ideal candidate for investigating the properties of the intra-cluster gas, from the cluster core to its outskirts within a filament.

\item RXCJ0528 has a well defined core, and it is located in a poor environment. A mild N-S elongation traces its main axis of accretion. It is populated by two small galaxy overdensities close to the centre, and by two other small substructures outside $R_{200}$. The red-sequence galaxies have a different dynamics to that of the blue population, most likely due to an infall along an axis close to the line of sight. The small amount of substructure found in RXCJ0528 does not affect the dynamical estimate of its mass. The quiet state of the cluster is also confirmed by the residual X-ray emission, whose main feature is a small shift between the position of the cool core and the centroid of the large-scale emission. In addition to the presence of a second BCG, it suggests, however, that the cluster has not yet fully reached dynamical equilibrium.

\item RXCJ2308 is part of a superstes-cluster. We found numerous structures located well beyond $R_{200}$ and distributed along two main axes of accretion. Owing to this very rich environment, the central component of RXCJ2308 will increase its mass by at least a factor three, which makes it a perfect example of the hierarchical growth of clusters. The core of RXCJ2308 also has a very complex configuration: an elongated morphology, an X-ray residual associated with a high-velocity component, and evidence for one or perhaps even two other high-velocity substructures.
\end{itemize}

In addition to the individual analysis of each cluster, we also note some general features:

(i) The largest substructures are found in the cluster outskirts, whereas the smaller ones are closer to the core. This trend is in good agreement with the expected tidal destruction of subhaloes in the high-density regions, a result that is also observed in numerical simulations;

(ii) The number of substructures in a cluster echoes the richness of its large-scale environment. On the one hand, RXCJ0225 has an estimated $\sim50\%$ of its mass contained in substructures, a value much higher than the $5\%-15\%$ typically found in simulations or in observational studies (e.g. \citealt{guennou14}). On the other hand, RXCJ0528 is found in a much poorer environment and does not have any significant substructures within its virial radius. RXCJ2308, which has a mass fraction of $\sim15\%-35\%$ in substructures, is located in an overdense region, similarly to RXCJ0225;

(iii) The three clusters host more than one BCG, which is in agreement with the other indicators pointing towards dynamically young objects;

(iv) We found a good correlation between the orientation of the main BCG, the overall cluster optical morphology, and the main axis of accretion, which suggests a collimated infall of matter on the cluster core.\\

These results indicate a cluster formation mainly driven by successive mergers of galaxy groups. The presence of several BCGs can be associated with relics of the past mass assembly; substructures trace recent/ongoing merging events; and groups located in elongated filamentary-like structures will provide the material for the future evolution into more massive systems.

The previous studies of DXL clusters by \cite{pierini08}, \cite{braglia09}, and \cite{ziparo12} have highlighted the influence of the cluster large-scale environment on the properties of its galaxy population and intra-cluster medium. Our work adds more evidence for such a link by connecting the level of substructure in a cluster to the matter distribution in its surroundings. However, the clusters analysed in this work were selected for an in-depth study based on their interesting X-ray and optical morphology. Therefore, our immediate objective is to analyse the remaining DXL clusters with the same methodology in order to estimate their level of substructure and quantify their large-scale environment. Studying the connection between small and large scales with better statistics will allow us to better quantify their role in cluster evolution within the context of $\Lambda$CDM. RXCJ0225, RXCJ2308 (this work), A2744 \citep{braglia09}, and A1300 \citep{ziparo12} are dynamically young objects. With the analysis of the remaining DXL clusters we will obtain more robust results regarding the fraction of such systems in the population of massive clusters at redshifts $z\sim0.3$.

\begin{acknowledgements}
The authors thank Mischa Schirmer for precious advice regarding the reduction of WFI images with THELI, and Feliberto Braglia for sharing data with us. We would like to acknowledge support from the Deutsche Forschungsgemeinschaft through the Transregio Program TR33 and through the Munich Excellence Cluster ``Structure and Evolution of the Universe'', as well as from Deutsches Zentrum f\"ur Luft- und Raumfahrt through grant No. 50OR1601.
\end{acknowledgements}

%

\section*{Appendix}
\setcounter{table}{0}
\renewcommand{\thetable}{A\arabic{table}}

Table \ref{table:app} contains the list of the photometric and spectroscopic cluster members used for this work. The full list is available at the CDS.

\bibliography{../references}

\begin{thebibliography}{142}
\expandafter\ifx\csname natexlab\endcsname\relax\def\natexlab#1{#1}\fi

\bibitem[{{Adami} {et~al.}(1998){Adami}, {Biviano}, \& {Mazure}}]{adami98b}
{Adami}, C., {Biviano}, A., \& {Mazure}, A. 1998, \aap, 331, 439

\bibitem[{{Aguerri} \& {S{\'a}nchez-Janssen}(2010)}]{aguerri10}
{Aguerri}, J.~A.~L. \& {S{\'a}nchez-Janssen}, R. 2010, \aap, 521, A28

\bibitem[{{Allen} {et~al.}(2011){Allen}, {Evrard}, \& {Mantz}}]{allen11}
{Allen}, S.~W., {Evrard}, A.~E., \& {Mantz}, A.~B. 2011, \araa, 49, 409

\bibitem[{{Altman}(1992)}]{altman92}
{Altman}, N.~S. 1992, j-AMER-STAT, 46, 175

\bibitem[{{Ashman} {et~al.}(1994){Ashman}, {Bird}, \& {Zepf}}]{ashman94}
{Ashman}, K.~M., {Bird}, C.~M., \& {Zepf}, S.~E. 1994, \aj, 108, 2348

\bibitem[{{Baade} {et~al.}(1999){Baade}, {Meisenheimer}, {Iwert}, {Alonso},
  {Augusteijn}, {Beletic}, {Bellemann}, {Benesch}, {B{\"o}hm}, {B{\"o}hnhardt},
  {Brewer}, {Deiries}, {Delabre}, {Donaldson}, {Dupuy}, {Franke}, {Gerdes},
  {Gilliotte}, {Grimm}, {Haddad}, {Hess}, {Ihle}, {Klein}, {Lenzen}, {Lizon},
  {Mancini}, {M{\"u}nch}, {Pizarro}, {Prado}, {Rahmer}, {Reyes}, {Richardson},
  {Robledo}, {Sanchez}, {Silber}, {Sinclaire}, {Wackermann}, \&
  {Zaggia}}]{baade99}
{Baade}, D., {Meisenheimer}, K., {Iwert}, O., {et~al.} 1999, The Messenger, 95,
  15

\bibitem[{{Balestra} {et~al.}(2016){Balestra}, {Mercurio}, {Sartoris},
  {Girardi}, {Grillo}, {Nonino}, {Rosati}, {Biviano}, {Ettori}, {Forman},
  {Jones}, {Koekemoer}, {Medezinski}, {Merten}, {Ogrean}, {Tozzi}, {Umetsu},
  {Vanzella}, {van Weeren}, {Zitrin}, {Annunziatella}, {Caminha}, {Broadhurst},
  {Coe}, {Donahue}, {Fritz}, {Frye}, {Kelson}, {Lombardi}, {Maier},
  {Meneghetti}, {Monna}, {Postman}, {Scodeggio}, {Seitz}, \&
  {Ziegler}}]{balestra16}
{Balestra}, I., {Mercurio}, A., {Sartoris}, B., {et~al.} 2016, \apjs, 224, 33

\bibitem[{{Barrena} {et~al.}(2009){Barrena}, {Girardi}, {Boschin}, \&
  {Das{\'{\i}}}}]{barrena09}
{Barrena}, R., {Girardi}, M., {Boschin}, W., \& {Das{\'{\i}}}, M. 2009, \aap,
  503, 357

\bibitem[{{Barrena} {et~al.}(2011){Barrena}, {Girardi}, {Boschin}, {de Grandi},
  {Eckert}, \& {Rossetti}}]{barrena11}
{Barrena}, R., {Girardi}, M., {Boschin}, W., {et~al.} 2011, \aap, 529, A128

\bibitem[{{Beers} {et~al.}(1990){Beers}, {Flynn}, \& {Gebhardt}}]{beers90}
{Beers}, T.~C., {Flynn}, K., \& {Gebhardt}, K. 1990, \aj, 100, 32

\bibitem[{{Beers} {et~al.}(1982){Beers}, {Geller}, \& {Huchra}}]{beers82}
{Beers}, T.~C., {Geller}, M.~J., \& {Huchra}, J.~P. 1982, \apj, 257, 23

\bibitem[{{Berlind} {et~al.}(2003){Berlind}, {Weinberg}, {Benson}, {Baugh},
  {Cole}, {Dav{\'e}}, {Frenk}, {Jenkins}, {Katz}, \& {Lacey}}]{berlind03}
{Berlind}, A.~A., {Weinberg}, D.~H., {Benson}, A.~J., {et~al.} 2003,
  Astrophys.~J., 593, 1

\bibitem[{{Bertin}(2006)}]{bertin06}
{Bertin}, E. 2006, in Astronomical Society of the Pacific Conference Series,
  Vol. 351, Astronomical Data Analysis Software and Systems XV, ed.
  C.~{Gabriel}, C.~{Arviset}, D.~{Ponz}, \& S.~{Enrique}, 112

\bibitem[{{Bertin}(2010)}]{bertin10}
{Bertin}, E. 2010, {SWarp: Resampling and Co-adding FITS Images Together},
  Astrophysics Source Code Library

\bibitem[{{Bertin} \& {Arnouts}(1996)}]{bertin96}
{Bertin}, E. \& {Arnouts}, S. 1996, Astron.~\&~Astrophys.s, 117, 393

\bibitem[{{Binney} \& {Tremaine}(1987)}]{binney87}
{Binney}, J. \& {Tremaine}, S. 1987, {Galactic dynamics}, ed. {Princeton
  University Press}

\bibitem[{{Biviano} {et~al.}(1992){Biviano}, {Girardi}, {Giuricin},
  {Mardirossian}, \& {Mezzetti}}]{biviano92}
{Biviano}, A., {Girardi}, M., {Giuricin}, G., {Mardirossian}, F., \&
  {Mezzetti}, M. 1992, \apj, 396, 35

\bibitem[{{Biviano} \& {Katgert}(2004)}]{biviano04}
{Biviano}, A. \& {Katgert}, P. 2004, \aap, 424, 779

\bibitem[{{Biviano} {et~al.}(2002){Biviano}, {Katgert}, {Thomas}, \&
  {Adami}}]{biviano02}
{Biviano}, A., {Katgert}, P., {Thomas}, T., \& {Adami}, C. 2002, \aap, 387, 8

\bibitem[{{Biviano} {et~al.}(2006){Biviano}, {Murante}, {Borgani}, {Diaferio},
  {Dolag}, \& {Girardi}}]{biviano06}
{Biviano}, A., {Murante}, G., {Borgani}, S., {et~al.} 2006, \aap, 456, 23

\bibitem[{{B{\"o}hringer} {et~al.}(2014){B{\"o}hringer}, {Chon}, \&
  {Collins}}]{bohringer14}
{B{\"o}hringer}, H., {Chon}, G., \& {Collins}, C.~A. 2014, \aap, 570, A31

\bibitem[{{B{\"o}hringer} {et~al.}(2010){B{\"o}hringer}, {Pratt}, {Arnaud},
  {Borgani}, {Croston}, {Ponman}, {Ameglio}, {Temple}, \&
  {Dolag}}]{boehringer10}
{B{\"o}hringer}, H., {Pratt}, G.~W., {Arnaud}, M., {et~al.} 2010, \aap, 514,
  A32

\bibitem[{{B{\"o}hringer} {et~al.}(2004){B{\"o}hringer}, {Schuecker}, {Guzzo},
  {Collins}, {Voges}, {Cruddace}, {Ortiz-Gil}, {Chincarini}, {De Grandi},
  {Edge}, {MacGillivray}, {Neumann}, {Schindler}, \& {Shaver}}]{bohringer04}
{B{\"o}hringer}, H., {Schuecker}, P., {Guzzo}, L., {et~al.} 2004,
  Astron.~\&~Astrophys., 425, 367

\bibitem[{{B{\"o}hringer} {et~al.}(2001){B{\"o}hringer}, {Schuecker}, {Guzzo},
  {Collins}, {Voges}, {Schindler}, {Neumann}, {Cruddace}, {De Grandi},
  {Chincarini}, {Edge}, {MacGillivray}, \& {Shaver}}]{bohringer01}
{B{\"o}hringer}, H., {Schuecker}, P., {Guzzo}, L., {et~al.} 2001,
  Astron.~\&~Astrophys., 369, 826

\bibitem[{{Boselli} \& {Gavazzi}(2006)}]{boselli06}
{Boselli}, A. \& {Gavazzi}, G. 2006, PASP, 118, 517

\bibitem[{{Braglia} {et~al.}(2007){Braglia}, {Pierini}, \&
  {B{\"o}hringer}}]{braglia07}
{Braglia}, F., {Pierini}, D., \& {B{\"o}hringer}, H. 2007, \aap, 470, 425

\bibitem[{{Braglia} {et~al.}(2009){Braglia}, {Pierini}, {Biviano}, \&
  {B{\"o}hringer}}]{braglia09}
{Braglia}, F.~G., {Pierini}, D., {Biviano}, A., \& {B{\"o}hringer}, H. 2009,
  \aap, 500, 947

\bibitem[{{Bryan} \& {Norman}(1998)}]{bryan98}
{Bryan}, G.~L. \& {Norman}, M.~L. 1998, Astrophys.~J., 495, 80

\bibitem[{{Caminha} {et~al.}(2016){Caminha}, {Grillo}, {Rosati}, {Balestra},
  {Mercurio}, {Vanzella}, {Biviano}, {Caputi}, {Delgado-Correal}, {Karman},
  {Lombardi}, {Meneghetti}, {Sartoris}, \& {Tozz}}]{caminha16}
{Caminha}, G.~B., {Grillo}, C., {Rosati}, P., {et~al.} 2016, ArXiv e-prints

\bibitem[{{Carlberg} {et~al.}(1997){Carlberg}, {Yee}, \&
  {Ellingson}}]{carlberg97}
{Carlberg}, R.~G., {Yee}, H.~K.~C., \& {Ellingson}, E. 1997, Astrophys.~J.,
  478, 462

\bibitem[{{Carlberg} {et~al.}(1996){Carlberg}, {Yee}, {Ellingson}, {Abraham},
  {Gravel}, {Morris}, \& {Pritchet}}]{carlberg96}
{Carlberg}, R.~G., {Yee}, H.~K.~C., {Ellingson}, E., {et~al.} 1996,
  Astrophys.~J., 462, 32

\bibitem[{{Carter} \& {Metcalfe}(1980)}]{carter80}
{Carter}, D. \& {Metcalfe}, N. 1980, \mnras, 191, 325

\bibitem[{{Chon} \& {B{\"o}hringer}(2012)}]{chon12}
{Chon}, G. \& {B{\"o}hringer}, H. 2012, \aap, 538, A35

\bibitem[{{Chon} {et~al.}(2012{\natexlab{a}}){Chon}, {B{\"o}hringer}, {Krause},
  \& {Tr{\"u}mper}}]{chon12c}
{Chon}, G., {B{\"o}hringer}, H., {Krause}, M., \& {Tr{\"u}mper}, J.
  2012{\natexlab{a}}, \aap, 545, L3

\bibitem[{{Chon} {et~al.}(2013){Chon}, {B{\"o}hringer}, \& {Nowak}}]{chon13}
{Chon}, G., {B{\"o}hringer}, H., \& {Nowak}, N. 2013, \mnras, 429, 3272

\bibitem[{{Chon} {et~al.}(2012{\natexlab{b}}){Chon}, {B{\"o}hringer}, \&
  {Smith}}]{chon12b}
{Chon}, G., {B{\"o}hringer}, H., \& {Smith}, G.~P. 2012{\natexlab{b}}, \aap,
  548, A59

\bibitem[{{Chon} {et~al.}(2015){Chon}, {B{\"o}hringer}, \& {Zaroubi}}]{chon15}
{Chon}, G., {B{\"o}hringer}, H., \& {Zaroubi}, S. 2015, \aap, 575, L14

\bibitem[{{Colberg} {et~al.}(1999){Colberg}, {White}, {Jenkins}, \&
  {Pearce}}]{colberg99}
{Colberg}, J.~M., {White}, S.~D.~M., {Jenkins}, A., \& {Pearce}, F.~R. 1999,
  \mnras, 308, 593

\bibitem[{{Colless} \& {Dunn}(1996)}]{colless96}
{Colless}, M. \& {Dunn}, A.~M. 1996, \apj, 458, 435

\bibitem[{{Contini} {et~al.}(2012){Contini}, {De Lucia}, \&
  {Borgani}}]{contini12}
{Contini}, E., {De Lucia}, G., \& {Borgani}, S. 2012, \mnras, 420, 2978

\bibitem[{{Cupani} {et~al.}(2008){Cupani}, {Mezzetti}, \&
  {Mardirossian}}]{cupani08}
{Cupani}, G., {Mezzetti}, M., \& {Mardirossian}, F. 2008, \mnras, 390, 645

\bibitem[{{Dahle} {et~al.}(2002){Dahle}, {Kaiser}, {Irgens}, {Lilje}, \&
  {Maddox}}]{dahle02}
{Dahle}, H., {Kaiser}, N., {Irgens}, R.~J., {Lilje}, P.~B., \& {Maddox}, S.~J.
  2002, \apjs, 139, 313

\bibitem[{{Danese} {et~al.}(1980){Danese}, {de Zotti}, \& {di
  Tullio}}]{danese80}
{Danese}, L., {de Zotti}, G., \& {di Tullio}, G. 1980, \aap, 82, 322

\bibitem[{{De Lucia} {et~al.}(2004){De Lucia}, {Kauffmann}, {Springel},
  {White}, {Lanzoni}, {Stoehr}, {Tormen}, \& {Yoshida}}]{delucia04}
{De Lucia}, G., {Kauffmann}, G., {Springel}, V., {et~al.} 2004, Mon. Not. R.
  Astron. Soc., 348, 333

\bibitem[{{den Hartog} \& {Katgert}(1996)}]{denhartog96}
{den Hartog}, R. \& {Katgert}, P. 1996, \mnras, 279, 349

\bibitem[{{Diemand} {et~al.}(2004){Diemand}, {Moore}, \& {Stadel}}]{diemand04}
{Diemand}, J., {Moore}, B., \& {Stadel}, J. 2004, \mnras, 352, 535

\bibitem[{{Dressler} \& {Shectman}(1988)}]{dressler88}
{Dressler}, A. \& {Shectman}, S.~A. 1988, \aj, 95, 985

\bibitem[{{Dubinski}(1998)}]{dubinsky98}
{Dubinski}, J. 1998, \apj, 502, 141

\bibitem[{{Dutton} \& {Macci{\`o}}(2014)}]{dutton14}
{Dutton}, A.~A. \& {Macci{\`o}}, A.~V. 2014, \mnras, 441, 3359

\bibitem[{{Eckert} {et~al.}(2015){Eckert}, {Jauzac}, {Shan}, {Kneib}, {Erben},
  {Israel}, {Jullo}, {Klein}, {Massey}, {Richard}, \& {Tchernin}}]{eckert15}
{Eckert}, D., {Jauzac}, M., {Shan}, H., {et~al.} 2015, \nat, 528, 105

\bibitem[{{Einasto} {et~al.}(2012){Einasto}, {Vennik}, {Nurmi}, {Tempel},
  {Ahvensalmi}, {Tago}, {Liivam{\"a}gi}, {Saar}, {Hein{\"a}m{\"a}ki},
  {Einasto}, \& {Mart{\'{\i}}nez}}]{einasto12}
{Einasto}, M., {Vennik}, J., {Nurmi}, P., {et~al.} 2012, \aap, 540, A123

\bibitem[{{Evrard} {et~al.}(2002){Evrard}, {MacFarland}, {Couchman}, {Colberg},
  {Yoshida}, {White}, {Jenkins}, {Frenk}, {Pearce}, {Peacock}, \&
  {Thomas}}]{evrard02}
{Evrard}, A.~E., {MacFarland}, T.~J., {Couchman}, H.~M.~P., {et~al.} 2002,
  Astrophys.~J., 573, 7

\bibitem[{{Evrard} {et~al.}(1993){Evrard}, {Mohr}, {Fabricant}, \&
  {Geller}}]{evrard93}
{Evrard}, A.~E., {Mohr}, J.~J., {Fabricant}, D.~G., \& {Geller}, M.~J. 1993,
  \apjl, 419, L9

\bibitem[{{Faltenbacher} {et~al.}(2008){Faltenbacher}, {Jing}, {Li}, {Mao},
  {Mo}, {Pasquali}, \& {van den Bosch}}]{faltenbacher08}
{Faltenbacher}, A., {Jing}, Y.~P., {Li}, C., {et~al.} 2008, \apj, 675, 146

\bibitem[{{Finoguenov} {et~al.}(2005){Finoguenov}, {B{\"o}hringer}, \&
  {Zhang}}]{finoguenov05}
{Finoguenov}, A., {B{\"o}hringer}, H., \& {Zhang}, Y.-Y. 2005, \aap, 442, 827

\bibitem[{{Flin} \& {Krywult}(2006)}]{flin06}
{Flin}, P. \& {Krywult}, J. 2006, \aap, 450, 9

\bibitem[{{Fo{\"e}x} {et~al.}(2013){Fo{\"e}x}, {Motta}, {Limousin}, {Verdugo},
  {More}, {Cabanac}, {Gavazzi}, \& {Mu{\~n}oz}}]{foex13}
{Fo{\"e}x}, G., {Motta}, V., {Limousin}, M., {et~al.} 2013, \aap, 559, A105

\bibitem[{{Gao} {et~al.}(2012){Gao}, {Navarro}, {Frenk}, {Jenkins}, {Springel},
  \& {White}}]{gao12}
{Gao}, L., {Navarro}, J.~F., {Frenk}, C.~S., {et~al.} 2012, \mnras, 425, 2169

\bibitem[{{Gao} {et~al.}(2004){Gao}, {White}, {Jenkins}, {Stoehr}, \&
  {Springel}}]{gao04}
{Gao}, L., {White}, S.~D.~M., {Jenkins}, A., {Stoehr}, F., \& {Springel}, V.
  2004, Mon. Not. R. Astron. Soc., 355, 819

\bibitem[{{Garilli} {et~al.}(2010){Garilli}, {Fumana}, {Franzetti}, {Paioro},
  {Scodeggio}, {Le F{\`e}vre}, {Paltani}, \& {Scaramella}}]{garilli10}
{Garilli}, B., {Fumana}, M., {Franzetti}, P., {et~al.} 2010, \pasp, 122, 827

\bibitem[{{Gebhardt} {et~al.}(1994){Gebhardt}, {Pryor}, {Williams}, \&
  {Hesser}}]{gebhardt94}
{Gebhardt}, K., {Pryor}, C., {Williams}, T.~B., \& {Hesser}, J.~E. 1994, \aj,
  107, 2067

\bibitem[{{Genel} {et~al.}(2010){Genel}, {Bouch{\'e}}, {Naab}, {Sternberg}, \&
  {Genzel}}]{genel10}
{Genel}, S., {Bouch{\'e}}, N., {Naab}, T., {Sternberg}, A., \& {Genzel}, R.
  2010, \apj, 719, 229

\bibitem[{{Ghigna} {et~al.}(2000){Ghigna}, {Moore}, {Governato}, {Lake},
  {Quinn}, \& {Stadel}}]{ghigna00}
{Ghigna}, S., {Moore}, B., {Governato}, F., {et~al.} 2000, Astrophys.~J., 544,
  616

\bibitem[{{Giocoli} {et~al.}(2008){Giocoli}, {Pieri}, \& {Tormen}}]{giocoli08}
{Giocoli}, C., {Pieri}, L., \& {Tormen}, G. 2008, \mnras, 387, 689

\bibitem[{{Giocoli} {et~al.}(2010){Giocoli}, {Tormen}, {Sheth}, \& {van den
  Bosch}}]{giocoli10}
{Giocoli}, C., {Tormen}, G., {Sheth}, R.~K., \& {van den Bosch}, F.~C. 2010,
  \mnras, 404, 502

\bibitem[{{Giodini} {et~al.}(2013){Giodini}, {Lovisari}, {Pointecouteau},
  {Ettori}, {Reiprich}, \& {Hoekstra}}]{giodini13}
{Giodini}, S., {Lovisari}, L., {Pointecouteau}, E., {et~al.} 2013, \ssr, 177,
  247

\bibitem[{{Girardi} {et~al.}(2006){Girardi}, {Boschin}, \&
  {Barrena}}]{girardi06}
{Girardi}, M., {Boschin}, W., \& {Barrena}, R. 2006, \aap, 455, 45

\bibitem[{{Girardi} {et~al.}(2010){Girardi}, {Boschin}, \&
  {Barrena}}]{girardi10}
{Girardi}, M., {Boschin}, W., \& {Barrena}, R. 2010, \aap, 517, A65

\bibitem[{{Girardi} {et~al.}(1997){Girardi}, {Escalera}, {Fadda}, {Giuricin},
  {Mardirossian}, \& {Mezzetti}}]{girardi97}
{Girardi}, M., {Escalera}, E., {Fadda}, D., {et~al.} 1997, \apj, 482, 41

\bibitem[{{Girardi} {et~al.}(1998){Girardi}, {Giuricin}, {Mardirossian},
  {Mezzetti}, \& {Boschin}}]{girardi98}
{Girardi}, M., {Giuricin}, G., {Mardirossian}, F., {Mezzetti}, M., \&
  {Boschin}, W. 1998, \apj, 505, 74

\bibitem[{{Girardi} {et~al.}(2015){Girardi}, {Mercurio}, {Balestra}, {Nonino},
  {Biviano}, {Grillo}, {Rosati}, {Annunziatella}, {Demarco}, {Fritz}, {Gobat},
  {Lemze}, {Presotto}, {Scodeggio}, {Tozzi}, {Bartosch Caminha}, {Brescia},
  {Coe}, {Kelson}, {Koekemoer}, {Lombardi}, {Medezinski}, {Postman},
  {Sartoris}, {Umetsu}, {Zitrin}, {Boschin}, {Czoske}, {De Lucia}, {Kuchner},
  {Maier}, {Meneghetti}, {Monaco}, {Monna}, {Munari}, {Seitz}, {Verdugo}, \&
  {Ziegler}}]{girardi15}
{Girardi}, M., {Mercurio}, A., {Balestra}, I., {et~al.} 2015, \aap, 579, A4

\bibitem[{{Grillo} {et~al.}(2015){Grillo}, {Suyu}, {Rosati}, {Mercurio},
  {Balestra}, {Munari}, {Nonino}, {Caminha}, {Lombardi}, {De Lucia}, {Borgani},
  {Gobat}, {Biviano}, {Girardi}, {Umetsu}, {Coe}, {Koekemoer}, {Postman},
  {Zitrin}, {Halkola}, {Broadhurst}, {Sartoris}, {Presotto}, {Annunziatella},
  {Maier}, {Fritz}, {Vanzella}, \& {Frye}}]{grillo15}
{Grillo}, C., {Suyu}, S.~H., {Rosati}, P., {et~al.} 2015, \apj, 800, 38

\bibitem[{{Guennou} {et~al.}(2014){Guennou}, {Adami}, {Durret}, {Lima Neto},
  {Ulmer}, {Clowe}, {LeBrun}, {Martinet}, {Allam}, {Annis}, {Basa}, {Benoist},
  {Biviano}, {Cappi}, {Cypriano}, {Gavazzi}, {Halliday}, {Ilbert}, {Jullo},
  {Just}, {Limousin}, {M{\'a}rquez}, {Mazure}, {Murphy}, {Plana}, {Rostagni},
  {Russeil}, {Schirmer}, {Slezak}, {Tucker}, {Zaritsky}, \&
  {Ziegler}}]{guennou14}
{Guennou}, L., {Adami}, C., {Durret}, F., {et~al.} 2014, \aap, 561, A112

\bibitem[{{Heisler} {et~al.}(1985){Heisler}, {Tremaine}, \&
  {Bahcall}}]{heisler85}
{Heisler}, J., {Tremaine}, S., \& {Bahcall}, J.~N. 1985, \apj, 298, 8

\bibitem[{{Huertas-Company} {et~al.}(2009){Huertas-Company}, {Foex}, {Soucail},
  \& {Pell{\'o}}}]{huertas09}
{Huertas-Company}, M., {Foex}, G., {Soucail}, G., \& {Pell{\'o}}, R. 2009,
  Astron.~\&~Astrophys., 505, 83

\bibitem[{{Ilbert} {et~al.}(2006){Ilbert}, {Arnouts}, {McCracken},
  {Bolzonella}, {Bertin}, {Le F{\`e}vre}, {Mellier}, {Zamorani}, {Pell{\`o}},
  {Iovino}, {Tresse}, {Le Brun}, {Bottini}, {Garilli}, {Maccagni}, {Picat},
  {Scaramella}, {Scodeggio}, {Vettolani}, {Zanichelli}, {Adami}, {Bardelli},
  {Cappi}, {Charlot}, {Ciliegi}, {Contini}, {Cucciati}, {Foucaud}, {Franzetti},
  {Gavignaud}, {Guzzo}, {Marano}, {Marinoni}, {Mazure}, {Meneux}, {Merighi},
  {Paltani}, {Pollo}, {Pozzetti}, {Radovich}, {Zucca}, {Bondi}, {Bongiorno},
  {Busarello}, {de La Torre}, {Gregorini}, {Lamareille}, {Mathez}, {Merluzzi},
  {Ripepi}, {Rizzo}, \& {Vergani}}]{ilbert06}
{Ilbert}, O., {Arnouts}, S., {McCracken}, H.~J., {et~al.} 2006,
  Astron.~\&~Astrophys., 457, 841

\bibitem[{{Jauzac} {et~al.}(2016){Jauzac}, {Eckert}, {Schwinn}, {Harvey},
  {Baugh}, {Robertson}, {Bose}, {Massey}, {Owers}, {Ebeling}, {Shan}, {Jullo},
  {Kneib}, {Richard}, {Atek}, {Cl{\'e}ment}, {Egami}, {Israel}, {Knowles},
  {Limousin}, {Natarajan}, {Rexroth}, {Taylor}, \& {Tchernin}}]{jauzac16}
{Jauzac}, M., {Eckert}, D., {Schwinn}, J., {et~al.} 2016, ArXiv e-prints

\bibitem[{{Jee} {et~al.}(2014){Jee}, {Hughes}, {Menanteau}, {Sif{\'o}n},
  {Mandelbaum}, {Barrientos}, {Infante}, \& {Ng}}]{jee14}
{Jee}, M.~J., {Hughes}, J.~P., {Menanteau}, F., {et~al.} 2014, \apj, 785, 20

\bibitem[{{Jeltema} {et~al.}(2005){Jeltema}, {Canizares}, {Bautz}, \&
  {Buote}}]{jeltema05}
{Jeltema}, T.~E., {Canizares}, C.~R., {Bautz}, M.~W., \& {Buote}, D.~A. 2005,
  \apj, 624, 606

\bibitem[{{Jiang} \& {van den Bosch}(2014)}]{jiang14}
{Jiang}, F. \& {van den Bosch}, F.~C. 2014, \mnras, 440, 193

\bibitem[{{Lambas} {et~al.}(1988){Lambas}, {Groth}, \& {Peebles}}]{lambas88}
{Lambas}, D.~G., {Groth}, E.~J., \& {Peebles}, P.~J.~E. 1988, \aj, 95, 996

\bibitem[{{Lemze} {et~al.}(2013){Lemze}, {Postman}, {Genel}, {Ford},
  {Balestra}, {Donahue}, {Kelson}, {Nonino}, {Mercurio}, {Biviano}, {Rosati},
  {Umetsu}, {Sand}, {Koekemoer}, {Meneghetti}, {Melchior}, {Newman}, {Bhatti},
  {Voit}, {Medezinski}, {Zitrin}, {Zheng}, {Broadhurst}, {Bartelmann},
  {Benitez}, {Bouwens}, {Bradley}, {Coe}, {Graves}, {Grillo}, {Infante},
  {Jimenez-Teja}, {Jouvel}, {Lahav}, {Maoz}, {Merten}, {Molino}, {Moustakas},
  {Moustakas}, {Ogaz}, {Scodeggio}, \& {Seitz}}]{lemze13}
{Lemze}, D., {Postman}, M., {Genel}, S., {et~al.} 2013, \apj, 776, 91

\bibitem[{{Limber} \& {Mathews}(1960)}]{limber60}
{Limber}, D.~N. \& {Mathews}, W.~G. 1960, \apj, 132, 286

\bibitem[{{Limousin} {et~al.}(2013){Limousin}, {Morandi}, {Sereno},
  {Meneghetti}, {Ettori}, {Bartelmann}, \& {Verdugo}}]{limousin13}
{Limousin}, M., {Morandi}, A., {Sereno}, M., {et~al.} 2013, \ssr, 177, 155

\bibitem[{{Mamon} {et~al.}(2010){Mamon}, {Biviano}, \& {Murante}}]{MBM10}
{Mamon}, G.~A., {Biviano}, A., \& {Murante}, G. 2010, \aap, 520, A30

\bibitem[{{Mann} \& {Ebeling}(2012)}]{mann12}
{Mann}, A.~W. \& {Ebeling}, H. 2012, \mnras, 420, 2120

\bibitem[{{Markevitch} {et~al.}(2004){Markevitch}, {Gonzalez}, {Clowe},
  {Vikhlinin}, {Forman}, {Jones}, {Murray}, \& {Tucker}}]{markevitch04}
{Markevitch}, M., {Gonzalez}, A.~H., {Clowe}, D., {et~al.} 2004, \apj, 606, 819

\bibitem[{{Martinet} {et~al.}(2016){Martinet}, {Clowe}, {Durret}, {Adami},
  {Acebr{\'o}n}, {Hernandez-Garc{\'{\i}}a}, {M{\'a}rquez}, {Guennou}, {Sarron},
  \& {Ulmer}}]{martinet16}
{Martinet}, N., {Clowe}, D., {Durret}, F., {et~al.} 2016, \aap, 590, A69

\bibitem[{{Maurogordato} {et~al.}(2011){Maurogordato}, {Sauvageot}, {Bourdin},
  {Cappi}, {Benoist}, {Ferrari}, {Mars}, \& {Houairi}}]{maurogordato11}
{Maurogordato}, S., {Sauvageot}, J.~L., {Bourdin}, H., {et~al.} 2011, \aap,
  525, A79

\bibitem[{{Merritt}(1985)}]{merritt85}
{Merritt}, D. 1985, Astrophys.~J., 289, 18

\bibitem[{{Merritt}(1988)}]{merritt88}
{Merritt}, D. 1988, in Astronomical Society of the Pacific Conference Series,
  Vol.~5, The Minnesota lectures on Clusters of Galaxies and Large-Scale
  Structure, ed. J.~M. {Dickey}, 175--196

\bibitem[{{Mohammed} {et~al.}(2016){Mohammed}, {Saha}, {Williams},
  {Liesenborgs}, \& {Sebesta}}]{mohammed16}
{Mohammed}, I., {Saha}, P., {Williams}, L.~L.~R., {Liesenborgs}, J., \&
  {Sebesta}, K. 2016, \mnras, 459, 1698

\bibitem[{{Mohr} {et~al.}(1995){Mohr}, {Evrard}, {Fabricant}, \&
  {Geller}}]{mohr95}
{Mohr}, J.~J., {Evrard}, A.~E., {Fabricant}, D.~G., \& {Geller}, M.~J. 1995,
  \apj, 447, 8

\bibitem[{{Moore} {et~al.}(1999){Moore}, {Quinn}, {Governato}, {Stadel}, \&
  {Lake}}]{moore99}
{Moore}, B., {Quinn}, T., {Governato}, F., {Stadel}, J., \& {Lake}, G. 1999,
  Mon. Not. R. Astron. Soc., 310, 1147

\bibitem[{{Munari} {et~al.}(2013){Munari}, {Biviano}, {Borgani}, {Murante}, \&
  {Fabjan}}]{munari13}
{Munari}, E., {Biviano}, A., {Borgani}, S., {Murante}, G., \& {Fabjan}, D.
  2013, \mnras, 430, 2638

\bibitem[{{Nagai} \& {Kravtsov}(2005)}]{nagai05}
{Nagai}, D. \& {Kravtsov}, A.~V. 2005, Astrophys.~J., 618, 557

\bibitem[{{Neumann} \& {Bohringer}(1997)}]{neumann97}
{Neumann}, D.~M. \& {Bohringer}, H. 1997, \mnras, 289, 123

\bibitem[{{Neumann} {et~al.}(2003){Neumann}, {Lumb}, {Pratt}, \&
  {Briel}}]{neumann03}
{Neumann}, D.~M., {Lumb}, D.~H., {Pratt}, G.~W., \& {Briel}, U.~G. 2003, \aap,
  400, 811

\bibitem[{{Newman} {et~al.}(2013){Newman}, {Treu}, {Ellis}, {Sand}, {Nipoti},
  {Richard}, \& {Jullo}}]{newman13}
{Newman}, A.~B., {Treu}, T., {Ellis}, R.~S., {et~al.} 2013, \apj, 765, 24

\bibitem[{{Niederste-Ostholt} {et~al.}(2010){Niederste-Ostholt}, {Strauss},
  {Dong}, {Koester}, \& {McKay}}]{niederste-ostholt10}
{Niederste-Ostholt}, M., {Strauss}, M.~A., {Dong}, F., {Koester}, B.~P., \&
  {McKay}, T.~A. 2010, \mnras, 405, 2023

\bibitem[{{Oegerle} \& {Hill}(2001)}]{oegerle01}
{Oegerle}, W.~R. \& {Hill}, J.~M. 2001, \aj, 122, 2858

\bibitem[{{Owers} {et~al.}(2009){Owers}, {Couch}, \& {Nulsen}}]{owers09}
{Owers}, M.~S., {Couch}, W.~J., \& {Nulsen}, P.~E.~J. 2009, \apj, 693, 901

\bibitem[{{Owers} {et~al.}(2011){Owers}, {Randall}, {Nulsen}, {Couch}, {David},
  \& {Kempner}}]{owers11}
{Owers}, M.~S., {Randall}, S.~W., {Nulsen}, P.~E.~J., {et~al.} 2011, \apj, 728,
  27

\bibitem[{{Panko} {et~al.}(2009){Panko}, {Juszczyk}, \& {Flin}}]{panko09}
{Panko}, E., {Juszczyk}, T., \& {Flin}, P. 2009, \aj, 138, 1709

\bibitem[{{Pierini} {et~al.}(2008){Pierini}, {Zibetti}, {Braglia},
  {B{\"o}hringer}, {Finoguenov}, {Lynam}, \& {Zhang}}]{pierini08}
{Pierini}, D., {Zibetti}, S., {Braglia}, F., {et~al.} 2008, \aap, 483, 727

\bibitem[{{Pinkney} {et~al.}(1996){Pinkney}, {Roettiger}, {Burns}, \&
  {Bird}}]{pinkney96}
{Pinkney}, J., {Roettiger}, K., {Burns}, J.~O., \& {Bird}, C.~M. 1996, \apjs,
  104, 1

\bibitem[{{Plionis} \& {Basilakos}(2002)}]{plionis02}
{Plionis}, M. \& {Basilakos}, S. 2002, \mnras, 329, L47

\bibitem[{{Ramella} {et~al.}(2007){Ramella}, {Biviano}, {Pisani}, {Varela},
  {Bettoni}, {Couch}, {D'Onofrio}, {Dressler}, {Fasano}, {Kj{\o}rgaard},
  {Moles}, {Pignatelli}, \& {Poggianti}}]{ramella07}
{Ramella}, M., {Biviano}, A., {Pisani}, A., {et~al.} 2007, \aap, 470, 39

\bibitem[{{Read} \& {Ponman}(2003)}]{read03}
{Read}, A.~M. \& {Ponman}, T.~J. 2003, \aap, 409, 395

\bibitem[{{Rines} {et~al.}(2013){Rines}, {Geller}, {Diaferio}, \&
  {Kurtz}}]{rines13}
{Rines}, K., {Geller}, M.~J., {Diaferio}, A., \& {Kurtz}, M.~J. 2013, \apj,
  767, 15

\bibitem[{{Rines} {et~al.}(2005){Rines}, {Geller}, {Kurtz}, \&
  {Diaferio}}]{rines05}
{Rines}, K., {Geller}, M.~J., {Kurtz}, M.~J., \& {Diaferio}, A. 2005, \aj, 130,
  1482

\bibitem[{{Rossetti} {et~al.}(2011){Rossetti}, {Eckert}, {Cavalleri},
  {Molendi}, {Gastaldello}, \& {Ghizzardi}}]{rossetti11}
{Rossetti}, M., {Eckert}, D., {Cavalleri}, B.~M., {et~al.} 2011, \aap, 532,
  A123

\bibitem[{{Ruel} {et~al.}(2014){Ruel}, {Bazin}, {Bayliss}, {Brodwin}, {Foley},
  {Stalder}, {Aird}, {Armstrong}, {Ashby}, {Bautz}, {Benson}, {Bleem},
  {Bocquet}, {Carlstrom}, {Chang}, {Chapman}, {Cho}, {Clocchiatti}, {Crawford},
  {Crites}, {de Haan}, {Desai}, {Dobbs}, {Dudley}, {Forman}, {George},
  {Gladders}, {Gonzalez}, {Halverson}, {Harrington}, {High}, {Holder},
  {Holzapfel}, {Hrubes}, {Jones}, {Joy}, {Keisler}, {Knox}, {Lee}, {Leitch},
  {Liu}, {Lueker}, {Luong-Van}, {Mantz}, {Marrone}, {McDonald}, {McMahon},
  {Mehl}, {Meyer}, {Mocanu}, {Mohr}, {Montroy}, {Murray}, {Natoli},
  {Nurgaliev}, {Padin}, {Plagge}, {Pryke}, {Reichardt}, {Rest}, {Ruhl},
  {Saliwanchik}, {Saro}, {Sayre}, {Schaffer}, {Shaw}, {Shirokoff}, {Song}, {{\v
  S}uhada}, {Spieler}, {Stanford}, {Staniszewski}, {Starsk}, {Story}, {Stubbs},
  {van Engelen}, {Vanderlinde}, {Vieira}, {Vikhlinin}, {Williamson}, {Zahn}, \&
  {Zenteno}}]{ruel14}
{Ruel}, J., {Bazin}, G., {Bayliss}, M., {et~al.} 2014, \apj, 792, 45

\bibitem[{{Santos} {et~al.}(2008){Santos}, {Rosati}, {Tozzi}, {B{\"o}hringer},
  {Ettori}, \& {Bignamini}}]{santos08}
{Santos}, J.~S., {Rosati}, P., {Tozzi}, P., {et~al.} 2008, \aap, 483, 35

\bibitem[{{Saro} {et~al.}(2013){Saro}, {Mohr}, {Bazin}, \& {Dolag}}]{saro13}
{Saro}, A., {Mohr}, J.~J., {Bazin}, G., \& {Dolag}, K. 2013, \apj, 772, 47

\bibitem[{{Schirmer}(2013)}]{schirmer13}
{Schirmer}, M. 2013, \apjs, 209, 21

\bibitem[{{Schuecker} {et~al.}(2001){Schuecker}, {B{\"o}hringer}, {Reiprich},
  \& {Feretti}}]{schuecker01}
{Schuecker}, P., {B{\"o}hringer}, H., {Reiprich}, T.~H., \& {Feretti}, L. 2001,
  \aap, 378, 408

\bibitem[{{Schwinn} {et~al.}(2016){Schwinn}, {Jauzac}, {Baugh}, {Bartelmann},
  {Eckert}, {Harvey}, {Natarajan}, \& {Massey}}]{schwinn16}
{Schwinn}, J., {Jauzac}, M., {Baugh}, C.~M., {et~al.} 2016, ArXiv e-prints

\bibitem[{{Scodeggio} {et~al.}(2005){Scodeggio}, {Franzetti}, {Garilli},
  {Zanichelli}, {Paltani}, {Maccagni}, {Bottini}, {Le Brun}, {Contini},
  {Scaramella}, {Adami}, {Bardelli}, {Zucca}, {Tresse}, {Ilbert}, {Foucaud},
  {Iovino}, {Merighi}, {Zamorani}, {Gavignaud}, {Rizzo}, {McCracken}, {Le
  F{\`e}vre}, {Picat}, {Vettolani}, {Arnaboldi}, {Arnouts}, {Bolzonella},
  {Cappi}, {Charlot}, {Ciliegi}, {Guzzo}, {Marano}, {Marinoni}, {Mathez},
  {Mazure}, {Meneux}, {Pell{\`o}}, {Pollo}, {Pozzetti}, \&
  {Radovich}}]{scodeggio05}
{Scodeggio}, M., {Franzetti}, P., {Garilli}, B., {et~al.} 2005, \pasp, 117,
  1284

\bibitem[{{Sebesta} {et~al.}(2016){Sebesta}, {Williams}, {Mohammed}, {Saha}, \&
  {Liesenborgs}}]{sebasta16}
{Sebesta}, K., {Williams}, L.~L.~R., {Mohammed}, I., {Saha}, P., \&
  {Liesenborgs}, J. 2016, \mnras, 461, 2126

\bibitem[{{Silverman}(1986)}]{silverman86}
{Silverman}, B.~W. 1986, {Density estimation for statistics and data analysis}

\bibitem[{{Smith} {et~al.}(2010){Smith}, {Khosroshahi}, {Dariush}, {Sanderson},
  {Ponman}, {Stott}, {Haines}, {Egami}, \& {Stark}}]{smith10}
{Smith}, G.~P., {Khosroshahi}, H.~G., {Dariush}, A., {et~al.} 2010, \mnras,
  409, 169

\bibitem[{{Smith} {et~al.}(2005){Smith}, {Kneib}, {Smail}, {Mazzotta},
  {Ebeling}, \& {Czoske}}]{smith05}
{Smith}, G.~P., {Kneib}, J.-P., {Smail}, I., {et~al.} 2005, \mnras, 359, 417

\bibitem[{{Solanes} {et~al.}(1999){Solanes}, {Salvador-Sol{\'e}}, \&
  {Gonz{\'a}lez-Casado}}]{solanes99}
{Solanes}, J.~M., {Salvador-Sol{\'e}}, E., \& {Gonz{\'a}lez-Casado}, G. 1999,
  \aap, 343, 733

\bibitem[{{Soucail} {et~al.}(2015){Soucail}, {Fo{\"e}x}, {Pointecouteau},
  {Arnaud}, \& {Limousin}}]{soucail15}
{Soucail}, G., {Fo{\"e}x}, G., {Pointecouteau}, E., {Arnaud}, M., \&
  {Limousin}, M. 2015, \aap, 581, A31

\bibitem[{{Springel} {et~al.}(2006){Springel}, {Frenk}, \&
  {White}}]{springel06}
{Springel}, V., {Frenk}, C.~S., \& {White}, S.~D.~M. 2006, \nat, 440, 1137

\bibitem[{{Springel} {et~al.}(2005){Springel}, {White}, {Jenkins}, {Frenk},
  {Yoshida}, {Gao}, {Navarro}, {Thacker}, {Croton}, {Helly}, {Peacock}, {Cole},
  {Thomas}, {Couchman}, {Evrard}, {Colberg}, \& {Pearce}}]{springel05}
{Springel}, V., {White}, S.~D.~M., {Jenkins}, A., {et~al.} 2005, Nature, 435,
  629

\bibitem[{{Stott} {et~al.}(2009){Stott}, {Pimbblet}, {Edge}, {Smith}, \&
  {Wardlow}}]{stott09}
{Stott}, J.~P., {Pimbblet}, K.~A., {Edge}, A.~C., {Smith}, G.~P., \& {Wardlow},
  J.~L. 2009, Mon. Not. R. Astron. Soc., 394, 2098

\bibitem[{{Taylor} \& {Babul}(2004)}]{taylorJE04}
{Taylor}, J.~E. \& {Babul}, A. 2004, \mnras, 348, 811

\bibitem[{{Treu} {et~al.}(2003){Treu}, {Ellis}, {Kneib}, {Dressler}, {Smail},
  {Czoske}, {Oemler}, \& {Natarajan}}]{treu03}
{Treu}, T., {Ellis}, R.~S., {Kneib}, J., {et~al.} 2003, Astrophys.~J., 591, 53

\bibitem[{{van den Bosch} {et~al.}(2005){van den Bosch}, {Tormen}, \&
  {Giocoli}}]{vandenbosch05}
{van den Bosch}, F.~C., {Tormen}, G., \& {Giocoli}, C. 2005, \mnras, 359, 1029

\bibitem[{{van der Marel} \& {Franx}(1993)}]{vandermarel93}
{van der Marel}, R.~P. \& {Franx}, M. 1993, \apj, 407, 525

\bibitem[{{Verdugo} {et~al.}(2016){Verdugo}, {Limousin}, {Motta}, {Mamon},
  {Fo{\"e}x}, {Gastaldello}, {Jullo}, {Biviano}, {Rojas}, {Mu{\~n}oz},
  {Cabanac}, {Maga{\~n}a}, {Fern{\'a}ndez-Trincado}, {Adame}, \& {De
  Leo}}]{verdugo16}
{Verdugo}, T., {Limousin}, M., {Motta}, V., {et~al.} 2016, ArXiv e-prints

\bibitem[{{Wen} \& {Han}(2013)}]{wen13}
{Wen}, Z.~L. \& {Han}, J.~L. 2013, \mnras, 436, 275

\bibitem[{{Wojtak} {et~al.}(2007){Wojtak}, {{\L}okas}, {Mamon},
  {Gottl{\"o}ber}, {Prada}, \& {Moles}}]{wojtak07}
{Wojtak}, R., {{\L}okas}, E.~L., {Mamon}, G.~A., {et~al.} 2007, \aap, 466, 437

\bibitem[{{Yahil} \& {Vidal}(1977)}]{yahil77}
{Yahil}, A. \& {Vidal}, N.~V. 1977, \apj, 214, 347

\bibitem[{{Zabludoff} {et~al.}(1993){Zabludoff}, {Franx}, \&
  {Geller}}]{zabludoff93}
{Zabludoff}, A.~I., {Franx}, M., \& {Geller}, M.~J. 1993, \apj, 419, 47

\bibitem[{{Zenteno} {et~al.}(2011){Zenteno}, {Song}, {Desai}, {Armstrong},
  {Mohr}, {Ngeow}, {Barkhouse}, {Allam}, {Andersson}, {Bazin}, {Benson},
  {Bertin}, {Brodwin}, {Buckley-Geer}, {Hansen}, {High}, {Lin}, {Lin}, {Liu},
  {Rest}, {Smith}, {Stalder}, {Stark}, {Tucker}, \& {Yang}}]{zenteno11}
{Zenteno}, A., {Song}, J., {Desai}, S., {et~al.} 2011, \apj, 734, 3

\bibitem[{{Zhang} {et~al.}(2005){Zhang}, {B{\"o}hringer}, {Finoguenov},
  {Ikebe}, {Matsushita}, {Schuecker}, {Guzzo}, \& {Collins}}]{zhang05}
{Zhang}, Y.-Y., {B{\"o}hringer}, H., {Finoguenov}, A., {et~al.} 2005, Advances
  in Space Research, 36, 667

\bibitem[{{Zhang} {et~al.}(2006){Zhang}, {B{\"o}hringer}, {Finoguenov},
  {Ikebe}, {Matsushita}, {Schuecker}, {Guzzo}, \& {Collins}}]{zhang06}
{Zhang}, Y.-Y., {B{\"o}hringer}, H., {Finoguenov}, A., {et~al.} 2006, \aap,
  456, 55

\bibitem[{{Zhang} {et~al.}(2004){Zhang}, {Finoguenov}, {B{\"o}hringer},
  {Ikebe}, {Matsushita}, \& {Schuecker}}]{zhang04b}
{Zhang}, Y.-Y., {Finoguenov}, A., {B{\"o}hringer}, H., {et~al.} 2004, \aap,
  413, 49

\bibitem[{{Ziparo} {et~al.}(2012){Ziparo}, {Braglia}, {Pierini}, {Finoguenov},
  {B{\"o}hringer}, \& {Bongiorno}}]{ziparo12}
{Ziparo}, F., {Braglia}, F.~G., {Pierini}, D., {et~al.} 2012, \mnras, 420, 2480

\end{thebibliography}

\begin{table*}
\centering
\begin{threeparttable}
\caption{Sky position and redshift of the cluster members.}
\label{table:app}
\begin{tabular}{l c c c c c c c}
\hline\hline\noalign{\smallskip}
Cluster & RA & DEC & $z_{s}$ & F$_{\mathrm{EZ}}$ & $z_{p}$ & $\delta z_{p}$ & F$_{\mathrm{RS}}$\\
& (J2000) & (J2000) & & & & \\
\noalign{\smallskip}\hline\noalign{\smallskip}
RXCJ0225 & 2:26:20.5 & -42:2:59.7 & 0.2241 & 4 & 0.2241 & - & 1\\
RXCJ0225 & 2:25:27.8 & -42:2:57.4 & - & - & 0.23 & 0.03 & 1\\
RXCJ0225 & 2:24:58.4 & -42:2:55.9 & - & - & 0.25 & 0.05 & 1\\
RXCJ0225 & 2:25:05.9 & -42:2:54.5 & - & - & 0.29 & 0.08 & 0\\
RXCJ0225 & 2:24:42.7 & -42:2:52.9 & - & - & 0.21 & 0.01 & 1\\
\noalign{\smallskip}\hline
\end{tabular}
    \begin{tablenotes}
      \small
      \item Columns: (1) Cluster host. (2,3) Equatorial coordinates of the galaxy. (4) VIMOS spectroscopic redshift. (5) EZ flag of the spectroscopic redshift estimate. (6) WFI photometric redshift, equal to the spectroscopic value when available. (7) Uncertainty of the photometric redshift. (8) Red-sequence membership flag. F$_{\mathrm{RS}}=1$ for the red-sequence galaxies, F$_{\mathrm{RS}}=0$ otherwise.
    \end{tablenotes}
  \end{threeparttable}
\end{table*}

\end{document}